\documentclass[aps,onecolumn,showpacs]{revtex4}
\usepackage{dcolumn}
\usepackage{graphicx}
\usepackage{amsmath}
\usepackage{amsfonts}
\usepackage{amssymb}
\usepackage{psfrag}
\usepackage{wrapfig}
\usepackage{subfigure}
\usepackage{makeidx}
\usepackage{bm}
\usepackage{epsf}
\usepackage{color}

 \newtheorem{lem}{Lemma}
 \newtheorem{prop}{Proposition}
\newtheorem{rem}{Remark}

\begin{document}

\title{General soliton solutions to a coupled Fokas-Lenells equation}

\author{Liming Ling$^{1}$}
\author{Bao-Feng Feng$^{2}$}
\author{Zuonong Zhu$^{3}$}

\affiliation{$^1$ School of Mathematics, South China University of Technology,
Guangzhou 510640, China}
\affiliation{$^2$ School of Mathematical and Statistical Sciences, The University of
Texas Rio Grande Valley, Edinburg Texas, 78541, USA}
\affiliation{$^3$ School of Mathematical Sciences, Shanghai Jiaotong University, Shanghai, China}

\begin{abstract}
In this paper, we firstly establish the multi-Hamiltonian structure and infinite many conservation laws for the vector Kaup-Newell hierarchy of the positive and negative orders.
The first nontrivial negative flow corresponds to a coupled Fokas-Lenells equation. By constructing a generalized Darboux transformation and using a limiting process, all kinds of one-soliton solutions are constructed including the bright-dark soliton, the dark-anti-dark soliton and the breather-like solutions. Furthermore, multi-bright and multi-dark soliton solutions are derived and their asymptotic behaviors are investigated.
\end{abstract}

\maketitle

\section{Introduction}
It is well known that the integrable nonlinear Schr\"odinger type equations, such as the classical
nonlinear Schr\"odinger (NLS) equation\cite{shabat-zakharov}, derivative-type NLS equation \cite{kuap-newell,chen-lee-liu,GVequation},
play an important role in the study of nonlinear wave propagation. Recently, a new integrable model, called the Fokas-Lenells (FL) equation \cite{lenells,lenells2}, was proposed in mono-mode optical fibers when certain higher order nonlinear effects were taken into account. The Hamiltonian structure and inverse scattering transformation for the FL equation were established by Fokas and Lenells in their original paper \cite{lenells}. The algebraic geometry solution was construct in \cite{fan13}. The rogue wave solutions for this system were constructed by  several authors \cite{chensihua,he1}. It is interesting that, unlike the NLS equation, the FL equation admits both the bright and dark soliton solutions without the sign change of nonlinear term. The bright soliton solutions were constructed by Hirota's bilinear method \cite{MatsunoFL1}, while the dark soliton solutions were constructed by Hirota's bilinear method \cite{Matsuno} and  by B\"acklund transformation \cite{Vekslerchik1}, respectively.

As pointed out by Lenells, the FL equation is related to the derivative NLS equation. As a matter of fact, there are lots of studies regarding the derivative NLS equation such as the Hamiltonian structure\cite{WenxiuMa}, the inverse scattering method \cite{ChenXJ,ChenXJ1,lenells1}, the Darboux/B\"acklund transformation\cite{Steudel,xiaoy,Veksler,guo2,wen} and infinite many conservation laws \cite{Luxing,lu1}. Meanwhile, there are quite a few works for the study of multi-component derivative NLS equations such as the Lax pair and its integrable properties \cite{fordy,Tsuchida,Tsuchida1} and exact solvable methods \cite{Morris,ling1,guo1}.
Similar to the case of the NLS equation \cite{Manakov1974}, it is necessary to consider the two-component or multi-component generalizations of the FL equation for describing the effects of polarization or anisotropy. Most recently, the coupled FL equation has been studied by several authors \cite{guo1,MXZhang,ZhangChow}.

Darboux transformation(DT) is a useful method to construct exact solutions for the integrable system \cite{Matveev,Gu,loop-group}. Actually, the DT is a method related to the inverse scattering method which can be used to solve the initial value problem of integrable equations. Recently, there are some progresses to use the DT to construct some more general analytical solutions for NLS-type equations \cite{ling1,ling1b,ling2,ling3,ling4,feng1}. However, in certain physical situations, two or more wave packets of different carrier frequencies appear simultaneously, and their interactions are governed by the coupled equations. Thus, a natural question is how to construct the generalized Darboux transformation and apply it to find exact solutions to the coupled integrable equations.

In this paper, we consider a coupled Fokas-Lenells equation
\begin{equation}\label{cFL}
  \begin{split}
     u_{1,xt}+u_1+{\rm i}(|u_1|^2+\frac{1}{2}\sigma |u_2|^2)u_{1,x}+\frac{{\rm i}}{2}\sigma u_1u_2^*u_{2,x}=&0,  \\
     u_{2,xt}+u_2+{\rm i}(\sigma |u_2|^2+\frac{1}{2}|u_1|^2)u_{2,x}+\frac{{\rm i}}{2}u_2u_1^*u_{1,x}=&0
  \end{split}
\end{equation}
where $\sigma=\pm 1$. This equation was first proposed by Guo and Ling \cite{guo1} with the matrix generalization of Lax pair. 
In \cite{MXZhang}, this coupled FL equation was reconstructed by the spectral gradient method. Zhang et. al constructed the bright, breather and first-order rogue wave solutions to a coupled FL equation, which is shown in \cite{guo1} to be  equivalent to the coupled FL equation (\ref{cFL}) via a gauge transformation. Since the FL equation admits both the bright and dark soliton solution simultaneously, the structure of the soliton solutions to the coupled FL equation (\ref{cFL}) is expected to be more complicated. It is the main objective of the present paper to explore the rich structure of soliton solutions to the coupled FL equation.

The Tu scheme \cite{tuguizhang,xiat,fan1} is an important technique to construct the Hamiltonian structure for the integrable hierarchy. In section \ref{par2}, we will apply the Tu scheme to construct the multi-Hamiltonian structure and the infinite many conservation laws for the coupled FL equation.
First, Hamiltonian structures for the vector Kaup-Newell spectral problem involving both the positive and negative hierarchy are constructed. The first negative flow of the hierarchy is nothing but the coupled FL equation. Based on the Lax pair, the infinite many conservation laws are constructed in both positive and negative orders. In section \ref{par3}, we construct a generalized DT for the coupled FL equation (\ref{cFL}). Then the general soliton solution to the coupled FL equation is constructed by the
DT method. In section \ref{par4}, based on the general soliton solution, a variety of single soliton solution from the zero and plane wave seed solutions are constructed and classified. These solutions include the bright soliton, the bright-dark soliton,
the bright-anti-dark soliton, the dark-dark (D-D) soliton, the dark-anti-dark (D-AD) soliton, the anti-dark-anti-dark (AD-AD) soliton, the rational D-D/D-AD/AD-AD soliton, and the breather-like solution with nonzero boundary conditions. In section \ref{par5}, we construct the multi-bright and multi-dark/anti-dark soliton solutions and perform its asymptotical analysis. The paper is concluded in Section VII by a brief summary and discussions.

\section{Hamiltonian structure and Conservation laws for the multi-component Kaup-Newell spectral problem}\label{par2}
It is well known that the important criterions for the integrable hierarchy are multi-Hamiltonian structure and the infinitely many conservation laws. Firstly, we consider the Hamiltonian structure for the multi-component Kaup-Newell hierarchy.

\subsection{Multi-Hamiltonian structure}
We consider the following vector Kaup-Newell spectral problem \cite{kuap-newell}
\begin{equation}\label{multi-KN0}
    \Psi_x=\left({\rm i}\lambda^{-2}\sigma_3+\lambda^{-1}Q\right)\Psi\equiv U(\lambda)\Psi,\,\,\, \Psi= \begin{pmatrix}
      \psi_1 \\
      \psi_2 \\
    \end{pmatrix},\,\, \sigma_3=\begin{pmatrix}
                                  1 & 0 \\
                                  0 & -I_{N} \\
                                \end{pmatrix},\,\, Q=\begin{pmatrix}
                                  0 & \mathbf{r}^{\mathrm{T}}\\
                                  \mathbf{q} & 0 \\
                                \end{pmatrix},
\end{equation}
where $\mathbf{q}=(q_1,q_2,\cdots,q_N)^{\mathrm{T}}$, $\mathbf{r}=(r_1,r_2,\cdots,r_N)^{\mathrm{T}}$, $I_N$ denotes a $N \times N$  identity matrix, $\lambda$ is the spectral parameter.
To obtain the soliton hierarchy and its Hamiltonian structure, we consider the stationary zero-curvature equation
\begin{equation}\label{station}
    W_x=[U,W],\,\,\,\, W=\begin{pmatrix}
        a & \mathbf{c}^{\mathrm{T}} \\
        \mathbf{b} & \mathbf{d} \\
      \end{pmatrix},
\end{equation}
where $\mathbf{b}$ and $\mathbf{c}$ are $N\times 1$ matrices, $\mathbf{d}$ is a $N\times N$ matrix.
The  equation \eqref{station} can be expanded as
\begin{eqnarray}
     a_x&=\lambda^{-1}(\mathbf{r}^{\mathrm{T}} \mathbf{b}-\mathbf{c}^{\mathrm{T}} \mathbf{q})\,\,\,\,\,\,\,\,\,\,\,\,\,\,\,\,\,\,\,& \mathbf{d}_x=
       \lambda^{-1}(\mathbf{q} \mathbf{c}^{\mathrm{T}}-\mathbf{b} \mathbf{r}^{\mathrm{T}}),\label{station1} \\
    \mathbf{b}_x&=-2{\rm i}\lambda^{-2}\mathbf{b}+\lambda^{-1}(a\mathbf{q}-\mathbf{d}\mathbf{q})\,\,,&\mathbf{c}_x=2{\rm i}\lambda^{-2}\mathbf{c}+\lambda^{-1}(\mathbf{d}^{\mathrm{T}}\mathbf{r}-
       a\mathbf{r}).\label{station2}
\end{eqnarray}
Firstly, we consider the positive hierarchy. 
To obtain the recursion operator for the positive hierarchy, we insert equations \eqref{station2} into equations \eqref{station1} to obtain
\begin{equation}\label{station11}
\begin{split}
    a_x & =\frac{{\rm i}}{2}\lambda(\mathbf{r}^{\mathrm{T}} \mathbf{b}_x+\mathbf{c}_x^{\mathrm{T}} \mathbf{q}),  \\
   \mathbf{d}_x&=-\frac{{\rm i}\lambda}{2}(\mathbf{q} \mathbf{c}_x^{\mathrm{T}}+\mathbf{b}_x\mathbf{r}^{\mathrm{T}})+\frac{{\rm i}}{2}
   \left((\mathbf{q}\mathbf{r}^{\mathrm{T}})\mathbf{d}-\mathbf{d}(\mathbf{q} \mathbf{r}^{\mathrm{T}})\right).
\end{split}
\end{equation}
Then inserting equation \eqref{station11} into equation \eqref{station2} yields
\begin{equation*}
    \begin{split}
       {\rm i}\lambda^{-2}\mathbf{b}=&-\frac{1}{2}\mathbf{b}_x+\frac{1}{2}\lambda^{-1}a \mathbf{q} -\frac{1}{2}\lambda^{-1}\mathbf{d} \mathbf{q},  \\
       {\rm i}\lambda^{-2}\mathbf{c}=&\frac{1}{2}\mathbf{c}_x-\frac{1}{2}\lambda^{-1}\mathbf{d}\mathbf{r}+
       \frac{1}{2}\lambda^{-1}a\mathbf{r}.
    \end{split}
\end{equation*}
Finally, supposing $W$ to be of the form
\begin{equation}\label{wexpand}
    \begin{split}
       a=&\sum_{i=0}^{\infty}a_i\lambda^{2i},\,\,\mathbf{d}=\sum_{i=0}^{\infty}\mathbf{d}_i\lambda^{2i},  \\
       \mathbf{b}=&\sum_{i=0}^{\infty}\mathbf{b}_i\lambda^{2i+1},\,\, \mathbf{c}=\sum_{i=0}^{\infty}\mathbf{c}_i\lambda^{2i+1},
    \end{split}
\end{equation}
we can obtain that
\begin{equation}
    \begin{split}
        &a_{0}=2\alpha_0,\,\, \mathbf{d}_{0}=-2\alpha_0I_N,\,\, b_0=\mathbf{q}\alpha_0,\,\, \mathbf{c}_{0}=\mathbf{r}\alpha_0, \\
        &a_{i+1,x} =\frac{{\rm i}}{2}(\mathbf{r}^{\mathrm{T}} \mathbf{b}_{i,x}+\mathbf{c}_{i,x}^{\mathrm{T}} \mathbf{q}),\,\, \mathrm{d}_{i+1,x}-\frac{{\rm i}}{2}\mathrm{ad}_{\mathbf{q}\mathbf{r}^{\mathrm{T}}}\mathbf{d}_{i+1} =-\frac{{\rm i}}{2}(\mathbf{b}_{i,x}\mathbf{r}^{\mathrm{T}} +\mathbf{q} \mathbf{c}_{i,x}^{\mathrm{T}}),\\
        &{\rm i}\mathbf{b}_{i+1}=-\frac{1}{2}\mathbf{b}_{i,x}+\frac{1}{2}(\mathbf{q} a_{i+1}-\mathbf{d}_{i+1} \mathbf{q}),\,\, {\rm i}\mathbf{c}_{i+1}=\frac{1}{2}\mathbf{c}_{i,x}-\frac{1}{2}(\mathbf{d}_{i+1}^{\mathrm{T}} \mathbf{r}-\mathbf{r} a_{i+1}),
    \end{split}
\end{equation}
where $\alpha_0$ is a complex constant, $\mathrm{ad}_{\mathbf{q}\mathbf{r}^{\mathrm{T}}}(\cdot)=[\mathbf{q}\mathbf{r}^{\mathrm{T}},\cdot]$. The term $-\frac{{\rm i}}{2}\mathrm{ad}_{\mathbf{q}\mathbf{r}^{\mathrm{T}}}\mathbf{d}_{i+1}$ does not occur in the scalar equation since it is commutative automatically. Based on the above equations, one can obtain the recursion relation
\begin{equation*}
    \begin{pmatrix}
      \mathbf{b}_{i+1} \\
      \mathbf{c}_{i+1} \\
    \end{pmatrix}=L_1 \partial_x\begin{pmatrix}
      \mathbf{b}_{i} \\
      \mathbf{c}_{i} \\
    \end{pmatrix},
\end{equation*}
where
\begin{equation*}
    L_1=\frac{{\rm i}}{2}\begin{pmatrix}
                                              I_N & 0 \\[5pt]
                                              0 & -I_{N} \\
                                            \end{pmatrix}+\frac{1}{4}\begin{pmatrix}
                                                                        \mathbf{q} \partial_x^{-1} \mathbf{r}^{\mathrm{T}}+A & \mathbf{q} \partial_x^{-1} \mathbf{q}^{\mathrm{T}}+C \\[5pt]
                                                                        \mathbf{r}\partial_x^{-1}\mathbf{r}^{\mathrm{T}}+D & \mathbf{r}\partial_x^{-1}\mathbf{q}^{\mathrm{T}}+B \\
                                                                     \end{pmatrix}\,,
\end{equation*}
and
\begin{equation*}
    \begin{split}
       A=&[(\partial_x-\frac{{\rm i}}{2}\mathrm{ad}_{\mathbf{q}\mathbf{r}^{\mathrm{T}}})^{-1}((\cdot)\mathbf{r}^{\mathrm{T}} )]\mathbf{q},\,\,\, B=[(\partial_x+\frac{{\rm i}}{2}\mathrm{ad}_{\mathbf{r}\mathbf{q}^{\mathrm{T}}})^{-1}((\cdot) \mathbf{q}^{\mathrm{T}})]\mathbf{r},  \\
      C= & [(\partial_x-\frac{{\rm i}}{2}\mathrm{ad}_{\mathbf{q}\mathbf{r}^{\mathrm{T}}})^{-1}(\mathbf{q} (\cdot)^{\mathrm{T}})]\mathbf{q},\,\,\, D=[(\partial_x+\frac{{\rm i}}{2}\mathrm{ad}_{\mathbf{r}\mathbf{q}^{\mathrm{T}}})^{-1}(\mathbf{r} (\cdot)^{\mathrm{T}})]\mathbf{r}\,.
    \end{split}
\end{equation*}
\begin{prop}
\begin{equation*}
  A^{\ast}=-B,\,\, C^*=-C,\,\, D^*=-D.
\end{equation*}
\end{prop}
\textbf{Proof:} By direct calculation, we have
\begin{equation*}
    \begin{split}
        &\int\mathbf{c}^{\mathrm{T}}[(\partial_x-\frac{{\rm i}}{2}\mathrm{ad}_{\mathbf{q}\mathbf{r}^{\mathrm{T}}})(\mathbf{b}\mathbf{r}^{\mathrm{T}} )]\mathbf{q}\mathrm{d}x   \\
       =&\int\mathbf{c}^{\mathrm{T}} [(\mathbf{b} \mathbf{r}^{\mathrm{T}} )_x-\frac{{\rm i}}{2}\mathrm{ad}_{\mathbf{q} \mathbf{r}^{\mathrm{T}}}(\mathbf{b} \mathbf{r}^{\mathrm{T}} )]  \mathbf{q}\mathrm{d}x\\
       =&\int\{-\mathbf{c}_x^{\mathrm{T}}  (\mathbf{b} \mathbf{r}^{\mathrm{T}} ) \mathbf{q}-\mathbf{c}^{\mathrm{T}}  (\mathbf{b} \mathbf{r}^{\mathrm{T}} ) \mathbf{q}_x -\frac{{\rm i}}{2}\mathbf{c}^{\mathrm{T}} [\mathrm{ad}_{\mathbf{q} \mathbf{r}^{\mathrm{T}}}(\mathbf{b} \mathbf{r}^{\mathrm{T}} )]  \mathbf{q}\}\mathrm{d}x\\
       =&\int\mathbf{b}^{\mathrm{T}}  [(-\partial_x-\frac{{\rm i}}{2}\mathrm{ad}_{\mathbf{r} \mathbf{q}^{\mathrm{T}}})(\mathbf{c} \mathbf{q}^{\mathrm{T}} )]  \mathbf{r}\mathrm{d}x\,.
    \end{split}
\end{equation*}
Thus, we can obtain that $A^{\ast}=-B$ formally. Other two cases can be proved in a similar way. $\square$

From the trace identity, one has
\begin{equation}\label{trace}
    \frac{\delta}{\delta\omega }\int \mathrm{tr}\left(W\frac{\partial U}{\partial \lambda}\right)\mathrm{d}x=\lambda^{-\gamma}\frac{\partial}{\partial \lambda}\left[\lambda^{\gamma}\mathrm{tr}(W\frac{\partial U}{\partial \omega})\right],\,\, \quad  \omega=\begin{pmatrix}
             \mathbf{q} \\
             \mathbf{r} \\
           \end{pmatrix}\,.
\end{equation}
Since
\begin{equation*}
\begin{split}
   \mathrm{tr}\left(W\frac{\partial U}{\partial \lambda}\right) & =-2{\rm i}\lambda^{-3}[a-\mathrm{tr}(\mathbf{d})]-\lambda^{-2}(\mathbf{c}^{\mathrm{T}} \mathbf{q}+\mathbf{r}^{\mathrm{T}} \mathbf{b})  \\
    & =-2{\rm i}[a_{0}-\mathrm{tr}(\mathbf{d}_{0})]-\sum_{i=0}^{\infty}\left[2{\rm i}[a_{i+1}-\mathrm{tr}(\mathbf{d}_{i+1})]+(\mathbf{c}_{i}^{\mathrm{T}} \mathbf{q}+\mathbf{r}^{\mathrm{T}}\mathbf{b}_{i}) \right]\lambda^{2i-1}\,,
\end{split}
\end{equation*}
and
\begin{equation*}
\begin{split}
  \mathrm{tr}(W\frac{\partial U}{\partial \omega}) & =\lambda^{-1}\begin{pmatrix}
             \mathbf{c} \\
             \mathbf{b} \\
           \end{pmatrix}\,,
\end{split}
\end{equation*}
it then follows that $$\begin{pmatrix}
             \mathbf{c}_0 \\
             \mathbf{b}_0 \\
           \end{pmatrix}=\frac{\delta}{\delta\omega}H_0^{+},\,\,H_0^{+}=\alpha_0 \int \mathbf{q}^{\mathrm{T}}\cdot \mathbf{r}\mathrm{d}x$$
and
\begin{equation*}
    \begin{pmatrix}
             \mathbf{c}_i \\
             \mathbf{b}_i \\
           \end{pmatrix}=\frac{\delta}{\delta\omega}H_i^{+},\,\, H_i^{+}=-\frac{1}{2i}\int\left[2{\rm i}(a_{i+1}-\mathrm{tr}(\mathbf{d}_{i+1}))+(\mathbf{c}_{i}^{\mathrm{T}} \mathbf{q}+\mathbf{r}^{\mathrm{T}}\mathbf{b}_{i})\right] \mathrm{d}x,\,\, i\geq 1\,.
\end{equation*}
Finally, setting $V(\lambda)=(\lambda^{-2k}W)_-$, we obtain the multi-Hamiltonian hierarchy
\begin{equation*}
   \omega_t=J\frac{\delta}{\delta\omega}H_n^{+}=JL^{n}\frac{\delta}{\delta\omega}H_0^{+}\,,
\end{equation*}
from the zero curvature condition $U_t-V_x+[U,V]=0$,  where
\begin{equation*}
    J=\begin{pmatrix}
        0 & \partial \\
        \partial & 0 \\
      \end{pmatrix},\,\, L=\begin{pmatrix}
        0 & I_N \\
        I_N & 0 \\
      \end{pmatrix}L_1\begin{pmatrix}
        0 & I_N \\
        I_N & 0 \\
      \end{pmatrix}\partial.
\end{equation*}
It is readily to see that the operators $J$ and $JL^{k}$ are skew symmetrical.

On the other hand, by assuming
\begin{equation*}
     \begin{split}
       a=&\sum_{i=0}^{\infty}a_i\lambda^{-2i},\,\,\mathbf{d}=\sum_{i=0}^{\infty}\mathbf{d}_i\lambda^{-2i},  \\
       \mathbf{b}=&\sum_{i=0}^{\infty}\mathbf{b}_i\lambda^{-2i-1},\,\, \mathbf{c}=\sum_{i=0}^{\infty}\mathbf{c}_i\lambda^{-2i-1}\,,
    \end{split}
\end{equation*}
one obtains the recursion relation
\begin{equation*}
    \begin{pmatrix}
      \mathbf{b}_i \\
      \mathbf{c}_i \\
    \end{pmatrix}_x=L_2\begin{pmatrix}
      \mathbf{b}_{i-1} \\
      \mathbf{c}_{i-1} \\
    \end{pmatrix},\,\, i\geq 1,\,\, \begin{pmatrix}
      \mathbf{b}_0 \\
      \mathbf{c}_0 \\
    \end{pmatrix}_x=\alpha_0\begin{pmatrix}
      \mathbf{q} \\
      \mathbf{r} \\
    \end{pmatrix}\,,
\end{equation*}
$a_{0}=\frac{1}{2}\alpha_0,$ $\mathbf{d}_{0}=-\frac{1}{2}\alpha_0I_N$, $a_i=\partial_x^{-1}(\mathbf{r}^{\mathrm{T}}\mathbf{b}_{i-1}-\mathbf{c}_{i-1}^{\mathrm{T}}\mathbf{q})$, $
\mathbf{d}_i=\partial_x^{-1}(\mathbf{q}\mathbf{c}_{i-1}^{\mathrm{T}}-\mathbf{b}_{i-1}\mathbf{r}^{\mathrm{T}})$ and
\begin{equation*}
    L_2=-2{\rm i}\begin{pmatrix}
                   I_N & 0 \\[5pt]
                   0 & -I_N \\
                 \end{pmatrix}+\begin{pmatrix}
                                  \mathbf{q} \partial_x^{-1}\mathbf{r}^{\mathrm{T}}+\left(\sum_{k=1}^{N}q_k\partial_x^{-1}r_k\right)I_N & -\mathbf{q} \partial_x^{-1}\mathbf{q}^{\mathrm{T}}-(\mathbf{q} \partial_x^{-1}\mathbf{q}^{\mathrm{T}})^{\mathrm{T}} \\[5pt]
                                  -\mathbf{r} \partial_x^{-1}\mathbf{r}^{\mathrm{T}}-(\mathbf{r} \partial_x^{-1}\mathbf{r}^{\mathrm{T}})^{\mathrm{T}} & \mathbf{r} \partial_x^{-1}\mathbf{q}^{\mathrm{T}}+
                                  \left(\sum_{k=1}^{N}r_k\partial_x^{-1}q_k\right)I_N \\
                               \end{pmatrix}.
\end{equation*}
Through the trace identity \eqref{trace}, we can obtain that
\begin{equation*}
    \begin{pmatrix}
      \mathbf{c}_{i} \\
      \mathbf{b}_{i} \\
    \end{pmatrix}=\frac{\delta}{\delta\omega} H_i^{-} ,\,\, H_i^{-}=\int\frac{1}{2(i+1)}\left[2{\rm i}(a_i-\mathrm{tr}(\mathbf{d}_i))+\mathbf{c}_{i}^{\mathrm{T}} \mathbf{q}+
    \mathbf{r}^{\mathrm{T}} \mathbf{b}_{i}\right]\mathrm{d}x\,.
\end{equation*}
Taking $V(\lambda)=(\lambda^{2n}W)_+$, we can obtain the negative hierarchy
\begin{equation*}
    \omega_t=-J\frac{\delta}{\delta\omega} H_n=-JK^n\frac{\delta}{\delta\omega} H_0
\end{equation*}
where $$K=\partial_x^{-1}\begin{pmatrix}
                           0 & I_N \\
                           I_N & 0 \\
                         \end{pmatrix}L_2\begin{pmatrix}
                           0 & I_N \\
                           I_N & 0 \\
                         \end{pmatrix}.
$$
It can be readily verified that the operator $JK^{i}$ is skew symmetric.
Consequently, both the positive and negative flows of the multi-component Kaup-Newell hierarchy are constructed by Tu scheme.
\subsection{Conservation laws}
We consider the multi-component Kaup-Newell spectral problem
\begin{equation}\label{multi-KN}
    \begin{pmatrix}
      \psi_1 \\
      \psi_2 \\
    \end{pmatrix}_x=\begin{pmatrix}
                      {\rm i}\lambda^{-2} & \lambda^{-1} \mathbf{v}_x^{\mathrm{T}} \\
                      \lambda^{-1} \mathbf{u}_x  & -{\rm i}\lambda^{-2} \\
                    \end{pmatrix}\begin{pmatrix}
      \psi_1 \\
      \psi_2 \\
    \end{pmatrix}
\end{equation}
and associated evolution equation
\begin{equation}\label{evolutionpart}
    \begin{pmatrix}
      \psi_1 \\
      \psi_2 \\
    \end{pmatrix}_t={\rm i}\begin{pmatrix}
                      \frac{1}{4}\lambda^{2}+\frac{1}{2}\mathbf{v}^{\mathrm{T}} \mathbf{u} & -\frac{\lambda}{2} \mathbf{v}^{\mathrm{T}} \\
                      \frac{\lambda}{2} \mathbf{u}  & -\lambda^{2}I_N-\frac{1}{2}\mathbf{u} \mathbf{v}^{\mathrm{T}} \\
                    \end{pmatrix}\begin{pmatrix}
      \psi_1 \\
      \psi_2 \\
    \end{pmatrix}.
\end{equation}

It follows that
\begin{equation}\label{conservation}
     \mathbf{v}_x^{\mathrm{T}} \left(\frac{\psi_{2}}{\psi_{1}}\right)_x=-2{\rm i}\lambda^{-2} \mathbf{v}_x^{\mathrm{T}} \frac{\psi_{2}}{\psi_{1}}+\lambda^{-1}\mathbf{v}_x^{\mathrm{T}} \mathbf{u}_x-\lambda^{-1} \left(\frac{\mathbf{v}_x^{\mathrm{T}} \psi_{2}}{\psi_{1}}\right)^2\,,
\end{equation}
and
\begin{equation}\label{conservationtpart}
     \mathbf{v}^{\mathrm{T}} \left(\frac{\psi_{2}}{\psi_{1}}\right)_t=-\frac{{\rm i}}{2}\lambda^{2} \mathbf{v}^{\mathrm{T}} \frac{\psi_{2}}{\psi_{1}}+\frac{{\rm i}}{2}\lambda\mathbf{v}^{\mathrm{T}} \mathbf{u}-{\rm i}(\mathbf{v}^{\mathrm{T}} \mathrm{u}) \mathbf{v}^{\mathrm{T}} \frac{\psi_{2}}{\psi_{1}}+\frac{{\rm i}\lambda}{2} \left(\frac{\mathbf{v}^{\mathrm{T}} \psi_{2}}{\psi_{1}}\right)^2.
\end{equation}
To find the conservation laws of negative orders, we substitute an expansion
\begin{equation}\label{expand1}
    \frac{\psi_2}{\psi_1}=\sum_{i=1}^{\infty}\mathbf{P}_i\lambda^{2i-1}
\end{equation}
into equation \eqref{conservation} and obtain
\begin{equation*}
    \begin{split}
        \mathbf{P}_0=&\mathbf{u},  \quad \mathbf{P}_1=-\frac{{\rm i}}{2}\mathbf{u}_x,  \\
       \mathbf{P}_{i+1}=&\frac{{\rm i}}{2}\left[\mathbf{P}_{i,x}+\sum_{j=1}^{i} \mathbf{P}_{j}
       \mathbf{v}_x^{\mathrm{T}}  \mathbf{P}_{i+1-j}\right].
    \end{split}
\end{equation*}
Then the conservation laws follow
\begin{equation*}
    (\ln \psi_1)_{xt}=\lambda^{-1}\left[\mathbf{v}_x^{\mathrm{T}}  \frac{\psi_{2}}{\psi_{1}}\right]_t={\rm i}\left[\frac{1}{2}\left(\mathbf{v}^{\mathrm{T}} \mathrm{u}\right)_x-\frac{\lambda}{2}
    \left(\mathbf{v}^{\mathrm{T}} \frac{\psi_{2}}{\psi_{1}}\right)_x\right]\,,
\end{equation*}
i.e.
\begin{equation*}
    {\rm i}\left[\mathbf{v}_x^{\mathrm{T}}  \mathbf{P}_i\right]_t =\frac{1}{2}
    \left(\mathbf{v}^{\mathrm{T}} \mathbf{P}_{i-1}\right)_x,\,\,\, i=1,2, \cdots\,.
\end{equation*}
These conservation laws are all local, among which the first two are listed below
\begin{equation*}
\begin{split}
   \left[\mathbf{v}_x^{\mathrm{T}}  \mathbf{u}_x\right]_t & =\frac{{\rm i}}{2}\left(\mathbf{v}^{\mathrm{T}} \mathbf{u}\right)_x,  \\
    \left[\frac{{\rm i}}{4}\mathbf{v}^{\mathrm{T}}  \mathbf{u}_{xx}+\frac{1}{8}(
    \mathbf{v}^{\mathrm{T}}  \mathbf{u}_{x})^2\right]_t&=\frac{1}{4}(\mathbf{v}^{\rm T}  \mathbf{u}_{x})_x\,.
\end{split}
\end{equation*}
On the other hand, substituting the following expansion
\begin{equation}\label{expand}
    \frac{\psi_2}{\psi_1}=\sum_{i=1}^{\infty}\mathbf{C}_i\lambda^{-2i+1}
\end{equation}
into equation \eqref{conservationtpart} and equating the coefficients of $\lambda^{-2i+1}$, one obtains
\begin{equation*}
    \begin{split}
      \lambda: &\,\,\, \mathbf{C}_1=\mathbf{u}, \\
      \lambda^{-2i+1}:  &\,\,\, \mathbf{v}^{\mathrm{T}} \mathbf{C}_{i+1}=2{\rm i}\mathbf{v}^{\mathrm{T}}  \mathbf{C}_{i,t}-2(\mathbf{v}^{\mathrm{T}} \mathbf{u})
      (\mathbf{v}^{\mathrm{T}} \mathbf{C}_i)+\sum_{j=1}^{i}(\mathbf{v}^{\mathrm{T}}  \mathbf{C}_j)(\mathbf{v}^{\mathrm{T}}  \mathbf{C}_{i+1-j})\,,
    \end{split}
\end{equation*}
which leads to the conservation laws
\begin{equation*}
    (\ln \psi_1)_{xt}=\lambda^{-1}\left[\mathbf{v}_x^{\mathrm{T}}  \frac{\psi_{2}}{\psi_{1}}\right]_t={\rm i}\left[\frac{1}{2}\left(\mathbf{v}^{\mathrm{T}} \mathbf{u}\right)_x-\frac{\lambda}{2}
    \left(\mathbf{v}^{\mathrm{T}} \frac{\psi_{2}}{\psi_{1}}\right)_x\right]\,,
\end{equation*}
i.e.
\begin{equation*}
    {\rm i}\left[\mathbf{v}_x^{\mathrm{T}}  \mathbf{C}_i\right]_t=\left[\frac{1}{2}\mathbf{v}^{\mathrm{T}}  \mathbf{C}_{i+1}\right]_x,\,\,\, i=1,2, \cdots\,.
\end{equation*}
The first conversation law is
\begin{equation*}
    {\rm i}\left[\mathbf{v}_x^{\mathrm{T}}  \mathbf{u}\right]_t=\left[{\rm i}\mathbf{v}^{\mathrm{T}}  \mathbf{u}_{t}-\frac{1}{2}(\mathbf{v}^{\mathrm{T}}  \mathbf{u})^2\right]_x.
\end{equation*}
In contrast to the conservation laws of negative orders, other conservation laws except the first one of positive orders are nonlocal.
\section{Spectral problem and Darboux transformation}\label{par3}
The coupled FL equation (\ref{cFL}) considered in the present paper
admits the following Lax pair
\begin{equation}\label{cFL-lax}
  \begin{split}
     \Phi_x=&U(x,t;\lambda)\Phi, \qquad U(x,t;\lambda)={\rm i}\lambda^{-2}\sigma_3+\lambda^{-1}Q_x, \\
     \Phi_t=&V(x,t;\lambda)\Phi, \qquad V(x,t;\lambda)={\rm i}\left(\frac{1}{4}\lambda^2\sigma_3+\frac{1}{2}\sigma_3(Q^2-\lambda Q)\right),
  \end{split}
\end{equation}
where
\begin{equation*}
  \sigma_3=\mathrm{diag}(1,-1,-1),\qquad Q=\begin{pmatrix}
                                             0 & u_1^*&\sigma u_2^* \\
                                             u_1 &0&0 \\
                                             u_2 &0&0
                                           \end{pmatrix},\qquad \sigma=\pm1.
\end{equation*}
So the solutions to equation \eqref{cFL} can be solved from its Lax pair \eqref{cFL-lax}. To the end, we consider the spectral problem \eqref{cFL-lax}
\begin{equation}\label{spectral}
  -{\rm i}\sigma_3\left[\partial_x-\lambda^{-1}Q_x\right]\Phi=\lambda^{-2}\Phi,
\end{equation}
which is an energy-dependent spectral problem, here $Q\in L^1_{loc}(\mathbb{R})$. The involution relation for the system \eqref{cFL-lax} can be concluded by the following Lemma:
\begin{lem}
  (a) The matrices $U,V$ are variant under the involution $\tau_1: A(\lambda)\mapsto \sigma_3 A(-\lambda)\sigma_3$ and $\tau_2:A(\lambda)\mapsto -J [A(\lambda^*)]^{\dag}J,$ where $J=\mathrm{diag}(1,-1,-\sigma).$
  (b) If the function $\Phi(x,t;\lambda)$ satisfies the system \eqref{cFL-lax} with initial data $\Phi(0,0;\lambda)=I_{3\times 3}$, then $\Phi(x,t;-\lambda)=\sigma_3\Phi(x,t;\lambda)\sigma_3,$ and $J[\Phi(x,t;\lambda)]^{\dag}J=[\Phi(x,t;\lambda^*)]^{-1}$.
\end{lem}
\textbf{Proof:}  The claim in (a) can be proved by direct calculation, which is omitted here. \\

b) Since $\Phi(x,t;\lambda)$ satisfies the system \eqref{cFL-lax} with initial data $\Phi(0,0;\lambda)=I_{3\times 3}$, and  $U,V$ are invariant under the involution $\tau_1$, we have
\begin{equation*}
  \begin{split}
     \frac{\partial}{\partial x}\Phi(x,t;-\lambda)=&U(x,t;-\lambda)\Phi(x,t;-\lambda),  \\
     \frac{\partial}{\partial t}\Phi(x,t;-\lambda)=&V(x,t;-\lambda)\Phi(x,t;-\lambda),
  \end{split}
\end{equation*}
and
\begin{equation*}
  \begin{split}
     \frac{\partial}{\partial x}[\sigma_3\Phi(x,t;\lambda)\sigma_3]=&U(x,t;-\lambda)[\sigma_3\Phi(x,t;\lambda)\sigma_3],  \\
     \frac{\partial}{\partial t}[\sigma_3\Phi(x,t;\lambda)\sigma_3]=&V(x,t;-\lambda)[\sigma_3\Phi(x,t;\lambda)\sigma_3]\,.
  \end{split}
\end{equation*}
 Based on the existence and uniqueness of ordinary differential equations (ODEs), we have $\Phi(x,t;-\lambda)=\sigma_3\Phi(x,t;\lambda)\sigma_3.$

Similarly, since $\Phi(x,t;\lambda)$ satisfies the system \eqref{cFL-lax} with initial data $\Phi(0,0;\lambda)=I_{3\times 3}$, and $U,V$ are invariant under the involution $\tau_2$, one has
\begin{equation*}
  \begin{split}
     -(J[\Phi(x,t;\lambda)]^{\dag}J)_x=&(J[\Phi(x,t;\lambda)]^{\dag}J)U(x,t;\lambda^*),  \\
     -(J[\Phi(x,t;\lambda)]^{\dag}J)_t=&(J[\Phi(x,t;\lambda)]^{\dag}J)V(x,t;\lambda^*)
  \end{split}
\end{equation*}
and
\begin{equation*}
  \begin{split}
     -([\Phi(x,t;\lambda)]^{-1})_x=&([\Phi(x,t;\lambda)]^{-1})U(x,t;\lambda^*),  \\
     -([\Phi(x,t;\lambda)]^{-1})_t=&([\Phi(x,t;\lambda)]^{-1})V(x,t;\lambda^*)\,,
  \end{split}
\end{equation*}
we then have $J[\Phi(x,t;\lambda)]^{\dag}J=[\Phi(x,t;\lambda^*)]^{-1}$.
$\square$

Therefore, the Darboux matrix $T(\lambda)$ converts the system \eqref{cFL-lax} into a new system of the form
\begin{equation}\label{cFL-lax-new}
  \begin{split}
     \Phi[1]_x=&U[1](x,t;\lambda)\Phi[1], \qquad U[1](x,t;\lambda)={\rm i}\lambda^{-2}\sigma_3+\lambda^{-1}Q[1]_x, \\
     \Phi[1]_t=&V[1](x,t;\lambda)\Phi[1], \qquad V[1](x,t;\lambda)={\rm i}\left(\frac{1}{4}\lambda^2\sigma_3+\frac{1}{2}\sigma_3(Q[1]^2-\lambda Q[1])\right).
  \end{split}
\end{equation}
where
\begin{equation}\label{potential}
  \begin{split}
     U[1]=&T_xT^{-1}+TUT^{-1},  \\
     V[1]=&T_tT^{-1}+TVT^{-1}.
  \end{split}
\end{equation}
\begin{lem}\label{lem2}
Assume that $T(\lambda)$ satisfies $T(\lambda)=\sigma_3 T(-\lambda) \sigma_3$ and $[T(\lambda)]^{-1}=J[T(\lambda^*)]^{\dag}J$, if the matrices $U$, $V$ are invariant under the involution $\tau_1$ and $\tau_2$, then the new potential functions keep invariant under the involution $\tau_1$ and $\tau_2$.
\end{lem}
\textbf{Proof:} From the relation \eqref{potential} and $T(\lambda)=\sigma_3 T(-\lambda) \sigma_3$, we have
\begin{equation*}
  \begin{split}
     \sigma_3 U[1](-\lambda)\sigma_3=&\sigma_3T(-\lambda)_x[T(-\lambda)]^{-1}\sigma_3+\sigma_3T(-\lambda)U(-\lambda)[T(-\lambda)]^{-1}\sigma_3=U[1](\lambda),  \\
     \sigma_3 V[1](-\lambda)\sigma_3=&\sigma_3T(-\lambda)_t[T(-\lambda)]^{-1}\sigma_3+\sigma_3T(-\lambda)V(-\lambda)[T(-\lambda)]^{-1}\sigma_3=V[1](\lambda).
  \end{split}
\end{equation*}
On the other hand, through the relation \eqref{potential} and $[T(\lambda)]^{-1}=J[T(\lambda^*)]^{\dag}J$, one obtains
\begin{equation*}
  \begin{split}
     J [U[1](\lambda^*)]^{\dag}J=&\sigma_3T(\lambda)_x[T(\lambda)]^{-1}\sigma_3+\sigma_3T(\lambda)U(\lambda)[T(\lambda)]^{-1}\sigma_3=U[1](\lambda),  \\
     \sigma_3 V[1](\lambda)\sigma_3=&\sigma_3T(-\lambda)_t[T(-\lambda)]^{-1}\sigma_3+\sigma_3T(-\lambda)V(-\lambda)[T(-\lambda)]^{-1}\sigma_3=V[1](\lambda).
  \end{split}
\end{equation*} $\square$

Due to the relation $T(\lambda)=\sigma_3 T(-\lambda) \sigma_3$,
the Darboux matrix can be constructed through the loop group method
\begin{equation}\label{dt}
  T(\lambda)=I+\frac{A_1}{\lambda-\lambda_1^*}-\frac{\sigma_3 A_1\sigma_3}{\lambda+\lambda_1^*}.
\end{equation}
Based on the Lemma \ref{lem2}, the inverse of Darboux matrix $T(\lambda)$ may be chosen as
\begin{equation}\label{dt-inverse}
  T(\lambda)^{-1}=J[T(\lambda^*)]^{\dag}J=I+\frac{J A_1^{\dag}J}{\lambda-\lambda_1}-\frac{\sigma_3 JA_1^{\dag}J\sigma_3}{\lambda+\lambda_1}.
\end{equation}
The $L^2(\mathbb{R})$ eigenfunction can be constructed from the Darboux matrix. Usually, we can construct two wave vector functions which satisfy
$\phi_{\pm}(\lambda_1)\rightarrow 0$ as $x\rightarrow \pm \infty$ with exponential decay, $\lambda_1\in \mathbb{C}/\{\mathbb{R}\cup {\rm i}\mathbb{R}\}$. The Darboux matrix satisfies the following proposition
\begin{equation}\label{kernel}
  T(\lambda_1)|y_1\rangle=0,\,\, |y_1\rangle=\phi_{1,+}+\gamma\phi_{1,-}\,,
\end{equation}
where  $\mathrm{rank}(A_1)$ is either $1$ or $2$. However, since the order for this spectral problem is three, 
 we can assume that $A_1=|x_1\rangle\langle z_1|J,$ where $\langle z_1|=|z_1\rangle^{\dag}.$ On the other hand, since $T(\lambda)[T(\lambda)]^{-1}=I$, one can obtain that $\mathrm{Res}_{\lambda=\lambda_1}(T(\lambda)[T(\lambda)]^{-1})=0$. It follows that
\begin{equation}\label{kernel1}
  \left[I+\frac{A_1}{\lambda_1-\lambda_1^*}-\frac{\sigma_3 A_1\sigma_3}{\lambda_1+\lambda_1^*}\right]|z_1\rangle=0.
\end{equation}
Together with equation \eqref{kernel}, one arrives at $|z_1\rangle =c_1|y_1\rangle$. For the sake of convenience, by setting $c_1=1$, equation \eqref{kernel1} is rewritten as
\begin{equation}\label{kernel2}
 |y_1\rangle+\frac{\langle y_1|J|y_1\rangle}{\lambda_1-\lambda_1^*}|x_1\rangle-\frac{\langle y_1|J\sigma_3|y_1\rangle}{\lambda_1+\lambda_1^*}\sigma_3|x_1\rangle=0.
\end{equation}
Denoting $|y_1\rangle=(\varphi_1,\psi_1,\chi_1)^{\mathrm{T}}$, one can solve
\begin{equation}\label{a-x}
  |x_1\rangle=\begin{pmatrix}
                \alpha^{-1} & 0 & 0 \\
                0 & \beta^{-1} & 0 \\
                0 & 0 & \beta^{-1} \\
              \end{pmatrix}|y_1\rangle,
\end{equation}
where
\begin{equation*}
  \alpha=\frac{2[\lambda_1^*|\varphi_1|^2-\lambda_1(|\psi_1|^2+\sigma|\chi_1|^2)]}{\lambda_1^{*2}-\lambda_1^2},\,\, \beta=
  \frac{2[\lambda_1|\varphi_1|^2-\lambda_1^*(|\psi_1|^2+\sigma|\chi_1|^2)]}{\lambda_1^{*2}-\lambda_1^2}.
\end{equation*}
Consequently, we obtain the transformation from potential function $Q$ to $Q[1]$
\begin{equation}\label{backlund}
  Q[1]=Q+(|x_1\rangle\langle y_1|J-\sigma_3|x_1\rangle\langle y_1|J\sigma_3)\,,
\end{equation}
or explicitly
\begin{equation}\label{backlund1}
  \begin{split}
     u_1[1]=&u_1+\frac{2}{\beta}\psi_1\varphi_1^*,  \\
     u_2[1]=&u_2+\frac{2}{\beta}\chi_1\varphi_1^*.
  \end{split}
\end{equation}
Generally, we could derive the following $N$-fold Darboux transformation
\begin{equation}\label{n-fold-dt}
  T_N=I+\sum_{i=1}^{N}\left[\frac{A_i}{\lambda-\lambda_i^*}-\frac{\sigma_3A_i\sigma_3}{\lambda+\lambda_i^*}\right]\,,
\end{equation}
where $A_i=|x_i\rangle \langle y_i|J$
\begin{equation*}\begin{split}
                    \left[|x_{1,1}\rangle,|x_{2,1}\rangle,\cdots,|x_{N,1}\rangle\right]&  =
  \left[|y_{1,1}\rangle,|y_{2,1}\rangle,\cdots,|y_{N,1}\rangle\right]B^{-1},\,\, B=(b_{ij})_{N\times N},\\
                    \left[|x_{1,k}\rangle,|x_{2,k}\rangle,\cdots,|x_{N,k}\rangle\right]&  =
  \left[|y_{1,k}\rangle,|y_{2,k}\rangle,\cdots,|y_{N,k}\rangle\right]M^{-1},\,\,M=(m_{ij})_{N\times N},\,\, k=2,3,
                 \end{split}
\end{equation*}
and
\begin{equation*}
  b_{ij}=\frac{\langle y_i|J|y_j\rangle}{\lambda_i^*-\lambda_j}+\frac{\langle y_i|J\sigma_3|y_j\rangle}{\lambda_i^*+\lambda_j},\,\, m_{ij}=\frac{\langle y_i|J|y_j\rangle}{\lambda_i^*-\lambda_j}-\frac{\langle y_i|J\sigma_3|y_j\rangle}{\lambda_i^*+\lambda_j}.
\end{equation*}
The transformation between old and new potential functions is
\begin{equation}\label{nfoldbt}
  Q[N]=Q+\sum_{i=1}^N(A_i-\sigma_3A_i\sigma_3).
\end{equation}
Moreover, we have
\begin{equation}\label{nfoldbt1}
  \begin{pmatrix}
    u_1[N] \\[8pt]
    u_2[N] \\
  \end{pmatrix}= \begin{pmatrix}
    u_1 \\[8pt]
    u_2 \\
  \end{pmatrix}+2Y_2M^{-1}Y_1^{\dag}
\end{equation}
where
\begin{equation*}
  Y_1=\begin{pmatrix}
        \varphi_1 & \varphi_2 & \cdots & \varphi_N
      \end{pmatrix},\,\,Y_2=\begin{pmatrix}
        \phi_1 & \phi_2 & \cdots & \phi_N \\
        \chi_1 & \chi_2 & \cdots & \chi_N \\
      \end{pmatrix}.
\end{equation*}

\section{Bright soliton solution with vanishing boundary condition}
\subsection{Single bright soliton solution}
Inserting the zero seed solution into Lax pair \eqref{cFL-lax} and introducing $z=1/\lambda^2$,
the fundamental solution to the system \eqref{cFL-lax} is $\Phi_1(\lambda)=\exp[{\rm i}(zx+\frac{1}{4z}t)\sigma_3]$.
Based on the formula \eqref{backlund}, one obtains the single soliton solution with vanishing boundary condition
 \begin{equation}\label{one-soliton-solution}
    \begin{split}
       u_s[1]=&\frac{z_1-z_1^*}{|z_1|^2}\frac{c_s{\rm e}^{\varpi_1^*}}{\lambda_1{\rm e}^{\varpi_1+\varpi_1^*}-\lambda_1^*(|c_1|^2+\sigma |c_2|^2)}.
    \end{split}
  \end{equation}
Let $\lambda_1=\lambda_{1R}+{\rm i}\lambda_{1I}$ and $\varpi_1=\varpi_{1R}+{\rm i}\varpi_{1I}={\rm i}(2z_1x+\frac{1}{2z_1}t)+\delta_1$, the single soliton solution can be represented as
  \begin{equation}\label{one-soliton}
    \begin{split}
       u_s[1]=&\frac{-2\lambda_{1R}\lambda_{1I}c_s{\rm i}{\rm e}^{-{\rm i}\varpi_{1I}}}{r_1\left[\lambda_{1R}\sinh(\varpi_{1R}-\ln(r_1))+{\rm i}\lambda_{1I}\cosh(\varpi_{1R}-\ln(r_1))\right]},
    \end{split}
  \end{equation}
for $c_1^2+\sigma c_2^2>0$ and
  \begin {equation}\label{one-soliton1}
    \begin{split}
       u_s[1]=&\frac{-2\lambda_{1R}\lambda_{1I}c_s{\rm i}{\rm e}^{-{\rm i}\varpi_{1I}}}{r_1\left[\lambda_{1R}\cosh(\varpi_{1R}-\ln(r_1))+{\rm i}\lambda_{1I}\sinh(\varpi_{1R}-\ln(r_1))\right]},
    \end{split}
  \end{equation}
for $c_1^2+\sigma c_2^2<0$. Here $r_1=(|c_1^2+\sigma c_2^2|)^{1/2}$ and
  \begin{equation*}
    \varpi_{1R}=\lambda_{1R}\lambda_{1I}\left[\frac{4x}{(\lambda_{1R}^2+\lambda_{1I}^2)^2}-t\right]+\delta_{1R},\,\, \varpi_{1I}=\frac{(\lambda_{1R}^2-\lambda_{1I}^2)}{2}\left[\frac{4x}{(\lambda_{1R}^2+\lambda_{1I}^2)^2}+t\right]+\delta_{1I},
  \end{equation*}
  where  $\delta_{1R},\delta_{1I}\in \mathbb{R}$.
  Basically, the soliton solution is the bright soliton solution of the bell shape.
  If $c_1^2+\sigma c_2^2>0$, the peak value for $|u_s[1]|^2$ is $4\lambda_{1R}^2c_s^2/r_1^2$ ( $s=1,2$). Whereas, if $c_1^2+\sigma c_2^2<0$, the peak value for $|u_s[1]|^2$ is $4\lambda_{1I}^2c_s^2/r_1^2$ ($s=1,2)$. 

  Particularly, if $\ln(r_1)=\delta_{1R}$, one can obtain a rational soliton solution of the form
     \begin{equation}\label{one-ration-soliton1}
    \begin{split}
       u_s[1]=&\frac{2c_s\lambda_{1R}{\rm e}^{-{\rm i}(2x/\lambda_{1R}^{2}+\lambda_{1R}^2t/2)}}{r_1\left[{\rm i}(4x/\lambda_{1R}^{2}-\lambda_{1R}^2t)-1\right]}.
    \end{split}
  \end{equation}
as $\lambda_{1I}\rightarrow 0$. On the other hand, if $\sigma=-1$, $\ln(r_1)=\delta_{1R}$, one has a rational soliton
 solution
     \begin{equation}\label{one-ration-soliton2}
    \begin{split}
       u_s[1]=&\frac{2c_s\lambda_{1I}{\rm i}{\rm e}^{{\rm i}(2x/\lambda_{1I}^{2}+\lambda_{1I}^2t/2)}}{r_1\left[{\rm i}(4x/\lambda_{1I}^{2}-\lambda_{1I}^2t)+1\right]}.
    \end{split}
  \end{equation}
as $\lambda_{1R}\rightarrow 0$.  
It can be shown that the rational soliton is also of the bell shape.
\subsection{Multi-bright soliton solutions}
Letting $|y_i\rangle=\Phi_1(\lambda_i)(1,c_{i,1},c_{i,2})^T$, we can obtain
\begin{equation*}
    m_{i,j}=\frac{2z_i^*z_j}{z_j-z_i^*}\left[\lambda_j{\rm e}^{\theta_i^*+\theta_j}+\lambda_i^*\gamma_{i,j}{\rm e}^{-\theta_i^*-\theta_j}\right],\,\, \gamma_{i,j}=-(c_{i,1}^{*}c_{j,1}+\sigma c_{i,2}^{*}c_{j,2})
\end{equation*}
and
\begin{equation*}
\begin{split}
   \varphi_i&={\rm e}^{\theta_i}, \,\,
    \phi_i  = c_{i,1}{\rm e}^{-\theta_i},\,\,
    \chi_i=c_{i,2}{\rm e}^{-\theta_i},\\
    \theta_i&={\rm i}\left(z_ix+\frac{1}{4z_i}t\right)\equiv -z_{i,I}(x-v_it)+{\rm i}z_{i,R}(x+v_it),\,\, v_i=\frac{1}{4|z_i|^2},\,\, z_i=z_{i,R}+{\rm i}z_{i,I},\,\, z_{i,I}<0.
\end{split}
\end{equation*}
Based on the formula \eqref{nfoldbt1}, we could derive the $N$-bright soliton solution formula:
\begin{equation}\label{n-bright}
    \begin{split}
       u_s[N]=& -2\frac{\det\begin{pmatrix}
                            M & Y_1^{\dag} \\
                            Y_2^{[s]} & 0 \\
                          \end{pmatrix}
       }{\det(M)},
    \end{split}
\end{equation}
where $M=(m_{i,j})_{1\leq i,j\leq N}$ and $Y_2^{[s]}$ represents the $s$-th row of matrix $Y_2$.
\begin{figure}[tbh]
\centering
\subfigure[$|u_1|^2$]{%
\includegraphics[height=50mm,width=65mm]{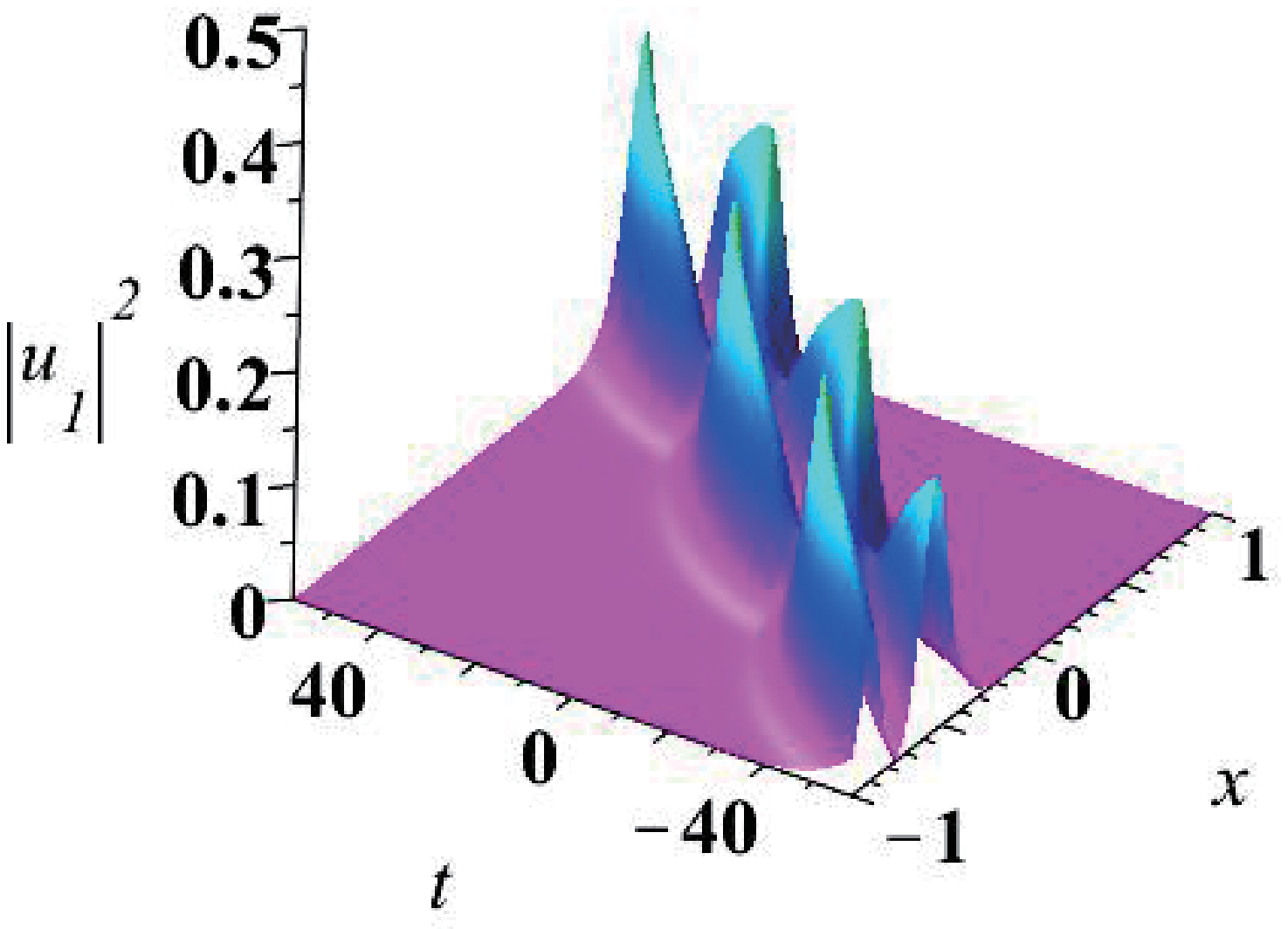}} \hfil
\subfigure[$|u_2|^2$]{%
\includegraphics[height=50mm,width=65mm]{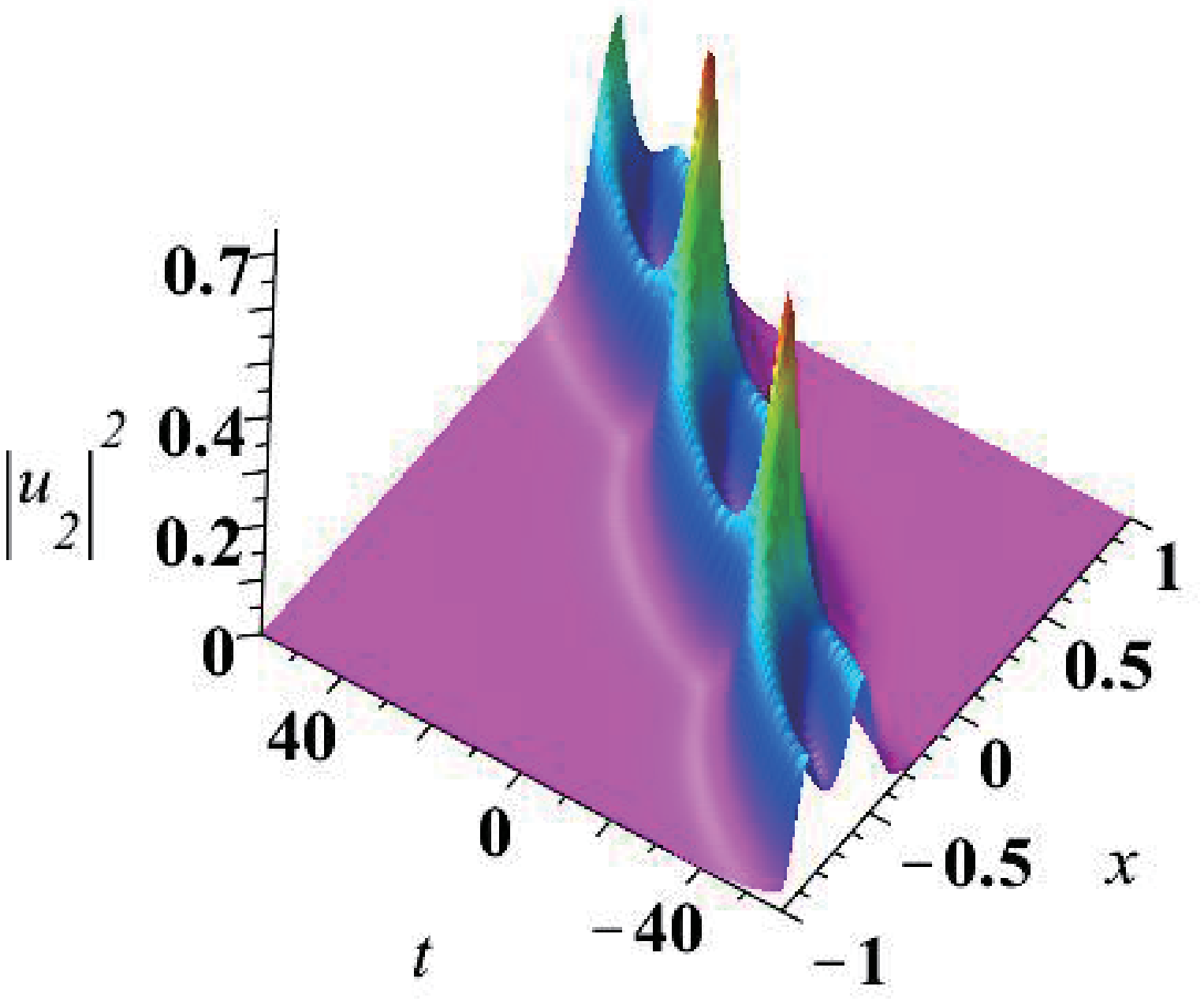}}
\caption{(color online): Breather solution with parameters: $\lambda_1=\frac{3}{10}+\frac{2}{5}{\rm i}$, $\lambda_2=\frac{2}{5}+\frac{3}{10}{\rm i}$, $c_{1,1}=-2$, $c_{2,1}=1$,
 $c_{1,2}=1$,  $c_{2,2}=-2$. }
\label{fig8}
\end{figure}

\begin{figure}[tbh]
\centering
\subfigure[$|u_1|^2$]{%
\includegraphics[height=50mm,width=65mm]{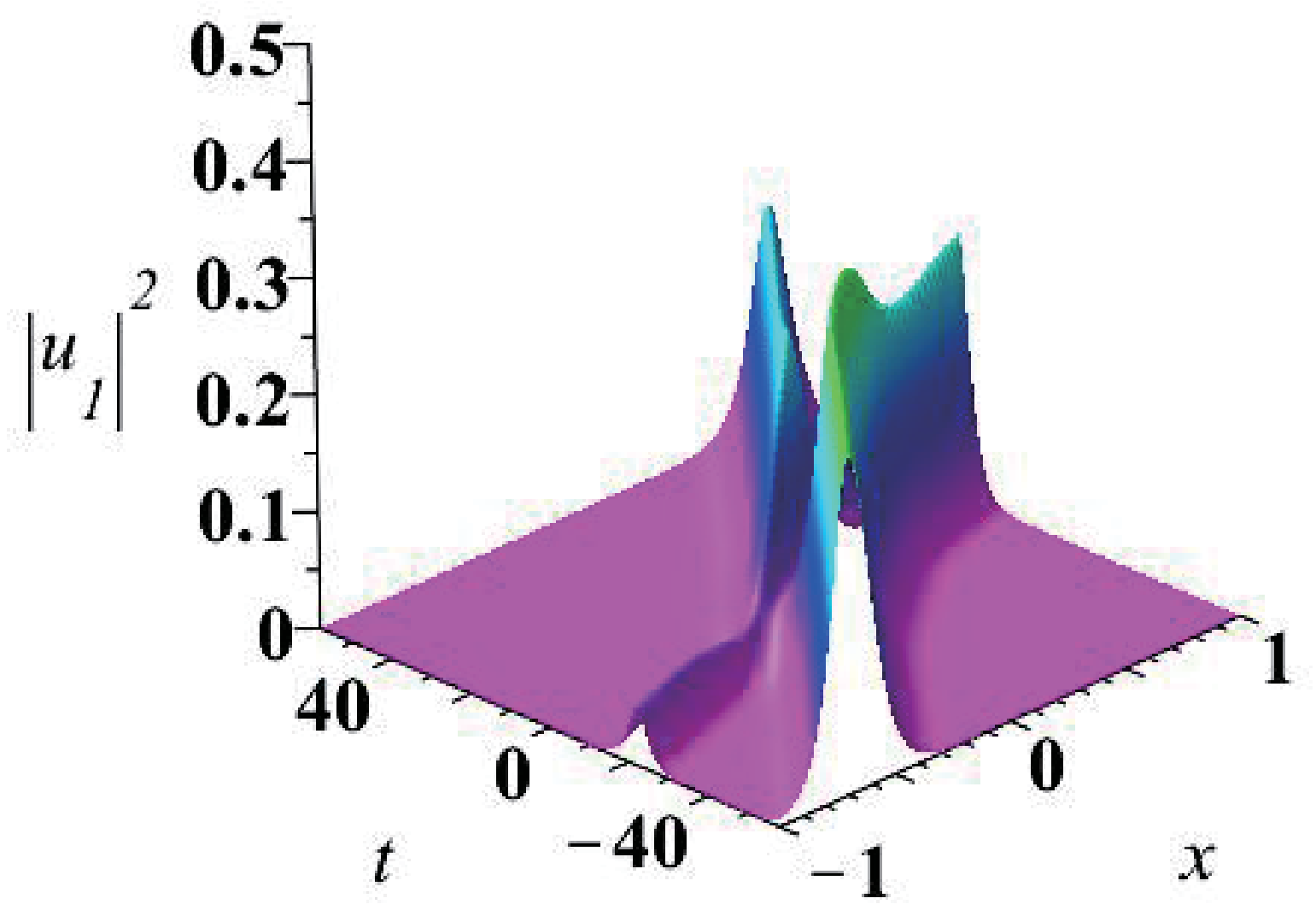}} \hfil
\subfigure[$|u_2|^2$]{%
\includegraphics[height=50mm,width=65mm]{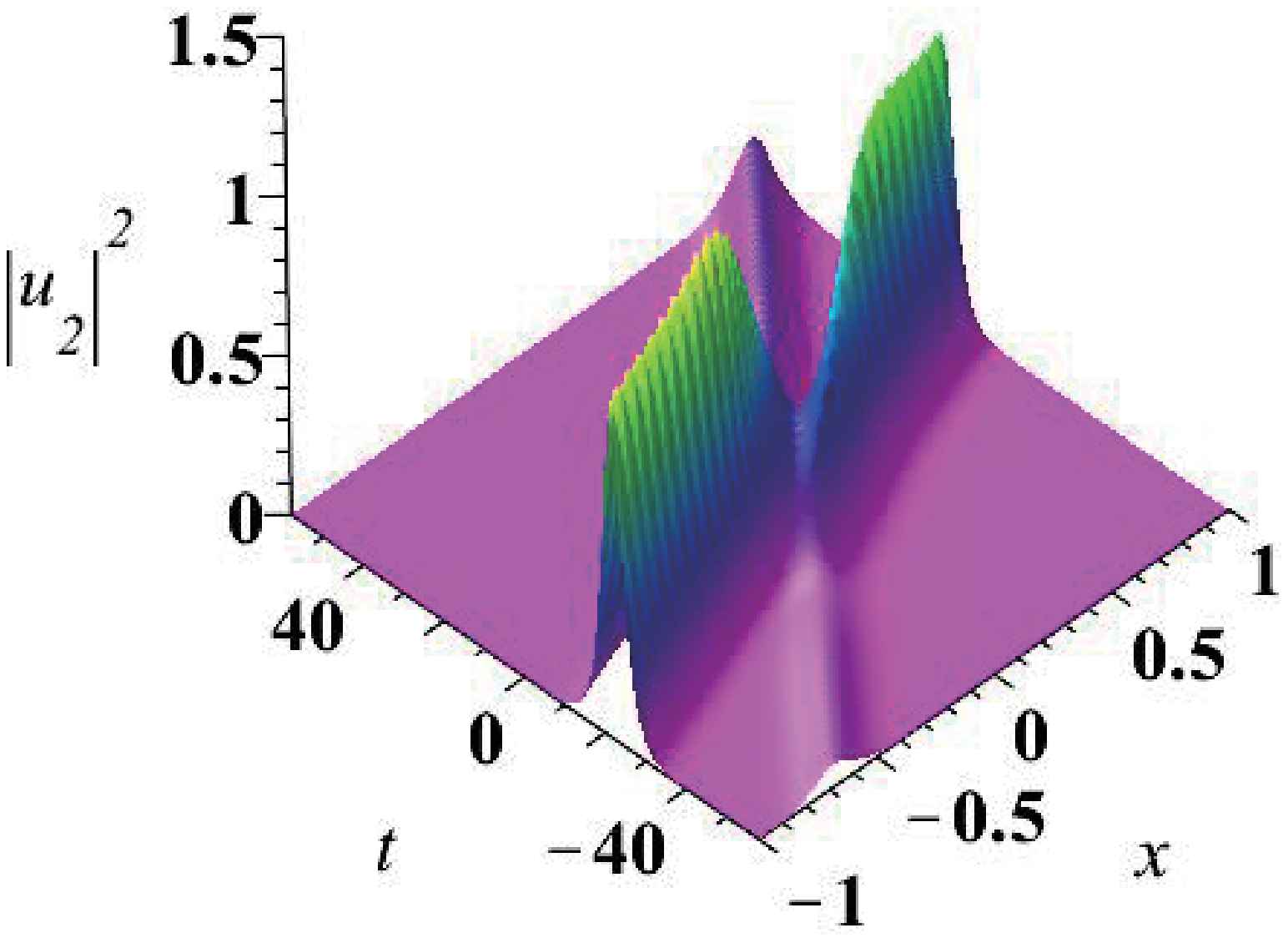}}
\caption{(color online): Two-soliton solution with parameters: $\lambda_1=\frac{3}{10}+\frac{2}{5}{\rm i}$, $\lambda_2=\frac{8}{15}+\frac{2}{5}{\rm i}$, $c_{1,1}=-2$, $c_{2,1}=1$,
 $c_{1,2}=1$,  $c_{2,2}=-2$. }
\label{fig9}
\end{figure}
To investigate the asymptotic behavior for the $N$-bright soliton solution with different velocity, we need to introduce the generalized Cauchy matrix
\begin{equation}\label{g-cauchy}
\begin{split}
   C(\Delta_k) &  =\left[\frac{c_{i}^{\dag}\Lambda c_{j}}{z_j-z_i^*}\right]_{i=1,j=1}^{k,k},\,\,  C(\Delta^k)  =\left[\frac{c_{i}^{\dag}\Lambda c_{j}}{z_j-z_i^*}\right]_{i=k,j=k}^{N,N},\,\, \Lambda=\rm{diag}(-1,-\sigma) \\
   \widehat{C}_s(\Delta_k)&= \begin{pmatrix}
                             \widehat{C}_{up} \\
                             \widehat{C}_{lower} \\
                           \end{pmatrix},\,\, \widehat{C}_{\rm up}=\left[\frac{c_{i}^{\dag}\Lambda c_{j}}{z_j-z_i^*}\right]_{i=1,j=1}^{k-1,k},\,\,\widehat{C}_{\rm lower}=\left(c_{1,s}, \cdots, c_{k-1,s}, c_{k,s}\right),\\
\widehat{C}_s(\Delta^k)&= \begin{pmatrix}
                           \widehat{C}^{up} \\
                             \widehat{C}^{lower} \\
                           \end{pmatrix},\,\, \widehat{C}^{\rm lower}=\left[\frac{c_{i}^{\dag}\Lambda c_{j}}{z_j-z_i^*}\right]_{i=k+1,j=k}^{N,N},\,\,\widehat{C}^{\rm up}=\left(c_{k,s}, \cdots, c_{N-1,s}, c_{N,s}\right),
\end{split}
\end{equation}
where
\begin{equation*}
    \begin{split}
     & C(z_{k}^{c},z_k^{c\dag})=\left[\frac{1}{z_j-z_i^*}\right]_{i=k+1,j=k+1}^{N,N}\,, \\
     & C(z_{k},z_k^{\dag})=\left[\frac{1}{z_j-z_i^*}\right]_{i=1,j=1}^{k,k}\,, \\
      &  \mathbf{c}_k=\left(c_i\right)_{i=1}^{k}\in \mathbb{C}^{k},\,\, c_i=z_i\begin{pmatrix}
                                                                              c_{i,1} \\
                                                                              c_{i,2} \\
                                                                            \end{pmatrix}.
    \end{split}
\end{equation*}

\begin{prop}
When $t\rightarrow \pm\infty$, the asymptotic of the $N$-soliton solution is
\begin{equation*}
    u_s[N]=\sum_{k=1}^{N} u_{s,\pm}^{[k]} +O({\rm e}^{-c|t|})
\end{equation*}
where $c=\mathrm{min}_{1,2,\cdots ,N}(|z_{i,I}|)\mathrm{min}_{i\neq
j}(|v_{i}-v_{j}|),$ the expressions of $u_{s,\pm}^{[k]}$ are given by equations \eqref{uk-} and \eqref{uk+}.
\end{prop}

\textbf{Proof:} Assuming $v_1<v_2<\cdots<v_N$, we have $\theta
_{1},\theta _{2},\cdots ,\theta _{k-1}\rightarrow -\infty $; $\theta
_{k+1},\theta _{k+2},\cdots ,\theta _{N}\rightarrow +\infty$ along the trajectory $x-v_{k}t=\mathrm{const}$ as $t\rightarrow -\infty $.  On the other
hand, $u_s[N]$ can be rewritten as
\begin{equation}
u_s[N]=-\frac{\det (\widehat{G}_s)}{\det (\widehat{M})},  \label{n-soliton}
\end{equation}%
where
\begin{equation*}
\begin{split}
\widehat{M}& =\left(\frac{z_i^*z_j}{z_j-z_i^*}\left[\lambda_j{\rm e}^{2(\theta_i^*+\theta_j)}+\lambda_i^*\gamma_{i,j}\right]\right) _{1\leq i,j\leq N},\,\,\widehat{G%
}_s=%
\begin{bmatrix}
\widehat{M} & \widehat{Y_{1}}^{\dag } \\
\widehat{Y_{2}}^{[s]} & 0 \\
\end{bmatrix}%
, \\
\widehat{Y_{1}}& =%
\begin{bmatrix}
\mathrm{e}^{2\theta _{1}} & \mathrm{e}^{2\theta _{2}} & \cdots & \mathrm{e}%
^{2\theta _{N}} \\
\end{bmatrix}%
,\,\,\widehat{Y_{2}}=%
\begin{bmatrix}
c_{1,1}&c_{2,1}&\cdots& c_{N,1}\\
c_{1,2}&c_{2,2}&\cdots& c_{N,2}\\
\end{bmatrix}%
.
\end{split}%
\end{equation*}%
It then follows
\begin{equation*}
\begin{split}
\det (\widehat{M})=& \mathrm{e}^{4\mathrm{Re}(\theta _{k+1}+\theta _{k+2}+\cdots
+\theta _{N})}\left[ \det (M_{k})+O(\mathrm{e}^{-c|t|})\right] , \\
\det (\widehat{G}_{s})=& \mathrm{e}^{4\mathrm{Re}(\theta _{k+1}+\theta _{k+2}+\cdots
+\theta _{N})}\left[ \det (G_{k}^{[s]})+O(\mathrm{e}^{-c|t|})\right] ,
\end{split}%
\end{equation*}%
where
\begin{equation*}
\begin{split}
M_{k}& =%
\begin{bmatrix}
\frac{|z_1|^2\lambda_1^*}{z _{1}-z_{1}^{\ast }}\gamma_{1,1} & \cdots & \frac{z_1^*z_{k-1}\lambda_{1}^*}{z_{k-1}-z_{1}^{\ast }}\gamma_{1,k-1} & \frac{z_1^*z_{k}\lambda_{1}^*}{z_{k}-z_{1}^{\ast }}\gamma_{1,k} &
0 & \cdots & 0 \\
\vdots & \ddots & \vdots & \vdots & \vdots & \ddots & \vdots \\
\frac{z_{k-1}^*z_1\lambda_{k-1}^*}{z _{1}-z_{k-1}^{\ast }}\gamma_{k-1,1} & \cdots & \frac{|z_{k-1}|^2\lambda_{k-1}^*}{z _{k-1}-z_{k-1}^{\ast }}\gamma_{k-1,k-1} & \frac{z_{k-1}^*z_{k}\lambda_{k-1}^*}{z_{k}-z_{k-1}^{\ast }}\gamma_{k-1,k} & 0 & \cdots & 0 \\
\frac{z_{k}^*z_1\lambda_{k}^*}{z_{1}-z_{k}^{\ast }}\gamma_{k,1} & \cdots & \frac{z_{k}^*z_{k-1}\lambda_{k}^*}{z_{k-1}-z_{k}^{\ast }}\gamma_{k,k-1} & \frac{|z_k|^2(\lambda_k\mathrm{e}^{2(\theta _{k}^{\ast
}+\theta _{k})}+\lambda_k^*\gamma_{k,k})}{z_{k}-z_{k}^{\ast}} & \frac{z_k^*z_{k+1}\lambda_{k+1}\mathrm{e}^{2\theta _{k}^{\ast }}}{z_{k+1}-z_{k}^{\ast }} & \cdots &
\frac{z_k^*z_{N}\lambda_{N}\mathrm{e}^{2\theta _{k}^{\ast }}}{z_{N}-z_{k}^{\ast }}
\\
0 & \cdots & 0 & \frac{z_{k+1}^{\ast
}z_{k}\lambda_k\mathrm{e}^{2\theta _{k}}}{z_{k}-z_{k+1}^{\ast
}} & \frac{|z_{k+1}|^2\lambda_{k+1}}{z_{k+1}-z_{k+1}^{\ast }} & \cdots &
\frac{z_{k+1}^{\ast
}z_{N}\lambda_N}{z_{N}-z_{k+1}^{\ast }} \\
\vdots & \ddots & \vdots & \vdots & \vdots & \ddots & \vdots \\
0 & \cdots & 0 & \frac{z_{N}^{\ast
}z_{k}\lambda_k\mathrm{e}^{2\theta_{k}}}{z_{k}-z_{N}^{\ast
}} & \frac{z_{N}^{\ast
}z_{k+1}\lambda_{k+1}}{z_{k+1}-z_{N}^{\ast }} & \cdots &
\frac{|z_{N}|^2\lambda_{N}}{z_{N}-z_{N}^{\ast }}
\end{bmatrix}%
, \\
G_{k}^{[s]}& =%
\begin{bmatrix}
\frac{|z_1|^2\lambda_1^*}{z _{1}-z_{1}^{\ast }}\gamma_{1,1} & \cdots & \frac{z_1^*z_{k-1}\lambda_{1}^*}{z_{k-1}-z_{1}^{\ast }}\gamma_{1,k-1} & \frac{z_1^*z_{k}\lambda_{1}^*}{z_{k}-z_{1}^{\ast }}\gamma_{1,k} &
0 & \cdots & 0 &0\\
\vdots & \ddots & \vdots & \vdots & \vdots & \ddots & \vdots& \vdots \\
\frac{z_{k-1}^*z_1\lambda_{k-1}^*}{z _{1}-z_{k-1}^{\ast }}\gamma_{k-1,1} & \cdots & \frac{|z_{k-1}|^2\lambda_{k-1}^*}{z _{k-1}-z_{k-1}^{\ast }}\gamma_{k-1,k-1} & \frac{z_{k-1}^*z_{k}\lambda_{k-1}^*}{z_{k}-z_{k-1}^{\ast }}\gamma_{k-1,k} & 0 & \cdots & 0&0 \\
\frac{z_{k}^*z_1\lambda_{k}^*}{z_{1}-z_{k}^{\ast }}\gamma_{k,1} & \cdots & \frac{z_{k}^*z_{k-1}\lambda_{k}^*}{z_{k-1}-z_{k}^{\ast }}\gamma_{k,k-1} & \frac{|z_k|^2(\lambda_k\mathrm{e}^{2(\theta _{k}^{\ast
}+\theta _{k})}-\lambda_k^*\gamma_{k,k})}{z_{k}-z_{k}^{\ast}} & \frac{z_k^*z_{k+1}\lambda_{k+1}\mathrm{e}^{2\theta _{k}^{\ast }}}{z_{k+1}-z_{k}^{\ast }} & \cdots &
\frac{z_k^*z_{N}\lambda_{N}\mathrm{e}^{2\theta _{k}^{\ast }}}{z_{N}-z_{k}^{\ast }}& \mathrm{e}^{2\theta _{k}^{\ast }}
\\
0 & \cdots & 0 & \frac{z_{k+1}^{\ast
}z_{k}\lambda_k\mathrm{e}^{2\theta _{k}}}{z_{k}-z_{k+1}^{\ast
}} & \frac{|z_{k+1}|^2\lambda_{k+1}}{z_{k+1}-z_{k+1}^{\ast }} & \cdots &
\frac{z_{k+1}^{\ast
}z_{N}\lambda_N}{z_{N}-z_{k+1}^{\ast }}&1 \\
\vdots & \ddots & \vdots & \vdots & \vdots & \ddots & \vdots& \vdots \\
0 & \cdots & 0 & \frac{z_{N}^{\ast
}z_{k}\lambda_k\mathrm{e}^{2\theta_{k}}}{z_{k}-z_{N}^{\ast
}} & \frac{z_{N}^{\ast
}z_{k+1}\lambda_{k+1}}{z_{k+1}-z_{N}^{\ast }} & \cdots &
\frac{|z_{N}|^2\lambda_{N}}{z_{N}-z_{N}^{\ast }}&1\\
c_{1,s} & \cdots & c_{k-1,s} & c_{k,s} & 0& \cdots &
0&0
\end{bmatrix}%
.
\end{split}%
\end{equation*}%
By direct calculation, we have
\begin{equation*}
\begin{split}
\det (G_{k}^{[s]})=& (-1)^{k+N+1}%
\begin{vmatrix}
\frac{|z_1|^2\lambda_1^*}{z _{1}-z_{1}^{\ast }}\gamma_{1,1} & \cdots & \frac{z_1^*z_{k-1}\lambda_{1}^*}{z_{k-1}-z_{1}^{\ast }}\gamma_{1,k-1} & \frac{z_1^*z_{k}\lambda_{1}^*}{z_{k}-z_{1}^{\ast }}\gamma_{1,k}
\\
\vdots & \ddots & \vdots & \vdots \\
\frac{z_{k-1}^*z_1\lambda_{k-1}^*}{z _{1}-z_{k-1}^{\ast }}\gamma_{k-1,1} & \cdots & \frac{|z_{k-1}|^2\lambda_{k-1}^*}{z _{k-1}-z_{k-1}^{\ast }}\gamma_{k-1,k-1} & \frac{z_{k-1}^*z_{k}\lambda_{k-1}^*}{z_{k}-z_{k-1}^{\ast }}\gamma_{k-1,k}\\
c_{1,s} & \cdots & c_{k-1,s} & c_{k,s}
\end{vmatrix}%
\begin{vmatrix}
\frac{z_k^*z_{k+1}\lambda_{k+1}\mathrm{e}^{2\theta _{k}^{\ast }}}{z_{k+1}-z_{k}^{\ast }} & \cdots &
\frac{z_k^*z_{N}\lambda_{N}\mathrm{e}^{2\theta _{k}^{\ast }}}{z_{N}-z_{k}^{\ast }}& \mathrm{e}^{2\theta _{k}^{\ast }}\\
 \frac{|z_{k+1}|^2\lambda_{k+1}}{z_{k+1}-z_{k+1}^{\ast }} & \cdots &
\frac{z_{k+1}^{\ast
}z_{N}\lambda_N}{z_{N}-z_{k+1}^{\ast }}&1 \\
\vdots & \ddots & \vdots & \vdots \\
\frac{z_{N}^{\ast
}z_{k+1}\lambda_{k+1}}{z_{k+1}-z_{N}^{\ast }} & \cdots &
\frac{|z_{N}|^2\lambda_{N}}{z_{N}-z_{N}^{\ast }}&1
\end{vmatrix}
\\
=&-\prod_{j=1}^{k-1}\lambda_j^*\det(\widehat{C}_s(\Delta_k))\prod_{j=k+1}^{N}\left(\lambda_jz_{j}^2\frac{z_j^*-z_k^*}{z_j-z_k^*}\right)\det(C(\mathbf{z}_{k}^{c},\mathbf{z}_k^{c\dag}))\mathrm{e}^{2\theta _{k}^{\ast }}\,,
\end{split}%
\end{equation*}%
and
\begin{equation*}
\begin{split}
& \det (M_{k}) \\
=&
\begin{bmatrix}
\frac{|z_1|^2\lambda_1^*}{z _{1}-z_{1}^{\ast }}\gamma_{1,1} & \cdots & \frac{z_1^*z_{k-1}\lambda_{1}^*}{z_{k-1}-z_{1}^{\ast }}\gamma_{1,k-1} &0 &
0 & \cdots & 0 \\
\vdots & \ddots & \vdots & \vdots & \vdots & \ddots & \vdots \\
\frac{z_{k-1}^*z_1\lambda_{k-1}^*}{z _{1}-z_{k-1}^{\ast }}\gamma_{k-1,1} & \cdots & \frac{|z_{k-1}|^2\lambda_{k-1}^*}{z _{k-1}-z_{k-1}^{\ast }}\gamma_{k-1,k-1} & 0  & 0 & \cdots & 0 \\
\frac{z_{k}^*z_1\lambda_{k}^*}{z_{1}-z_{k}^{\ast }}\gamma_{k,1} & \cdots & \frac{z_{k}^*z_{k-1}\lambda_{k}^*}{z_{k-1}-z_{k}^{\ast }}\gamma_{k,k-1} & \frac{|z_k|^2\lambda_k\mathrm{e}^{2(\theta _{k}^{\ast
}+\theta _{k})}}{z_{k}-z_{k}^{\ast}} & \frac{z_k^*z_{k+1}\lambda_{k+1}\mathrm{e}^{2\theta _{k}^{\ast }}}{z_{k+1}-z_{k}^{\ast }} & \cdots &
\frac{z_k^*z_{N}\lambda_{N}\mathrm{e}^{2\theta _{k}^{\ast }}}{z_{N}-z_{k}^{\ast }}
\\
0 & \cdots & 0 & \frac{z_{k+1}^{\ast
}z_{k}\lambda_k\mathrm{e}^{2\theta _{k}}}{z_{k}-z_{k+1}^{\ast
}} & \frac{|z_{k+1}|^2\lambda_{k+1}}{z_{k+1}-z_{k+1}^{\ast }} & \cdots &
\frac{z_{k+1}^{\ast
}z_{N}\lambda_N}{z_{N}-z_{k+1}^{\ast }} \\
\vdots & \ddots & \vdots & \vdots & \vdots & \ddots & \vdots \\
0 & \cdots & 0 & \frac{z_{N}^{\ast
}z_{k}\lambda_k\mathrm{e}^{2\theta_{k}}}{z_{k}-z_{N}^{\ast
}} & \frac{z_{N}^{\ast
}z_{k+1}\lambda_{k+1}}{z_{k+1}-z_{N}^{\ast }} & \cdots &
\frac{|z_{N}|^2\lambda_{N}}{z_{N}-z_{N}^{\ast }}
\end{bmatrix}\\
&+\begin{bmatrix}
\frac{|z_1|^2\lambda_1^*}{z _{1}-z_{1}^{\ast }}\gamma_{1,1} & \cdots & \frac{z_1^*z_{k-1}\lambda_{1}^*}{z_{k-1}-z_{1}^{\ast }}\gamma_{1,k-1} & \frac{z_1^*z_{k}\lambda_{1}^*}{z_{k}-z_{1}^{\ast }}\gamma_{1,k} &
0 & \cdots & 0 \\
\vdots & \ddots & \vdots & \vdots & \vdots & \ddots & \vdots \\
\frac{z_{k-1}^*z_1\lambda_{k-1}^*}{z _{1}-z_{k-1}^{\ast }}\gamma_{k-1,1} & \cdots & \frac{|z_{k-1}|^2\lambda_{k-1}^*}{z _{k-1}-z_{k-1}^{\ast }}\gamma_{k-1,k-1} & \frac{z_{k-1}^*z_{k}\lambda_{k-1}^*}{z_{k}-z_{k-1}^{\ast }}\gamma_{k-1,k} & 0 & \cdots & 0 \\
\frac{z_{k}^*z_1\lambda_{k}^*}{z_{1}-z_{k}^{\ast }}\gamma_{k,1} & \cdots & \frac{z_{k}^*z_{k-1}\lambda_{k}^*}{z_{k-1}-z_{k}^{\ast }}\gamma_{k,k-1} & \frac{|z_k|^2\lambda_k^*\gamma_{k,k}}{z_{k}-z_{k}^{\ast}} & \frac{z_k^*z_{k+1}\lambda_{k+1}\mathrm{e}^{2\theta _{k}^{\ast }}}{z_{k+1}-z_{k}^{\ast }} & \cdots &
\frac{z_k^*z_{N}\lambda_{N}\mathrm{e}^{2\theta _{k}^{\ast }}}{z_{N}-z_{k}^{\ast }}
\\
0 & \cdots & 0 & 0 & \frac{|z_{k+1}|^2\lambda_{k+1}}{z_{k+1}-z_{k+1}^{\ast }} & \cdots &
\frac{z_{k+1}^{\ast
}z_{N}\lambda_N}{z_{N}-z_{k+1}^{\ast }} \\
\vdots & \ddots & \vdots & \vdots & \vdots & \ddots & \vdots \\
0 & \cdots & 0 & 0 & \frac{z_{N}^{\ast
}z_{k+1}\lambda_{k+1}}{z_{k+1}-z_{N}^{\ast }} & \cdots &
\frac{|z_{N}|^2\lambda_{N}}{z_{N}-z_{N}^{\ast }}
\end{bmatrix}
\\
=&\det(C(\mathbf{z}_{k}^{c},\mathbf{z}_{k}^{c\dag}))\prod_{j=1}^{k-1}\lambda_j^*\prod_{j=k+1}^{N}\lambda_j|z_j|^2\left[\frac{|z_k|^2\lambda_k\det(C(\Delta_{k-1}))}{z_k-z_k^*}\prod_{j=k+1}^{N}\left|\frac{z_k-z_j}{z_k-z_j^*}\right|^2\mathrm{e}^{2(\theta _{k}^{\ast
}+\theta_{k})}+\lambda_k^*\det(C(\Delta_{k}))\right]\,.
\end{split}%
\end{equation*}%
Thus, along the trajectory $x-v_{k}t=\mathrm{const}$, we have
\begin{equation*}
\begin{split}
       u_s[N]&=\frac{{\displaystyle \det(\widehat{C}_s(\Delta_k))\prod_{j=k+1}^{N}\left(\frac{z_{j}^2}{|z_j|^2}\frac{z_j^*-z_k^*}{z_j-z_k^*}\right)\mathrm{e}^{2\theta _{k}^{\ast}}}}{{\displaystyle \frac{|z_k|^2\lambda_k}{z_k-z_k^*}\det(C(\Delta_{k-1}))\prod_{j=k+1}^{N}\left|\frac{z_k-z_j}{z_k-z_j^*}\right|^2\mathrm{e}^{2(\theta _{k}^{\ast
}+\theta_{k})}+\lambda_k^*\det(C(\Delta_{k}))}}+O(\mathrm{e}^{-c|t|})\\
&=u_{s,-}^{[k]}
+O(\mathrm{e}^{-c|t|})
\end{split}
\end{equation*}
as $t\rightarrow -\infty$, where
\begin{equation}\label{uk-}
    \begin{split}
    u_{s,-}^{[k]}=&\hat{c}_{k,s}^{-}\frac{z_k-z_k^*}{|z_k|^2}\frac{\mathrm{e}^{2\widehat{\theta}_{k,-}^{\ast }}}{\lambda_k\mathrm{e}^{2({\widehat{\theta}_{k,-}}^{\ast }+\widehat{\theta}_{k,-})}+\lambda_k^*\delta_{k}^{-}} \\
    \widehat{\theta}_{k,-}=&\theta_k+\frac{1}{2}\sum_{j=k+1}^{N}\ln\left(\frac{z_j-z_k}{z_j^*-z_k}\right),  \\
       \hat{c}_{k,s}^{-}=& \frac{\det(\widehat{C}_s(\Delta_k))}{\det(C(\Delta_{k-1}))}\prod_{j=k+1}^{N}\frac{z_{j}^2}{|z_j|^2}, \\
       \delta_{k}^{-}=& \frac{z_k-z_k^*}{|z_k|^2}\frac{\det(C(\Delta_k))}{\det(C(\Delta_{k-1}))}.
    \end{split}
\end{equation}
For the general
case $x-vt=\mathrm{const}$, $v\neq v_{k}(k=1,2,\cdots ,N)$, we have $u_s[N]=O(%
\mathrm{e}^{-c|t|})$. Thus the asymptotic behavior is analyzed as $t\rightarrow -\infty$.

By the same procedure as above, we can obtain the asymptotical behavior %
as $t\rightarrow +\infty$. To be specific, along the trajectory $x-v_{k}t=\mathrm{const}$, one has 
\begin{equation*}
\begin{split}
       u_s[N]&=\frac{{\displaystyle \det(\widehat{C}_s(\Delta^k))\prod_{j=1}^{k-1}\left(\frac{z_{j}^2}{|z_j|^2}\frac{z_j^*-z_k^*}{z_j-z_k^*}\right)\mathrm{e}^{2\theta _{k}^{\ast}}}}{{\displaystyle \frac{|z_k|^2\lambda_k}{z_k-z_k^*}\det(C(\Delta^{k+1}))\prod_{j=1}^{k-1}\left|\frac{z_k-z_j}{z_k-z_j^*}\right|^2\mathrm{e}^{2(\theta _{k}^{\ast
}+\theta_{k})}+\lambda_k^*\det(C(\Delta^{k}))}}+O(\mathrm{e}^{-c|t|})\\
&=u_{s,+}^{[k]}
+O(\mathrm{e}^{-c|t|})
\end{split}
\end{equation*}
as $t\rightarrow +\infty$, where
\begin{equation}\label{uk+}
    \begin{split}
    u_{s,+}^{[k]}=&\hat{c}_{k,s}^{+}\frac{z_k-z_k^*}{|z_k|^2}\frac{\mathrm{e}^{2\widehat{\theta}_{k,+}^{\ast }}}{\lambda_k\mathrm{e}^{2({\widehat{\theta}_{k,+}}^{\ast }+\widehat{\theta}_{k,+})}+\lambda_k^*\delta_{k}^{+}} \\
    \widehat{\theta}_{k,+}=&\theta_k+\frac{1}{2}\sum_{j=1}^{k-1}\ln\left(\frac{z_j-z_k}{z_j^*-z_k}\right)  \\
       \hat{c}_{k,s}^{+}=& \frac{\det(\widehat{C}_s(\Delta^k))}{\det(C(\Delta^{k+1}))}\prod_{j=1}^{k-1}\frac{z_{j}^2}{|z_j|^2}, \\
       \delta_{k}^{+}=& \frac{z_k-z_k^*}{|z_k|^2}\frac{\det(C(\Delta^k))}{\det(C(\Delta^{k+1}))}.
    \end{split}
\end{equation}
Consequently, the asymptotic behaviors of $N$-bright soliton's are analyzed. $\square$
\section{The soliton solutions with nonvanishing boundary condition}\label{par4}
In this section, we construct all kinds of soliton solutions from the plane wave seed solution. These soliton solutions of  coupled FL equation are similar to the ones of coupled nonlinear Schr\"odinger (CNLS) equation, but they have much richer structures than the ones of the CNLS equation. For instance, the bright-anti-dark soliton and dark-anti-dark soliton solutions, which do not occur in the CNLS equation, exist in the coupled FL equation.  We remark here that the soliton is called dark if its amplitude is lower than the background, and the soliton is called anti-dark if its amplitude is  higher than the background.

To this end, we start with the plane wave solution for
equation \eqref{cFL}:
\begin{equation}\label{seed}
  u_i[0]=a_i{\rm e}^{{\rm i}\omega_i},\,\, i=1,2
\end{equation}
 where
\begin{equation*}
            \begin{split}
              \omega_1=&b_1x-\frac{1}{2}\left(2a_1^2+\sigma a_2^2-\frac{2}{b_1}+\sigma a_2^2\frac{b_2}{b_1}\right)t,  \\
              \omega_2=&b_2x-\frac{1}{2}\left(2\sigma a_2^2+a_1^2-\frac{2}{b_2}+a_1^2\frac{b_1}{b_2}\right)t,
           \end{split}
\end{equation*}
$a_i\in \mathbb{R}$ and $b_1,b_2$ are nonzero real parameters. Assuming $a_1\neq 0$ and introducing $z=1/\lambda^2$, we have
the fundamental solution to the system \eqref{cFL-lax}
  \begin{equation}\label{fund2}
    \Phi_2(\lambda)=D_1E_1\mathrm{diag}\left({\rm e}^{\vartheta_1},{\rm e}^{\vartheta_2},{\rm e}^{\vartheta_3}\right),\,\, D_1=\mathrm{diag}\left(1,{\rm e}^{{\rm i}\omega_1},{\rm e}^{{\rm i}\omega_2}\right),
  \end{equation}
where
  \begin{equation*}
  \begin{split}
      \vartheta_i&={\rm i}(\kappa_i-z)\left(x+\frac{1}{2b_1b_2z}(\kappa_i-z+b_1+b_2)t\right),\,\, i=1,2,3.\\
  \end{split}
  \end{equation*}
Then we have the following conclusions
\begin{itemize}
    \item If $b_1=b_2$, then
  \begin{equation*}
  \begin{split}
      E_1&=\begin{pmatrix}
             1 & 1 & 0 \\
             \frac{a_1b_1}{\lambda(\kappa_1+b_1)} & \frac{a_1b_1}{\lambda(\kappa_2+b_1)} & -\frac{a_2}{a_1} \\
             \frac{a_2b_1}{\lambda(\kappa_1+b_1)} & \frac{a_2b_1}{\lambda(\kappa_2+b_1)} & 1 \\
           \end{pmatrix},
  \end{split}
  \end{equation*}
 where  $\kappa_i, (i=1,2)$ satisfies the following equation
  \begin{equation}\label{chara-1}
    (\kappa/z-2)(\kappa+b_1)+(a_1^2+\sigma a_2^2)b_1^2=0\,,
  \end{equation}
  and $\kappa_3=-b_1.$
 \item If $b_1\neq b_2$ and $a_2\neq 0$, then
  \begin{equation*}
  \begin{split}
      E_1&=\begin{pmatrix}
             1 & 1 & 1 \\
             \frac{a_1b_1}{\lambda(\kappa_1+b_1)} & \frac{a_1b_1}{\lambda(\kappa_2+b_1)} & \frac{a_1b_1}{\lambda(\kappa_3+b_1)} \\
             \frac{a_2b_2}{\lambda(\kappa_1+b_2)} & \frac{a_2b_2}{\lambda(\kappa_2+b_2)} & \frac{a_2b_2}{\lambda(\kappa_3+b_2)} \\
           \end{pmatrix},
  \end{split}
  \end{equation*}
  where $\kappa_i, (i=1,2,3)$ satisfies the following equation
  \begin{equation}\label{chara-2}
    (\kappa/z-2)(\kappa+b_1)(\kappa+b_2)+a_1^2 b_1^2(\kappa+b_2)+\sigma a_2^2b_2^2(\kappa+b_1)=0.
  \end{equation}
\end{itemize}
Based on the roots of characteristic equation \eqref{chara-1} and \eqref{chara-2}, we can classify the parameters into four cases. Here
we denote $z=z_1=\lambda_1^{-2}$.
\begin{description}
  \item[Case (a):] $\Omega_1\equiv \{a_1,a_2,b_1,b_2| b_1=b_2,\,\, a_1\neq 0,\,\, 2b_1-(a_1^2+\sigma a_2^2)b_1^2=0\}$. In this case, the root for the characteristic equation \eqref{chara-1} can be solved as
  \begin{equation*}
      \kappa_1=2z_1-b_1,\,\, \kappa_2=0,\,\, \kappa_3=-b_1.
\end{equation*}
  \item[Case (b):] $\Omega_2\equiv \{a_1,a_2,b_1,b_2| b_1=b_2,\,\, a_1\neq 0,\,\, 2b_1-(a_1^2+\sigma a_2^2)b_1^2\neq 0\}$. In this case, to avoid the radical expression for the root, we introduce a variable $\xi_1\in \mathbb{C}$.
 Denote $\gamma_1=b_1-(a_1^2+\sigma a_2^2)b_1^2$, $\delta_1=|b_1|\sqrt{|a_1^2+\sigma a_2^2||\gamma_1+b_1|}$, we then have
 \begin{equation}\label{xi1}
    \begin{split}
     z_1=& \left\{
\begin{array}{l}
\displaystyle \frac{\delta_1}{4}(\xi_1-\xi_1^{-1})-\frac{\gamma_1}{2}, \quad \text{if} (\gamma_1+b_1)(a_1^2+\sigma a_2^2)>0\,, \\
\displaystyle\frac{\delta_1}{4}(\xi_1+ \xi_1^{-1})-\frac{\gamma_1}{2}, \quad \text{if} (\gamma_1+b_1)(a_1^2+\sigma a_2^2)< 0 \,,%
\end{array}%
\right. \\
      \kappa_1=&-\frac{b_1+\gamma_1}{2}+\frac{\delta_1}{2}\xi_1,\\
      \kappa_2=&-\frac{b_1+\gamma_1}{2}\mp\frac{\delta_1}{2}\xi_1^{-1}, \quad    \kappa_3=-b_1.
    \end{split}
 \end{equation}
  \item[Case (c):] $\Omega_3\equiv \{a_1,a_2,b_1,b_2| a_1,a_2\neq 0,\,\, \sigma b_2=2a_2^{-2}-b_1a_1^{2}a_2^{-2}\}$. In this case, we introduce a variable $\xi_2\in \mathbb{C}$ for the same reason as case (b).
 Denote $\delta_2=\frac{a_1^2|(a_1^2+\sigma a_2^2)b_1-2|}{a_2^2}\sqrt{|b_1(2/a_1^{2}-b_1)|}$, $\gamma_2=-\frac{(1-a_1^2b_1)((a_1^2+\sigma a_2^2)b_1-2)}{2a_2^2}$, it then follows 
 \begin{equation}\label{xi2}
    \begin{split}
       z_1= &
    \left\{
\begin{array}{l}
\displaystyle \frac{\delta_2}{4}(\xi_2-\xi_2^{-1})+\gamma_2,\quad \text{if } b_1(2/a_1^{2}-b_1)>0\,, \\
\displaystyle \frac{\delta_2}{4}(\xi_2+\xi_2^{-1})+\gamma_2, \quad \text{if } b_1(2/a_1^{2}-b_1)<0\,,%
\end{array}%
\right. \\
        \\
       \kappa_1=&-\frac{b_1+b_2}{2}+\gamma_2+\frac{\delta_2}{2}\xi_2,\\
       \kappa_2=&-\frac{b_1+b_2}{2}+\gamma_2\mp\frac{\delta_2}{2}\xi_2^{-1},\,\quad  \kappa_3=0.
    \end{split}
 \end{equation}
  \item[Case (d):] $\Omega_4\equiv \{a_1,a_2,b_1,b_2| a_1,a_2\neq 0,\,\, \sigma b_2\neq 2a_2^{-2}-b_1a_1^{2}a_2^{-2},\, b_1\neq b_2\}$. In this case, the  radical expression for the roots is we unavoidable. A direct way of obtaining the roots is to use the Kardan formula. However, one often turns to find numerical solutions for the roots to avoid the complicated formula.
\end{description}
\begin{rem}
The relations between the roots and coefficients for the characteristic equation \eqref{chara-2} are
\begin{equation}\label{root-coeff}
    \begin{split}
       \sum_{i=1}^3\kappa_i+b_1+b_2=&2z_1,  \\
       \prod_{i=1}^3(\kappa_i+b_1)=&a_1^2b_1^2(b_1-b_2)(\sum_{i=1}^3\kappa_i+b_1+b_2),     \\
       \prod_{i=1}^3(\kappa_i+b_2)=&-\sigma a_2^2b_2^2(b_1-b_2)(\sum_{i=1}^3\kappa_i+b_1+b_2).
    \end{split}
\end{equation}
\end{rem}
In what follows, we will use the formulas \eqref{backlund1}, \eqref{seed} and \eqref{fund2} to construct single soliton solutions. Generally, the single soliton solutions can be represented as
\begin{equation}\label{localizedwave}
  u_1[1]=a_1\left[\frac{\beta+\frac{2\psi_1\varphi_1^*}{a_1}{\rm e}^{-{\rm i}\omega_1}}{\beta}\right]{\rm e}^{{\rm i}\omega_1},\,\, u_2[1]=a_2\left[\frac{\frac{\beta+2\chi_1\varphi_1^*}{a_2}{\rm e}^{-{\rm i}\omega_2}}{\beta}\right]{\rm e}^{{\rm i}\omega_2}.
\end{equation}
Prior to calculating the explicit form for $\beta$, we need the following proposition.
\begin{prop}\label{prop3}
If we choose a special solution $|y_1\rangle=(\varphi_1,\psi_1,\chi_1)^{\mathrm{T}}\equiv \Phi_2(\lambda_1)(c_1,c_2,c_3)^{\mathrm{T}}$, then
\begin{equation}\label{beta}
         \beta=2\lambda_1^{-1}M_1,\,\,
          M_1=Z_{l,m}:c_l^*c_m{\rm e}^{\vartheta_l^*+\vartheta_m}\equiv\sum_{l=1}^{3}\sum_{m=1}^{3}Z_{l,m}c_l^*c_m{\rm e}^{\vartheta_l^*+\vartheta_m}\,,
      \end{equation}
where $Z_{l,m}$ are of the following forms depending on four cases mentioned previously.
\begin{itemize}
   \item  If $a_1,a_2,b_1,b_2\in \Omega_1$, then
      \begin{equation}\label{beta1}
          Z_{l,m}=\left\{
                 \begin{array}{ll}
                   \frac{\kappa_l^*}{\kappa_m-\kappa_l^*} &  \text{if } 1\leq l\leq 2,\,\, 1\leq m\leq 2,\,\, l+m<4, \\[8pt]
                   \frac{|z_1|^2((a_1^2+\sigma a_2^2)z_1-1)}{|z_1|^2-z_1^2} & \text{if } l=m=2, \\
                   \frac{(\sigma+\frac{a_2^2}{a_1^2})|z_1|}{z_1^*-z_1} & \text{if } l=m=3, \\
                   0& \text{otherwise}\,.
                 \end{array}
               \right.
      \end{equation}
      \item If $a_1,a_2,b_1,b_2\in \Omega_2$, then
      \begin{equation}\label{beta2}
       Z_{l,m}=\left\{
                 \begin{array}{ll}
                   \frac{\kappa_l^*}{\kappa_m-\kappa_l^*} & \text{if } 1\leq l\leq 2,\,\, 1\leq m\leq 2, \\[8pt]
                   \frac{(\sigma+\frac{a_2^2}{a_1^2})|z_1|}{z_1^*-z_1} & \text{if } l=m=3, \\
                   0 & \text{otherwise}\,.
                 \end{array}
               \right.
      \end{equation}
   \item If $a_1,a_2,b_1,b_2\in \Omega_3$, then
      \begin{equation}\label{beta3}
       Z_{l,m}=\left\{
                 \begin{array}{ll}
                 \frac{|z_1|^2((a_1^2+\sigma a_2^2)z_1-1)}{|z_1|^2-z_1^2} & \text{if } l=m=3,\\[8pt]
                  \frac{\kappa_l^*}{\kappa_m-\kappa_l^*} & \text{otherwise}\,.
                 \end{array}
               \right.
      \end{equation}
  \item If $a_1,a_2,b_1,b_2\in \Omega_4$, then
     \begin{equation}\label{beta4}
     Z_{l,m}=\frac{\kappa_l^*}{\kappa_m-\kappa_l^*}.
      \end{equation}
\end{itemize}
\end{prop}
\textbf{Proof:} Firstly, we have
\begin{equation}\label{equation1}
  \frac{\kappa_m}{z_1}-2+\frac{a_1^2b_1^2}{\kappa_m+b_1}+\frac{\sigma a_2^2b_2^2}{\kappa_m+b_2}=0\,,
\end{equation}
and its complex conjugate
\begin{equation}\label{equation2}
  \frac{\kappa_l^*}{z_1^*}-2+\frac{a_1^2b_1^2}{\kappa_l^*+b_1}+\frac{\sigma a_2^2b_2^2}{\kappa_l^*+b_2}=0.
\end{equation}
Subtracting  equation \eqref{equation2} from \eqref{equation1} yields
\begin{equation}\label{equation3}
  \frac{\lambda_1^2\kappa_m-\lambda_1^{*2}\kappa_l^*}{(\kappa_l^*-\kappa_m)}+\frac{a_1^2b_1^2}{(\kappa_m+b_1)(\kappa_l^*+b_1)}+\frac{\sigma a_2^2b_2^2}{(\kappa_m+b_2)(\kappa_l^*+b_2)}=0.
\end{equation}
The coefficient of ${\rm e}^{\vartheta_m+\vartheta_l^*}$ 
can be simplified as
\begin{equation}\begin{split}
                    &\frac{2 c_lc_m^*}{\lambda_1^{*2}-\lambda_1^2}\left[\lambda_1-\frac{\lambda_1^*}{\lambda_1\lambda_1^*}\left(\frac{a_1^2b_1^2}{(\kappa_m+b_1)(\kappa_l^*+b_1)}+\frac{\sigma a_2^2b_2^2}{(\kappa_m+b_2)(\kappa_l^*+b_2)}\right)\right]  \\
                    =&\frac{2 c_lc_m^*}{\lambda_1^{*2}-\lambda_1^2}\left[\frac{\lambda_1^*(\lambda_1^2\kappa_m-\lambda_1^{*2}\kappa_l^*)+\lambda_1^2\lambda_1^*(\kappa_l^*-\kappa_m)}{
                    \lambda_1\lambda_1^*(\kappa_l^*-\kappa_m)}\right]\\
                    =&\frac{2 c_lc_m^*\kappa_l^*}{
                    \lambda_1(\kappa_m-\kappa_l^*)}
                 \end{split}
\end{equation}
by referring equation \eqref{equation3}. The four cases can be proved by analyzing above relation by tedious work, which is omitted here.

We comment here that if $\kappa_i\in \mathbb{R}$, we can not use the above relation. Instead, we have to use the definition to derive the relation directly.
$\square$
 \subsection{Bright-dark/Bright-anti-dark solution}
 First, we consider $a_1,a_2,b_1,b_2\in \Omega_1$ and $a_2=0$. Under this case, one can obtain the following general solution by formula \eqref{localizedwave}
\begin{equation*}
    \begin{split}
       u_1[1]=&a_1\left(\frac{N_1^{[1]}}{M_1}\right){\rm e}^{{\rm i}\omega_1},  \,\,
       u_2[1]=\frac{N_2^{[1]}}{M_1}{\rm e}^{{\rm i}\omega_1},
    \end{split}
  \end{equation*}
where
\begin{equation*}
  \begin{split}
     N_1^{[1]}=&{\displaystyle \sum_{l,m=1,\,\, l+m<4}^{2,2}\frac{\kappa_l^*+b_1}{\kappa_m+b_1}\frac{c_mc_l^*\kappa_m{\rm e}^{\vartheta_m+\vartheta_l^*}}{\kappa_m-\kappa_l^*}
        +\frac{|c_2|^2z_1^2(a_1^2z_1^*-1)}{|z_1|^2-z_1^2}{\rm e}^{\vartheta_2+\vartheta_2^*}+\frac{\sigma|c_3|^2|z_1|}{(z_1^*-z_1)}{\rm e}^{\vartheta_3+\vartheta_3^*}},  \\
      N_2^{[1]}=&c_3\lambda_1(c_1^*{\rm e}^{\vartheta_3+\vartheta_1^*}+c_2^*{\rm e}^{\vartheta_3+\vartheta_2^*})\,.
  \end{split}
\end{equation*}
If $c_1=0$, $c_2c_3\neq 0$, the solution $|u_1[1]|^2=a_1^2$ and $|u_2[1]|^2$ is a bright soliton along the line $t=\text{const}$. If $c_2=0$, $c_1c_3\neq 0$, the solution is a bright-dark soliton.
If $c_3=0$, $c_1c_2\neq 0$, then the solution $|u_1[1]|^2$ is a breather solution and $u_2[1]=0$.
If $c_1c_2c_3\neq 0$, the solution is a resonant bright-dark-breather solution which can be viewed as the nonlinear superposition of above three types of solutions.
\begin{figure}[tbh]
\centering
\subfigure[$|u_1|^2$]{%
\includegraphics[height=50mm,width=65mm]{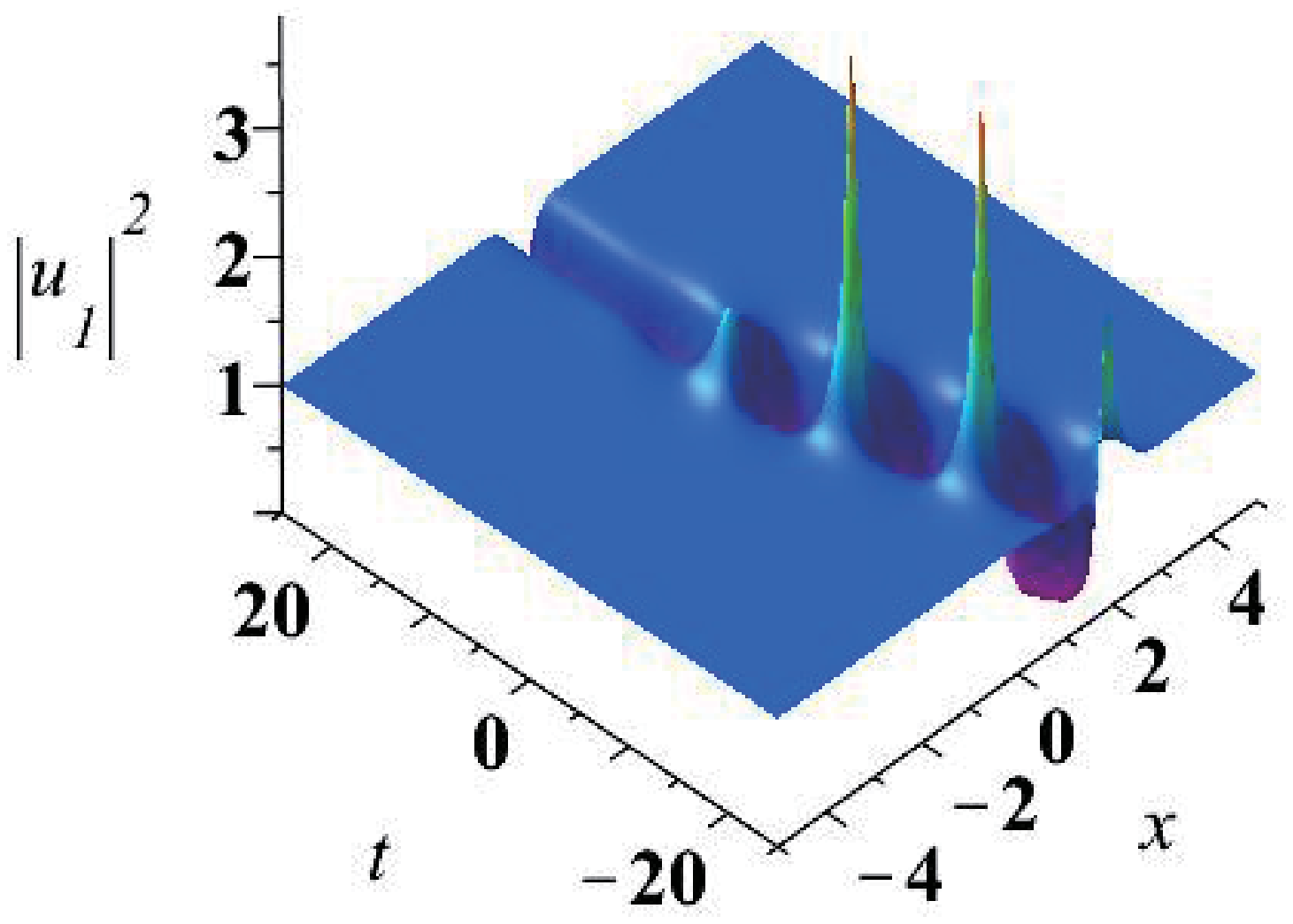}} \hfil
\subfigure[$|u_2|^2$]{%
\includegraphics[height=50mm,width=65mm]{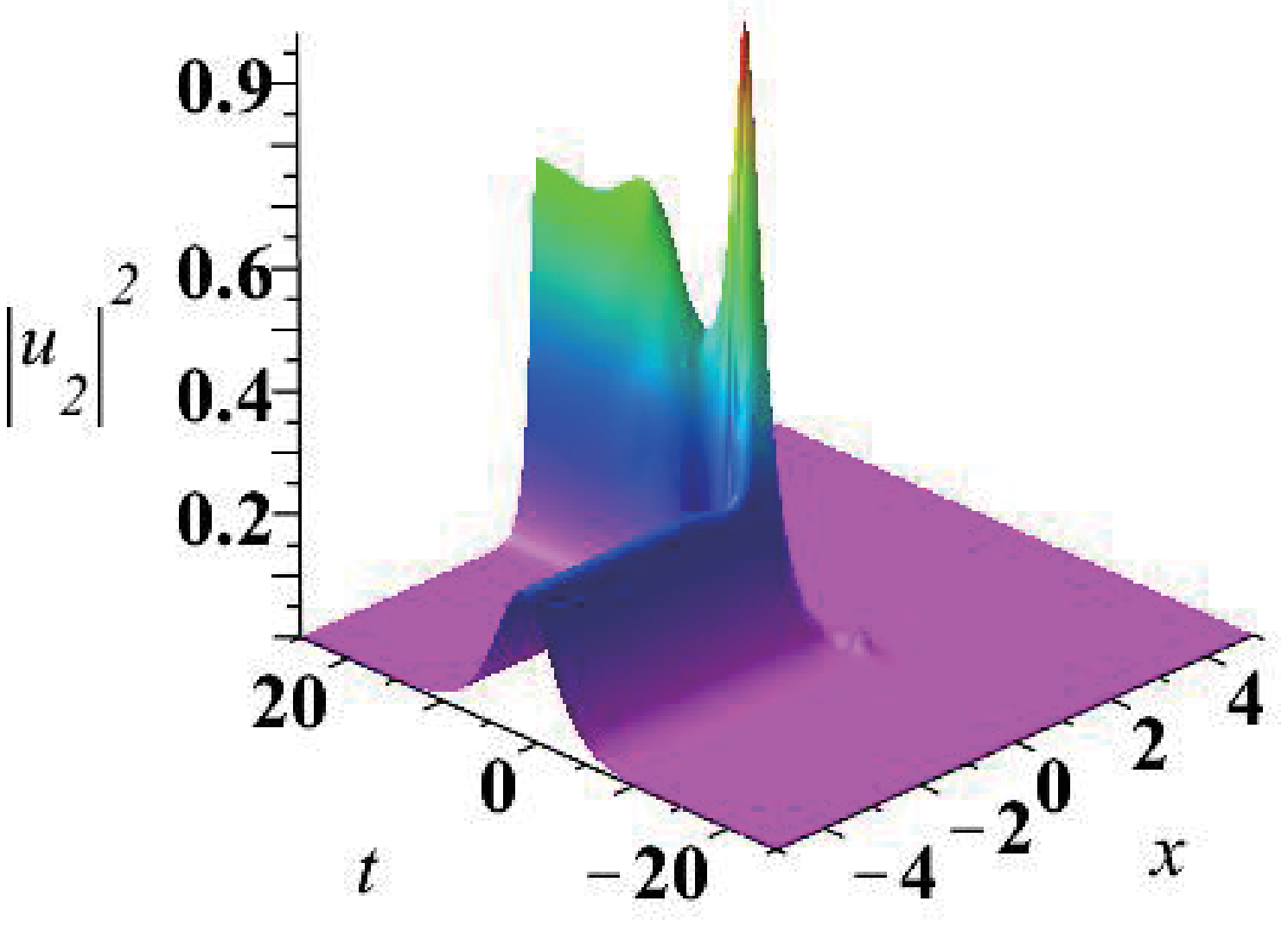}}
\caption{(color online): Resonant bright-dark-breather solution with parameters: $a_1=1$, $b_1=b_2=2$, $\lambda_1=\frac{1}{2}(1+{\rm i})$, $z_1=-2{\rm i}$, $\kappa_1=-2-4{\rm i}$, $\kappa_2=0$, $c_1=c_2=c_3=1$.}
\label{fig1}
\end{figure}
Then we consider the case $a_1,a_2,b_1,b_2\in \Omega_2$ and $a_2=0$, under which we can obtain the following general solution by formula \eqref{localizedwave}
  \begin{equation*}
    \begin{split}
       u_1[1]=&a_1\left(\frac{N_1^{[2]}}{M_1}\right){\rm e}^{{\rm i}\omega_1},  \\
       u_2[1]=&\left(\frac{N_2^{[2]}}{M_1}\right){\rm e}^{{\rm i}\omega_1},
    \end{split}
  \end{equation*}
 where
 \begin{equation*}
   \begin{split}
      N_1^{[2]}=&\sum_{l,m=1}^{2}\frac{\kappa_l^*+b_1}{\kappa_m+b_1}\frac{c_mc_l^*\kappa_m{\rm e}^{\vartheta_m+\vartheta_l^*}}{\kappa_m-\kappa_l^*}
        +\frac{\sigma|c_3|^2|z_1|}{(z_1^*-z_1)}{\rm e}^{\vartheta_3+\vartheta_3^*} \\
      N_2^{[2]}=&c_3\lambda_1(c_1^*{\rm e}^{\vartheta_3+\vartheta_1^*}+c_2^*{\rm e}^{\vartheta_3+\vartheta_2^*})
   \end{split}
 \end{equation*}
 and
 \begin{equation*}
      \vartheta_i={\rm i}(\kappa_i-z_1)\left[x+\frac{z_1^*}{2b_1b_2|z_1|^2}(\kappa_i-z_1+2b_1)t\right].
 \end{equation*}
It can be shown that if $c_1=0$, $c_2c_3\neq 0$ ($c_2=0$, $c_1c_3\neq 0$), the solution is either a bright-dark soliton or a bright-anti-dark soliton. Particularly,  when $\mathrm{sign}(\mathrm{Re}(\kappa_1)/\mathrm{Re}(z_1))>0$, the solution is a bright-dark soliton. The peak for $|u_1[1]|^2$ is along the line
 $\mathrm{Re}(\vartheta_3-\vartheta_1)+\ln\left(\frac{|f_1||z_1|\mathrm{Im}(\kappa_1)}{|d_1||\kappa_1|\mathrm{Im}(z_1)}\right)=0$, and the peak values are $a_1^2\left[1-\frac{2\mathrm{Im}(\kappa_1)\mathrm{Im}(\frac{\kappa_1}{\kappa_1+b_1})}{|\kappa_1|-\mathrm{Re}(\kappa_1)}\right]$ 
 for $|u_1[1]|^2$ and $\frac{2\left|\mathrm{Im}(z_1)\mathrm{Im}(\kappa_1)\right|}{|z_1|^2(|\kappa_1|-\mathrm{Re}(\kappa_1))}$ for $|u_2[1]|^2$ . When $\mathrm{sign}(\mathrm{Re}(\kappa_1)/\mathrm{Re}(z_1))<0$, the solution is a bright-anti-dark soliton. The peak values are $a_1^2\left[1+\frac{2\left|\mathrm{Im}(\kappa_1)\mathrm{Im}(\frac{\kappa_1}{\kappa_1+b_1})\right|}{|\kappa_1|+\mathrm{Re}(\kappa_1)}\right]$ $|u_1[1]|^2$ and $\frac{2\left|\mathrm{Im}(z_1)\mathrm{Im}(\kappa_1)\right|}{|z_1|^2(|\kappa_1|+\mathrm{Re}(\kappa_1))}$ for $|u_2[1]|^2$.

 On the other hand, as $c_1c_2\neq 0$, $c_3=0$, one obtains a breather solution, as $c_1c_2c_3\neq 0$, one arrives at a resonant bright-dark-breather solution. 
\begin{figure}[tbh]
\centering
\subfigure[$|u_1|^2$]{%
\includegraphics[height=50mm,width=65mm]{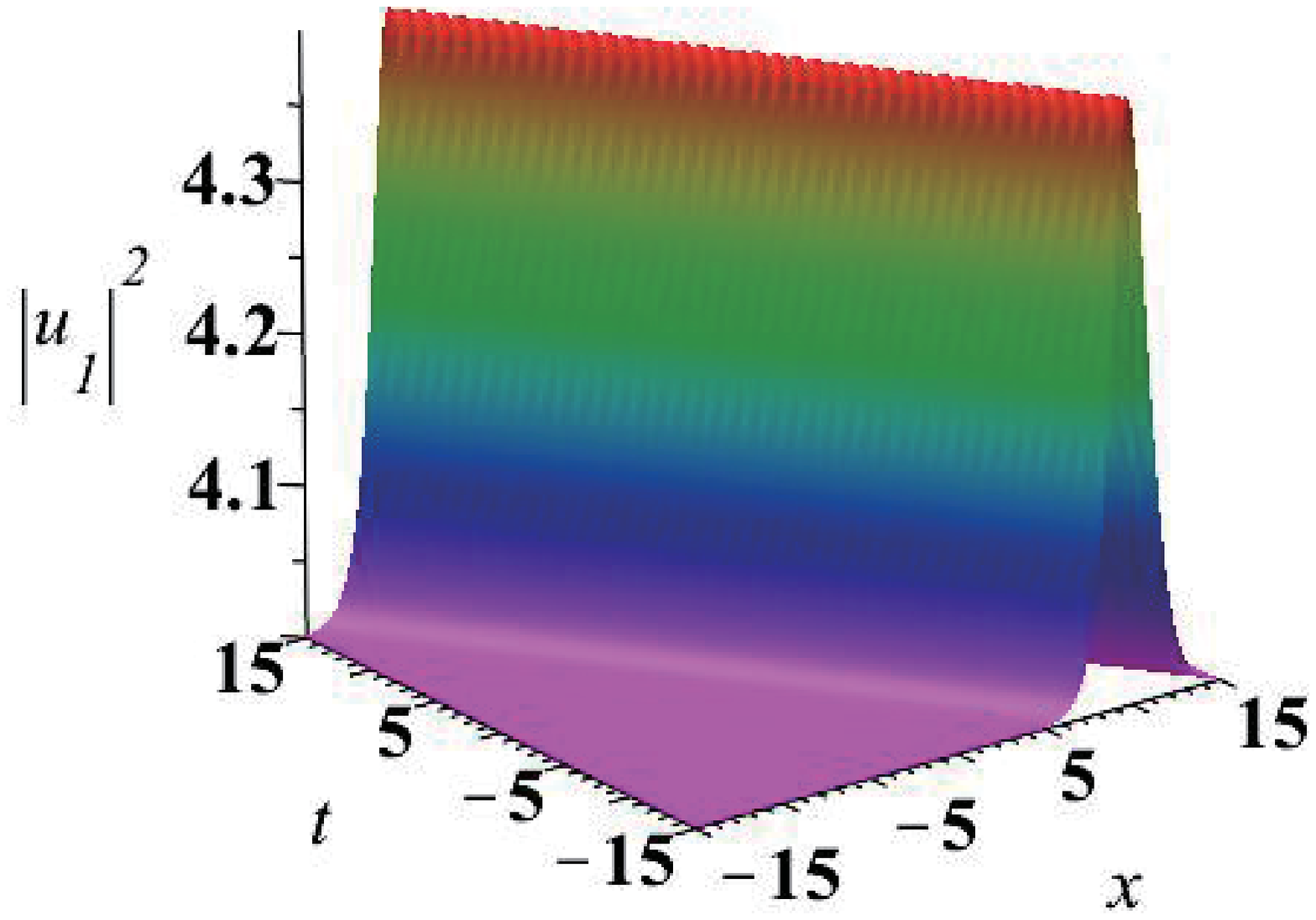}} \hfil
\subfigure[$|u_2|^2$]{%
\includegraphics[height=50mm,width=65mm]{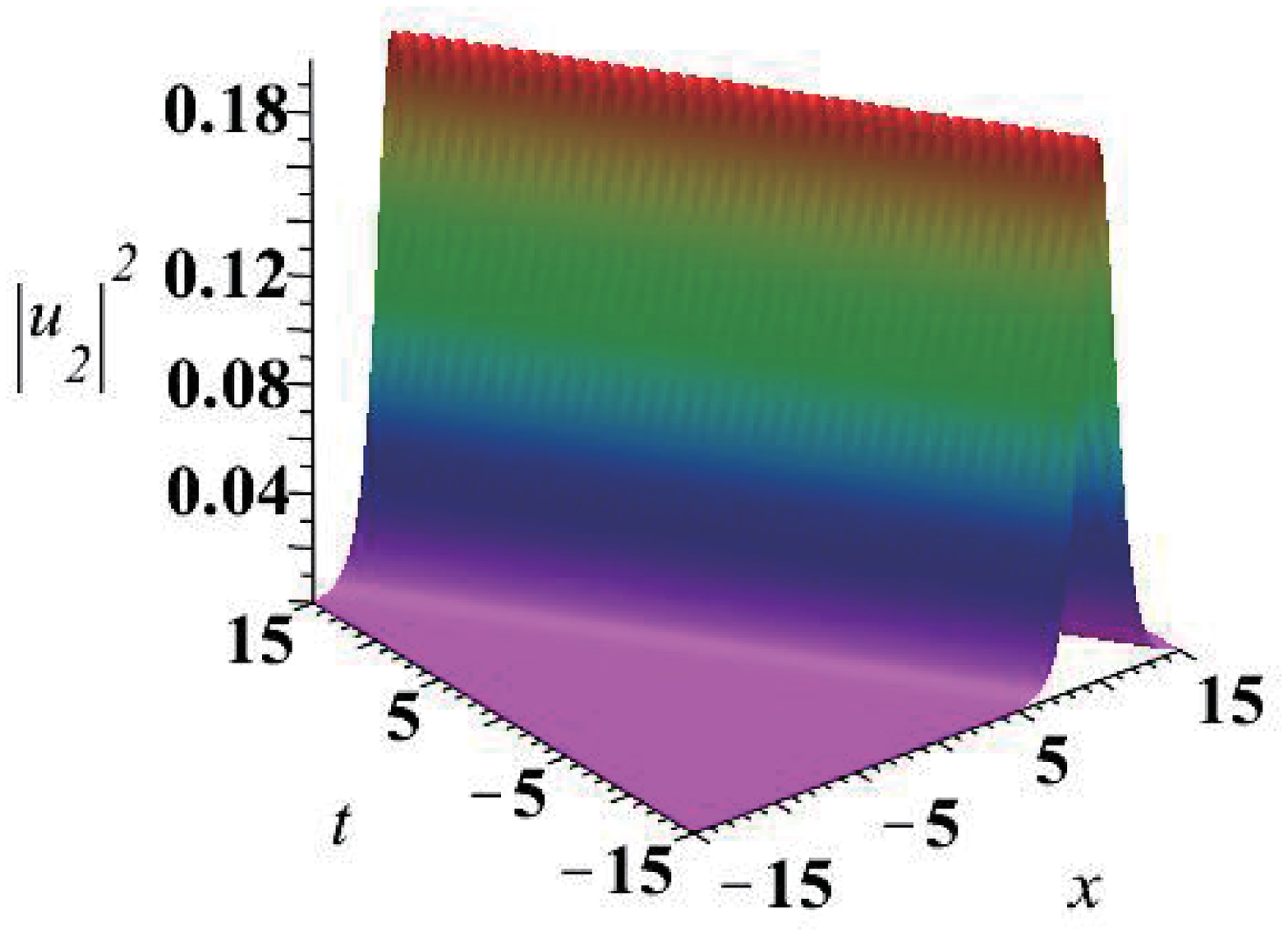}}
\caption{(color online): Bright-anti-dark soliton solution with parameters: $a_1=2$, $b_1=b_2=1$, $z_1=\frac{3}{2}-\frac{3}{4}\sqrt{2}{\rm i}$, $\kappa_1=1-2\sqrt{2}{\rm i}$,
$\kappa_2=1+\frac{\sqrt{2}}{2}{\rm i}$, $c_1=0$, $c_2=c_3=1$. The peaks value for $|u_1|^2$ is $4.399$ and for $|u_2|^2$ is $0.199$.}
\label{fig2}
\end{figure}

\begin{figure}[tbh]
\centering
\subfigure[$|u_1|^2$]{%
\includegraphics[height=50mm,width=65mm]{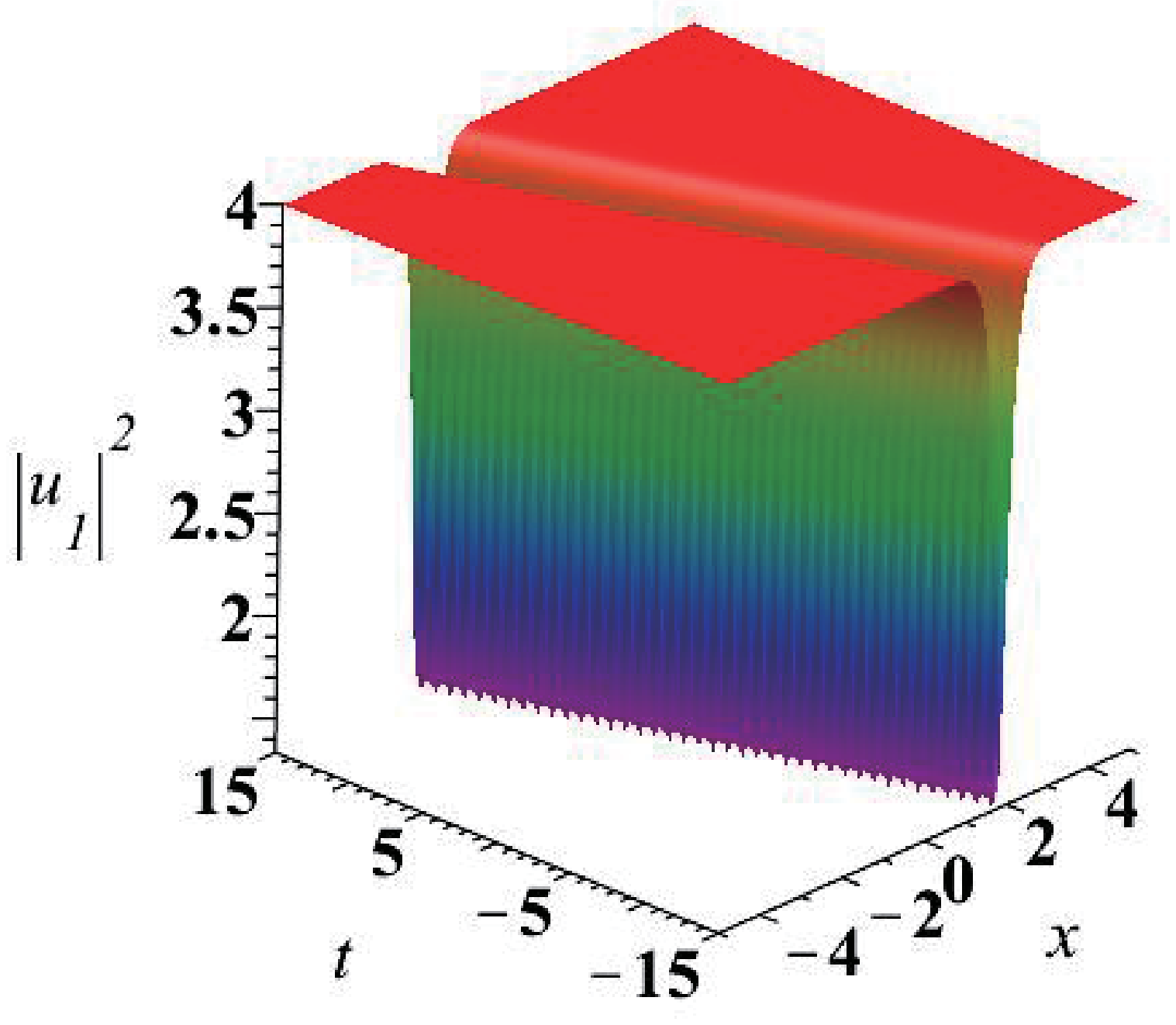}} \hfil
\subfigure[$|u_2|^2$]{%
\includegraphics[height=50mm,width=65mm]{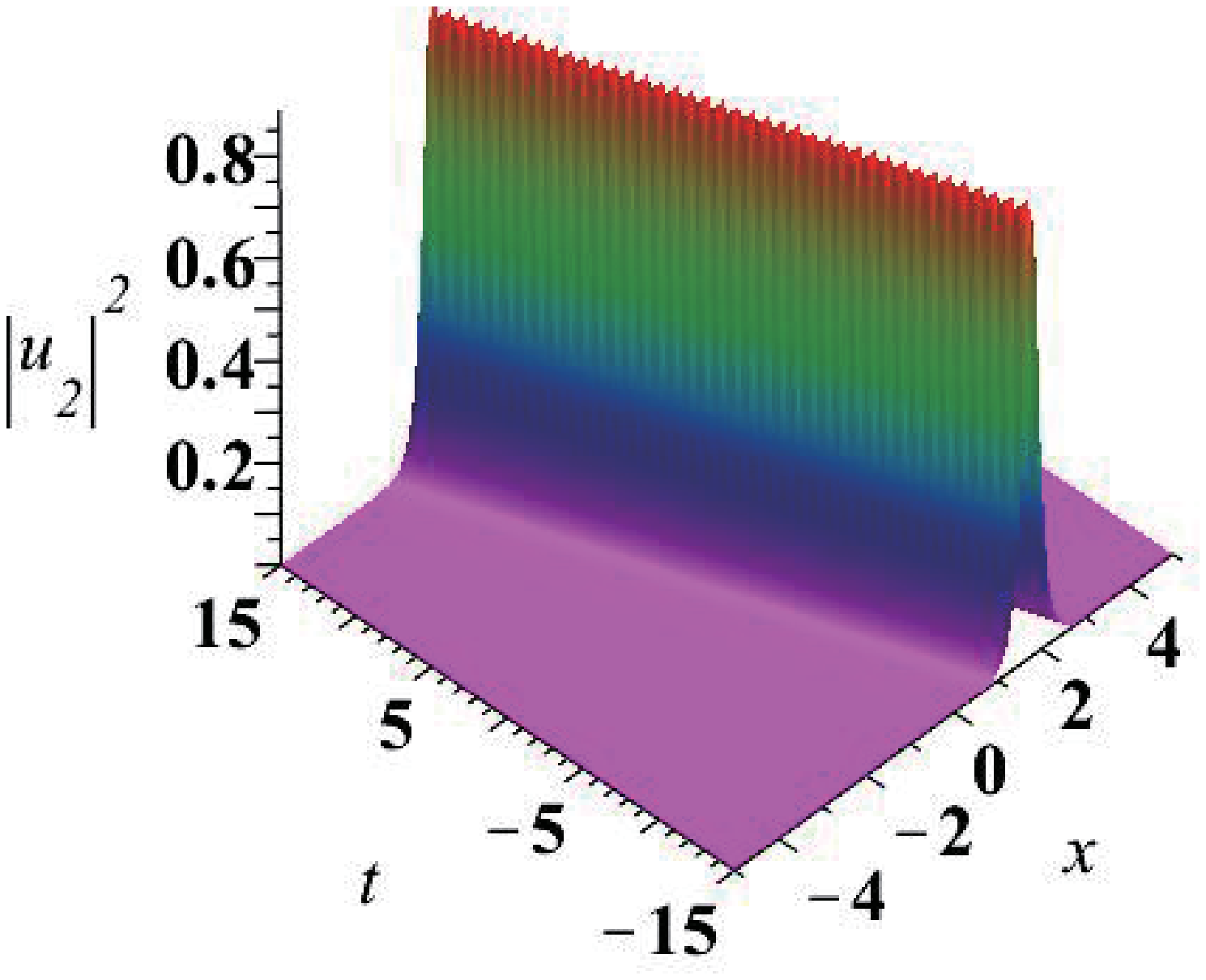}}
\caption{(color online): Bright-dark soliton solution with parameters: $a_1=2$, $b_1=b_2=1$, $z_1=\frac{3}{2}-\frac{3}{4}\sqrt{2}{\rm i}$, $\kappa_1=1-2\sqrt{2}{\rm i}$,
$\kappa_2=1+\frac{\sqrt{2}}{2}{\rm i}$, $c_2=0$, $c_1=c_3=1$. The peaks value for $|u_1|^2$ is $\frac{4}{3}$ and for $|u_2|^2$ is $\frac{8}{9}$.}
\label{fig3}
\end{figure}

 \subsection{Dark/anti-dark soliton}
In CNLS equation, there only exists dark soliton solution. Whereas there exists dark and anti-dark soliton solutions in the coupled FL equation \eqref{cFL}. If we apply the formula \eqref{localizedwave} for a complex spectral parameter $\lambda_1$, the singularity occurs. With the aid of technique in  \cite{ling2}, we can derive the dark/anti-dark soliton to equation \eqref{cFL}.

Choosing $\lambda_1\in \mathbb{R}\cup {\rm i}\mathbb{R}$ such that the characteristic equations \eqref{chara-1},\eqref{chara-2} possess a pair of conjugate complex roots. 
Further, by taking a special solution
  \begin{equation*}
    |y_1\rangle =D_1\left[\begin{pmatrix}
                            1 \\
                            \frac{a_1b_1}{\lambda_1(\kappa_1+b_1)} \\
                             \frac{a_1b_1}{\lambda_1(\kappa_1+b_1)} \\
                          \end{pmatrix}{\rm e}^{\vartheta_1}+\alpha_1(\lambda_1^2-\lambda_1^{*2})\begin{pmatrix}
                            1 \\
                            \frac{a_1b_1}{\lambda_1(\kappa_2+b_1)} \\
                             \frac{a_1b_1}{\lambda_1(\kappa_2+b_1)} \\
                          \end{pmatrix}{\rm e}^{\vartheta_2}
    \right]
  \end{equation*}
  and combining with the limit $\lambda_1\rightarrow \pm \lambda_1^*$, $\kappa_2\rightarrow \kappa_1^*$ for an appropriate $\alpha_1$, we obtain 
  \begin{equation*}
  \begin{split}
    \beta & \rightarrow \frac{2\lambda_1^{-1}}{\kappa_1-\kappa_1^*}\left[\kappa_1^*{\rm e}^{\vartheta_1+\vartheta_1^*}\pm |\kappa_1|\right],\\
    \psi_1\varphi_1^* &\rightarrow\frac{a_1b_1\lambda_1^{-1}}{\kappa_1+b_1}{\rm e}^{\vartheta_1+\vartheta_1^*},\\
    \chi_1\varphi_1^* &\rightarrow\frac{a_2b_2\lambda_1^{-1}}{\kappa_1+b_1}{\rm e}^{\vartheta_1+\vartheta_1^*}.
  \end{split}
  \end{equation*}
  Finally, we can obtain a soliton solution of either dark or anti-dark type
  \begin{equation}\label{dark-soliton}
    \begin{split}
      u_s[1]=&a_s\left[\frac{\kappa_1\exp(\vartheta_1+\vartheta_1^*+{\rm i}\tau_s)+\varsigma_1|\kappa_1|}{\kappa_1^*\exp(\vartheta_1+\vartheta_1^*)+\varsigma_1|\kappa_1|}\right]{\rm e}^{{\rm i}\omega_s}\,,
    \end{split}
  \end{equation}
  where $\exp({\rm i}\tau_{s})=\frac{\kappa_1^*+b_s}{\kappa_1+b_s}$ and $\varsigma_1=\pm1$. The peak values of $|u_s[1]|^2$ are
  $$a_i^2\left[1+\frac{2b_i\kappa_{1I}^2}{(\kappa_{1R}+\varsigma_1 |\kappa_1|)|\kappa_1+b_i|^2}\right]$$.
  It can be shown by direct calculation that if $\varsigma_1b_i>0$, the solution \eqref{dark-soliton} is an anti-dark soliton; if $\varsigma_1b_i>0$, it is a dark soliton.
The velocity of the dark/anti-dark soliton is $v=\frac{-1}{b_1b_2z_1}\left(\kappa_{1R}-z_1+\frac{b_1+b_2}{2}\right)$.
To obtain the stationary solution, we must choose the parameters such that $b_1+b_2=0$, $\kappa_{1R}=z_1>0,$ $\kappa_1=\kappa_{1R}+{\rm i}\kappa_{1I},$ and
  \begin{equation*}
    a_1=\pm\frac{\sqrt{2}}{2|b_1\lambda_1|}\sqrt{(b_1\lambda_1^2+1)^2+\kappa_{1I}^2\lambda_1^4},\,\, a_2=\pm\frac{\sqrt{2}}{2|b_1\lambda_1|}\sqrt{(b_1\lambda_1^2-1)^2+\kappa_{1I}^2\lambda_1^4}.
  \end{equation*}
\begin{figure}[tbh]
\centering
\subfigure[$|u_1|^2$]{%
\includegraphics[height=50mm,width=65mm]{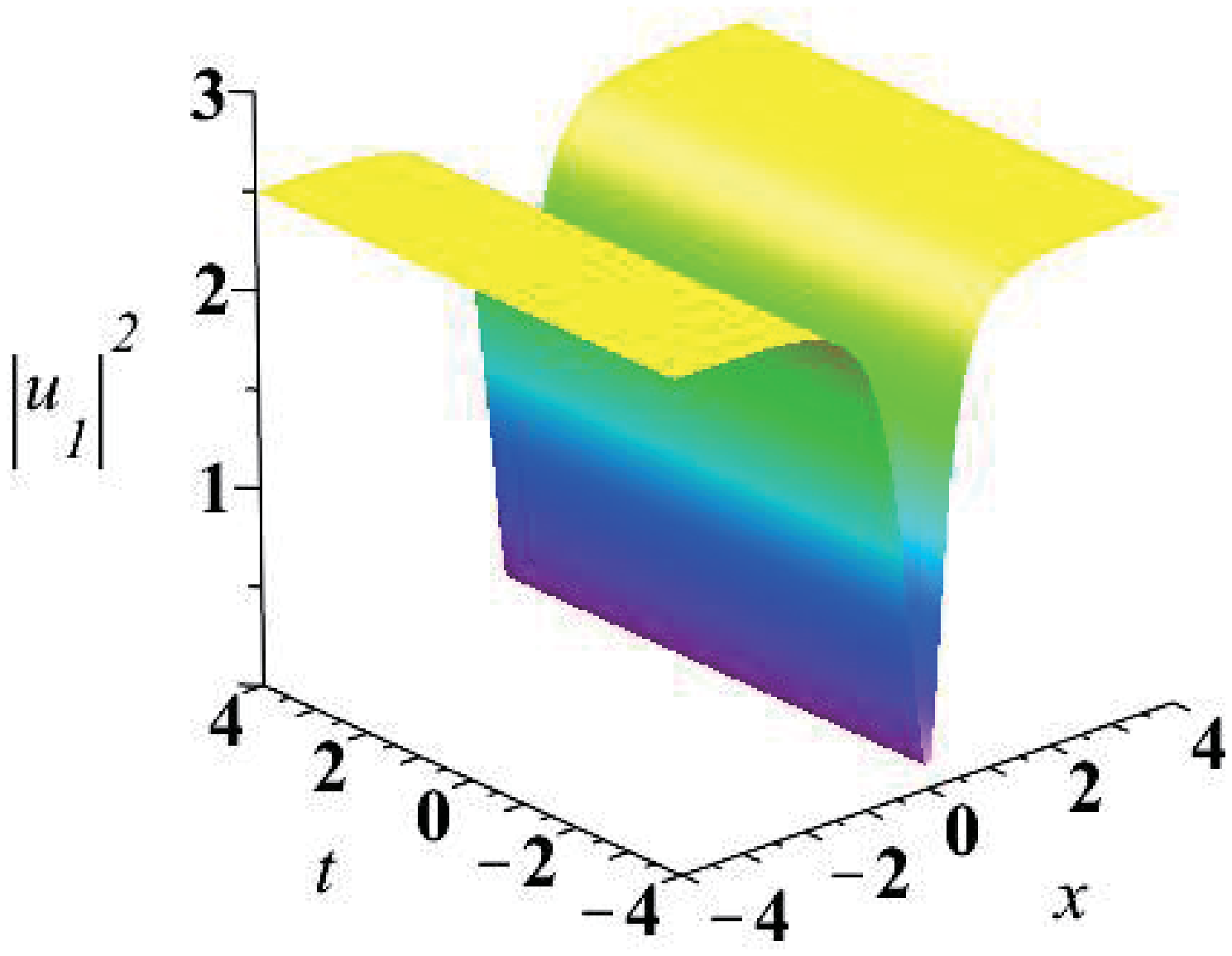}} \hfil
\subfigure[$|u_2|^2$]{%
\includegraphics[height=50mm,width=65mm]{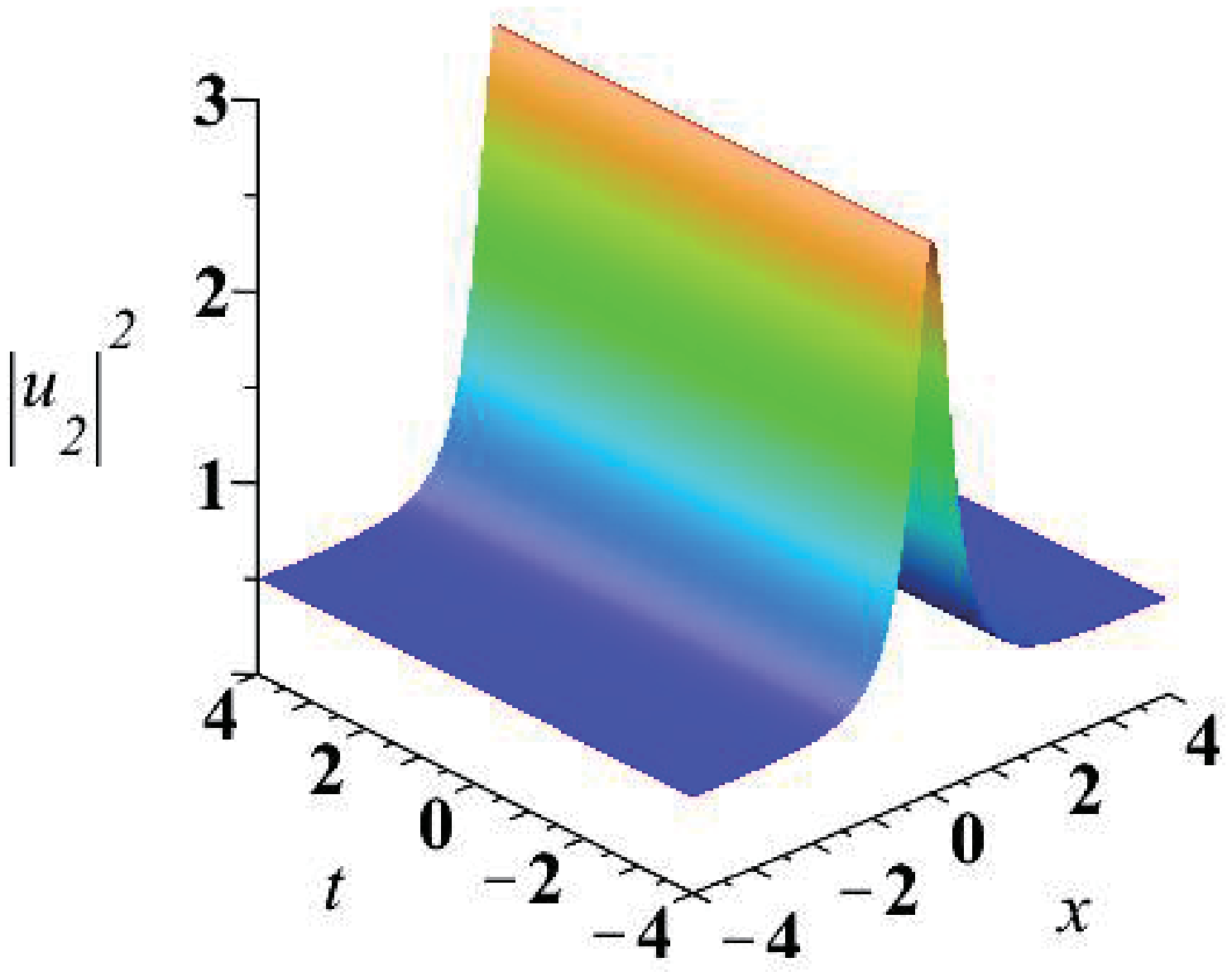}}
\caption{(color online): Dark-anti-dark soliton solution with parameters: $a_1=\frac{\sqrt{2}}{2}$, $a_2=\frac{\sqrt{10}}{2}$, $b_1=-b_2=1$, $\lambda_1=1$, $z_1=1$, $\kappa_1=1+{\rm i}$, $\kappa_2=1-{\rm i}$. The peaks value for $|u_1|^2$ is $0.085$ and for $|u_2|^2$ is $2.914$.}
\label{fig4}
\end{figure}

If $\kappa_1$ is a repeated real root to the characteristic equation \eqref{chara-1} or \eqref{chara-2}, choosing the appropriate parameter $\varsigma_1$ in the formula \eqref{dark-soliton},
we can obtain the following rational solution through the limit technique
  \begin{equation*}
    \begin{split}
       u_s[1]=&a_s\left[\frac{2\kappa_1\left(x+\frac{1}{b_1b_2z_1}[\kappa_1-z_1+\frac{b_1+b_2}{2}]t\right)+{\rm i}\frac{\kappa_1-b_s}{\kappa_1+b_s}
       }{2\kappa_1\left(x+\frac{1}{b_1b_2z_1}[\kappa_1-z_1+\frac{b_1+b_2}{2}]t\right)+{\rm i}}\right]{\rm e}^{{\rm i}\omega_s}.
    \end{split}
  \end{equation*}
It can be easily shown that the peak values for $|u_s[1]|^2$ are $a_s^2\frac{(\kappa_1-b_s)^2}{(\kappa_1+b_s)^2}$.
What we should point out that the rational solution obtained here is a soliton solution, not a rogue wave solution.
\begin{figure}[tbh]
\centering
\subfigure[$|u_1|^2$]{%
\includegraphics[height=50mm,width=65mm]{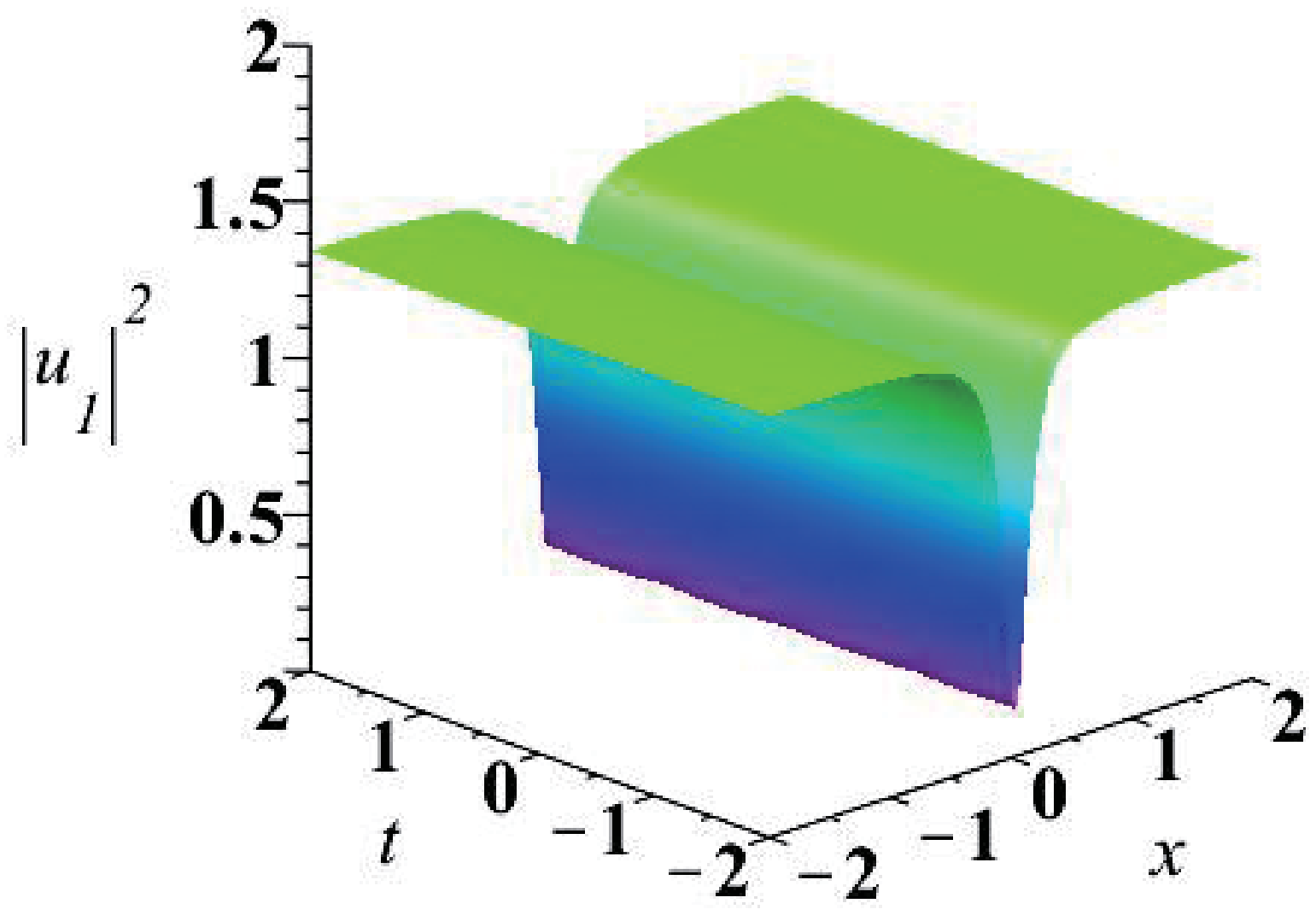}} \hfil
\subfigure[$|u_2|^2$]{%
\includegraphics[height=50mm,width=65mm]{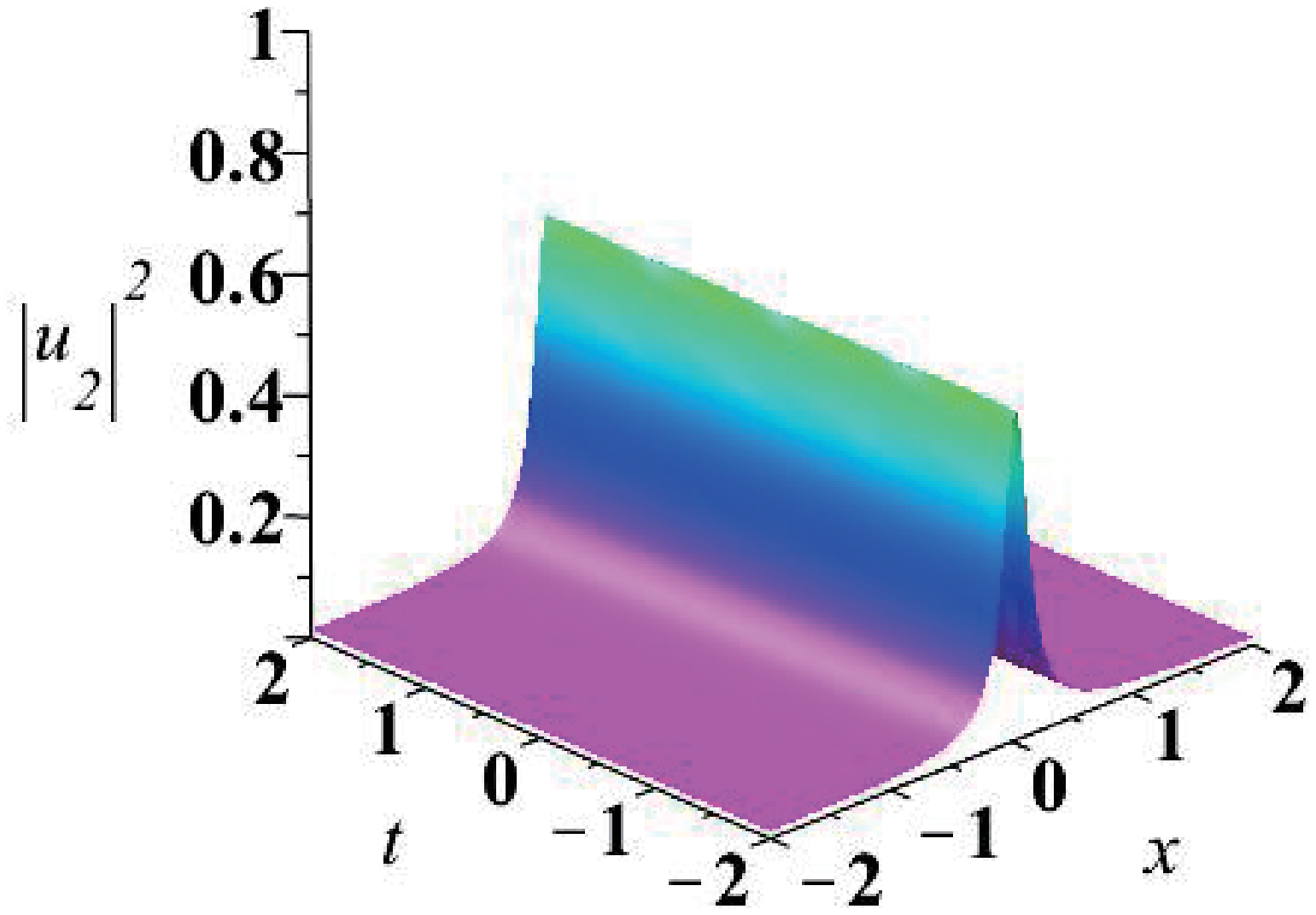}}
\caption{(color online): Dark-anti-dark soliton solution with parameters: $a_1=\frac{3\sqrt{15}}{10}$, $a_2=\frac{\sqrt{10}}{30}$, $b_1=2$, $b_2=-3$, $\lambda_1=\frac{1}{2}$,
 $z_1=4$, $\kappa_1=4$. The peaks value for $|u_1|^2$ is $\frac{3}{20}$ and for $|u_2|^2$ is $\frac{49}{90}$.}
\label{fig5}
\end{figure}

\subsection{Breather-like solution with nonvanishing boundary condition}
 To gain other types of solutions, it is necessary to provide an expression in the numerator of the formula \eqref{localizedwave}. We conclude it by the following proposition.
\begin{prop}\label{prop4}
If we choose a special solution $|y_1\rangle=(\varphi_1,\psi_1,\chi_1)^{\mathrm{T}}\equiv \Phi_2(\lambda_1)(c_1,c_2,c_3)^{\mathrm{T}}$, then
\begin{equation}\label{betab}
         \beta+\frac{2\psi_1\varphi_1^*}{a_1}{\rm e}^{-{\rm i}\omega_1}=2\lambda_1^{-1}N_1^{[3]},\,\,\beta+\frac{2\chi_1\varphi_1^*}{a_2}{\rm e}^{-{\rm i}\omega_2}=2\lambda_1^{-1}N_2^{[3]},
      \end{equation}
where
\begin{equation*}
  \begin{split}
     N_s^{[3]}=K_{l,m}^{[s]}:c_mc_l^*{\rm e}^{\vartheta_m+\vartheta_l^*}\equiv &{\displaystyle \sum_{l,m=1}^{3,3}K_{l,m}^{[s]}c_mc_l^*{\rm e}^{\vartheta_m+\vartheta_l^*}},\,\, s=1,2,
  \end{split}
\end{equation*}
and
\begin{equation*}
   \begin{split}
      \vartheta_i=&{\rm i}(\kappa_i-z_1)\left[x+\frac{z_1^*}{2b_1b_2|z_1|^2}(\kappa_i-z_1+b_1+b_2)t\right],\,\, i=1,2,3.
   \end{split}
\end{equation*}
\begin{itemize}
   \item  If $a_1,a_2,b_1,b_2\in \Omega_1$ and $a_2\neq 0$, then
      \begin{equation}\label{beta1b}
          K_{l,m}^{[i]}=\left\{
                 \begin{array}{ll}
                   \frac{\kappa_l^*+b_1}{\kappa_m+b_1}\frac{\kappa_m}{\kappa_m-\kappa_l^*} & \mbox{if } 1\leq l\leq 2,\,\, 1\leq m\leq 2,\,\, l+m<4, \\[8pt]
                   \frac{|c_2|^2z_1^2((a_1^2+\sigma a_2^2)z_1^*-1)}{|z_1|^2-z_1^2} & \mbox{if }l=m=2, \\
                   \frac{(\sigma+\frac{a_2^2}{a_1^2})|z_1|}{z_1^*-z_1} & \mbox{if }l=m=3, \\[8pt]
                    -\frac{a_2}{a_1^2}\lambda_1&  \mbox{if }l=1,2,\,\,m=3,\,\, i=1, \\[8pt]
                    \frac{\lambda_1}{a_2}&  \mbox{if }l=1,2,\,\,m=3,\,\, i=2, \\[8pt]
                   0& \mbox{otherwise}\,.\\
                 \end{array}
               \right.
      \end{equation}
      \item If $a_1,a_2,b_1,b_2\in \Omega_2$, then
      \begin{equation}\label{beta2b}
       K^{[i]}_{l,m}=\left\{
                 \begin{array}{ll}
                   \frac{\kappa_l^*+b_1}{\kappa_m+b_1}\frac{\kappa_m}{\kappa_m-\kappa_l^*} & \hbox{if } 1\leq l\leq 2,\,\, 1\leq m\leq 2, \\[8pt]
                   \frac{(\sigma+\frac{a_2^2}{a_1^2})|z_1|}{z_1^*-z_1} & \hbox{if }l=m=3, \\[8pt]
                    -\frac{a_2}{a_1^2}\lambda_1&  \hbox{if }l=1,2,\,\,m=3,\,\, i=1 \\[8pt]
                    \frac{\lambda_1}{a_2}&  \hbox{if }l=1,2,\,\,m=3,\,\, i=2 \\[8pt]
                   0& \hbox{otherwise}\,.
                 \end{array}
               \right.
      \end{equation}
   \item If $a_1,a_2,b_1,b_2\in \Omega_3$, then
      \begin{equation}\label{beta3b}
       K^{[i]}_{l,m}=\left\{
                 \begin{array}{ll}
                 \frac{|c_2|^2z_1^2((a_1^2+\sigma a_2^2)z_1^*-1)}{|z_1|^2-z_1^2}, & \hbox{if }l=m=3,\\[8pt]
                 \frac{\kappa_l^*+b_i}{\kappa_m+b_i}\frac{\kappa_m}{\kappa_m-\kappa_l^*}  , & \hbox{others,}
                 \end{array}
               \right.
      \end{equation}
  \item If $a_1,a_2,b_1,b_2\in \Omega_4$, then
     \begin{equation}\label{beta4b}
     K^{[i]}_{l,m}=\frac{\kappa_l^*+b_i}{\kappa_m+b_i}\frac{\kappa_m}{\kappa_m-\kappa_l^*} .
      \end{equation}
\end{itemize}
\end{prop}
Proposition \ref{prop4} can be proved in the similar way as Proposition \ref{prop3}, thus we omit the proof here. 

Based on Propositions \ref{prop3} and \ref{prop4}, we can obtain
the general solution
\begin{equation*}
    \begin{split}
       u_s[1]=&a_s\left(\frac{N_s^{[3]}}{M_1}\right){\rm e}^{{\rm i}\omega_s}.
    \end{split}
  \end{equation*}
In what follows, we present the dynamics for the general solution. When $a_1,a_2,b_1,b_2\in \Omega_1\cup \Omega_3$, we have a  stationary soliton solution.
When $a_1,a_2,b_1,b_2\in \Omega_1$ and $c_1c_3\neq 0$, $c_2=0$, then we obtain a very complicated soliton solution, whose dynamics is difficult to be analyzed.
The denominator of $|u_s[1]|^2$ is
\begin{equation*}
    |M_1|^2=O\left(1+\frac{(z_1^*-b_1/2)|c_1|^2}{|z_1|(1+\sigma a_2^2/a_1^2)|c_3|^2}{\rm e}^{\vartheta_1-\vartheta_3+\vartheta_1^*-\vartheta_3^*}\right),
\end{equation*}
and the center of this soliton turns out to be
$$x=\frac{1}{4\mathrm{Im}(z_1)}\ln\left|\frac{(z_1^*-b_1/2)|c_1|^2}{|z_1|(1+\sigma a_2^2/a_1^2)|c_3|^2}\right|.$$
When $a_1,a_2,b_1,b_2\in \Omega_3$ and $c_1c_2\neq 0$, $c_3=0$, then one has a similar complicated soliton solution, whose center is located at
$$x=\frac{1}{\delta_1\mathrm{Im}(\xi_2\pm\xi_2^{-1})}\ln\left|\frac{\kappa_1^*(\kappa_2^*-\kappa_2)|c_1|^2}{\kappa_2^*(\kappa_1^*-\kappa_1)|c_2|^2}\right|,$$
here $\xi_2$ is given in the formula \eqref{xi2}.
\begin{figure}[tbh]
\centering
\subfigure[$|u_1|^2$]{%
\includegraphics[height=50mm,width=65mm]{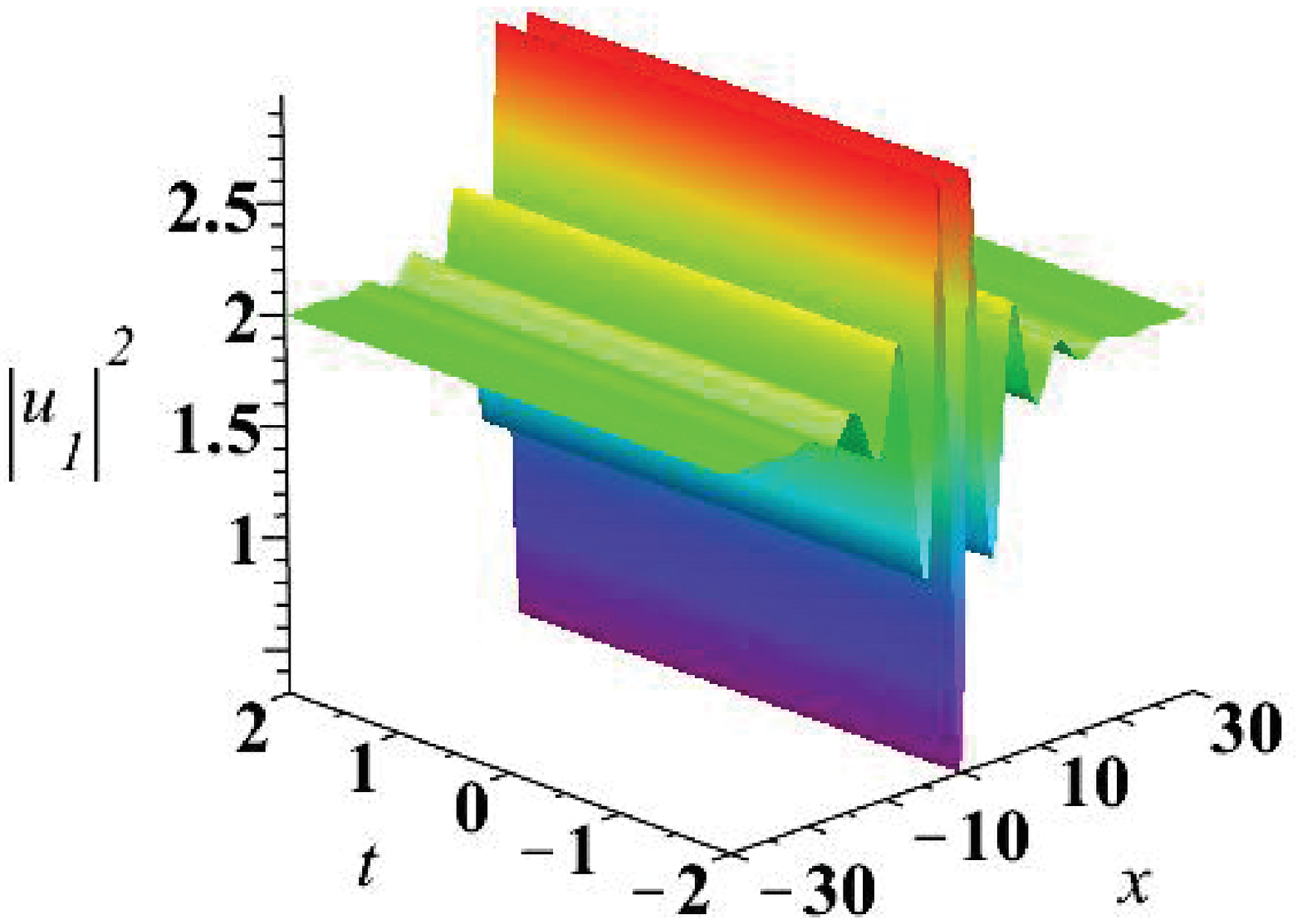}} \hfil
\subfigure[$|u_2|^2$]{%
\includegraphics[height=50mm,width=65mm]{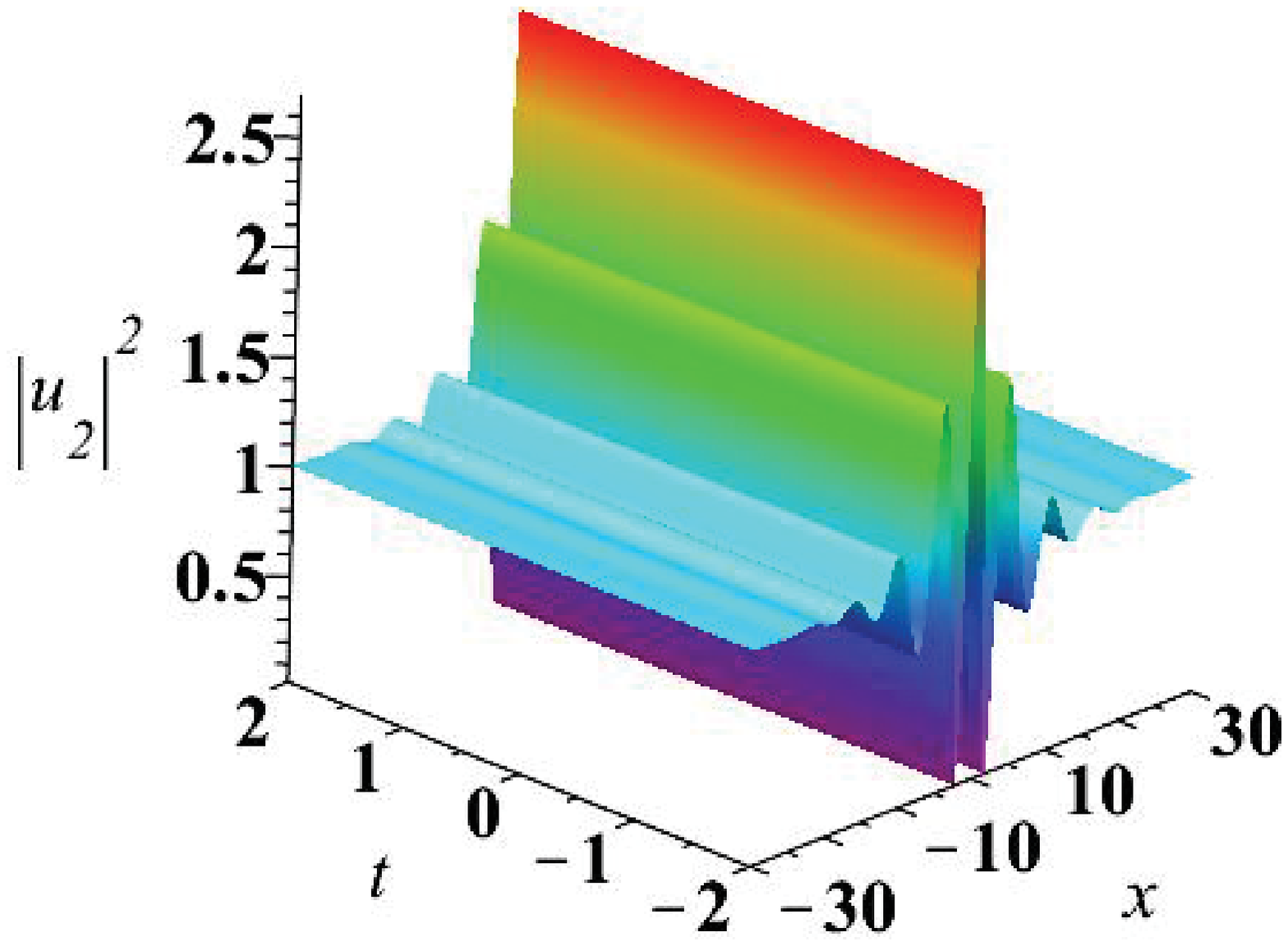}}
\caption{(color online): A resonant breather solution with parameters: $a_1=\sqrt{2}$, $a_2=1$, $b_1=\frac{1}{2}$, $b_2=1$, $z_1=\frac{8}{17}+\frac{9}{68}{\rm i}$,
$\kappa_1=\frac{1}{4}+\frac{1}{4}{\rm i}$, $\kappa_2=-\frac{55}{68}+\frac{{\rm i}}{68}$, $\kappa_3=0$, $c_1=1$, $c_2=\frac{1}{6}$, $c_3=0$.}
\label{fig6}
\end{figure}

Other cases lead to the breather-like solution with nonvanishing boundary condition. 
Particularly, if $c_1c_2c_3\neq 0$, a resonant breather solution can be obtained (see Fig. \ref{fig6}). Actually, this type of breather solution also exists in the CNLS equation.
\begin{figure}[tbh]
\centering
\subfigure[$|u_1|^2$]{%
\includegraphics[height=50mm,width=65mm]{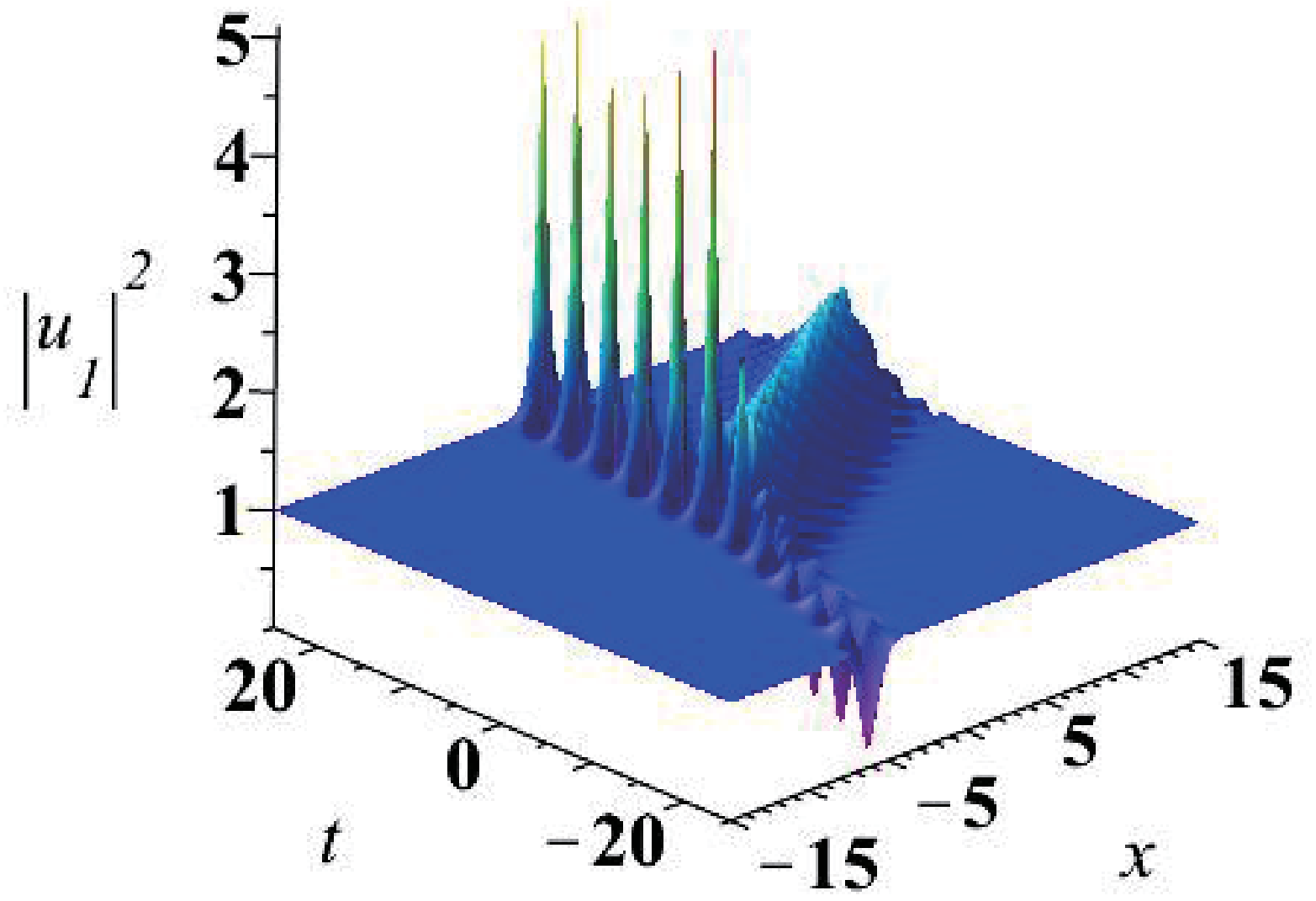}} \hfil
\subfigure[$|u_2|^2$]{%
\includegraphics[height=50mm,width=65mm]{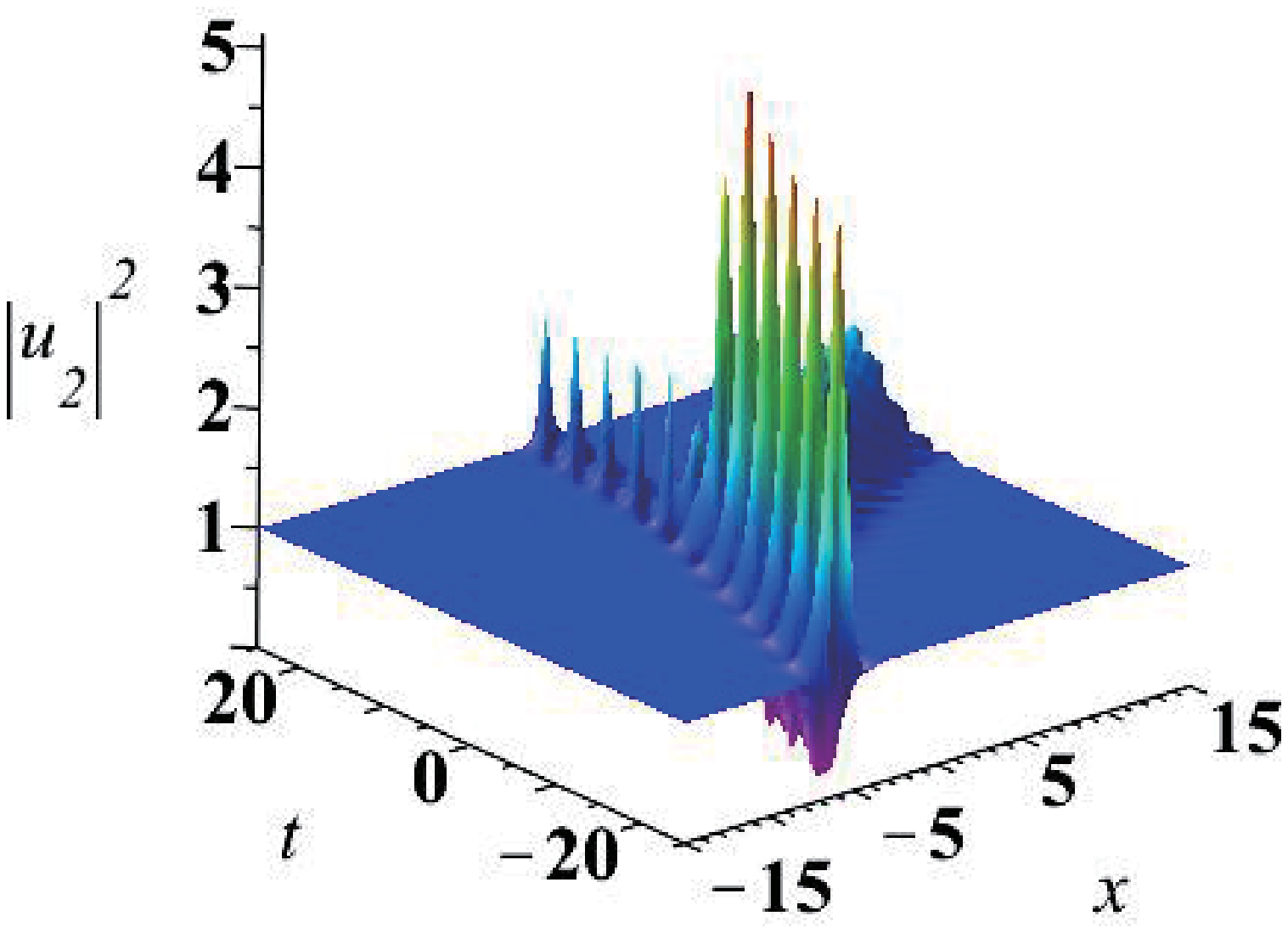}}
\caption{(color online): Resonant breather solution with parameters: $a_1=a_2=1$, $b_1=-b_2=1$, $z_1={\rm i}$,
$\kappa_1=1.359959341-.2573179718{\rm i}$, $\kappa_2=-.7352886592+2.169427339{\rm i}$, $\kappa_3=-.6246706814+0.087890633{\rm i}$, $c_1=1$, $c_2=1$, $c_3=1$. }
\label{fig7}
\end{figure}

\section{Multi-dark/anti-dark soliton solutions}\label{par5}
In this section, we give the muti-dark/anti-dark soliton solution and its asymptotical analysis. 
To this end, we take a special solution
  \begin{equation*}
    |y_i\rangle =D_1\left[\begin{pmatrix}
                            1 \\
                            \frac{a_1b_1}{\lambda_i(\kappa_{i,1}+b_1)} \\
                             \frac{a_1b_1}{\lambda_i(\kappa_{i,1}+b_1)} \\
                          \end{pmatrix}{\rm e}^{\vartheta_{i,1}}+\alpha_i(\lambda_i^2-\lambda_i^{*2})\begin{pmatrix}
                            1 \\
                            \frac{a_1b_1}{\lambda_i(\kappa_{i,2}+b_1)} \\
                             \frac{a_1b_1}{\lambda_i(\kappa_{i,2}+b_1)} \\
                          \end{pmatrix}{\rm e}^{\vartheta_{i,2}}
    \right]\,,
  \end{equation*}
  \begin{equation*}
  \begin{split}
      \vartheta_{i,l}&={\rm i}(\kappa_{i,l}-z_i)\left(x+\frac{1}{2b_1b_2z_i}(\kappa_{i,l}-z_i+b_1+b_2)t\right),\,\, l=1,2,\\
  \end{split}
  \end{equation*}
where $\kappa_{i,l} (l=1,2)$ satisfies the following equation
  \begin{equation}\label{chara-3}
    (z_i^{-1}\kappa_i-2)(\kappa_i+b_1)(\kappa_i+b_2)+a_1^2 b_1^2(\kappa_i+b_2)+\sigma a_2^2b_2^2(\kappa_i+b_1)=0.
  \end{equation}
It is noted that as $z_i\rightarrow z_i^*$, $\kappa_{i,2}\rightarrow \kappa_{i,1}^*$. Based on above equations, one obtains
\begin{equation*}
\begin{split}
  M=(m_{ij})_{1\leq i,j\leq N} ,\,\,\, m_{ij} &=\frac{2\lambda_j^{-1}}{\kappa_{j,1}-\kappa_{i,1}^*}
   \left[\kappa_{i,1}^*{\rm e}^{\vartheta_{i,1}+\vartheta_{i,1}^*}+\delta_{i,j}\varsigma_i|\kappa_{i,1}|\right],  \\
  H_s=(h_{ij}^{[s]})_{1\leq i,j\leq N} ,\,\,\,  h_{ij}^{[s]} & =\frac{2\lambda_j^{-1}}{\kappa_{j,1}-\kappa_{i,1}^*}
   \left[\frac{\kappa_{i,1}^*+b_s}{\kappa_{j,1}+b_s}\kappa_{j,1}{\rm e}^{\vartheta_{i,1}+\vartheta_{i,1}^*}+\delta_{i,j}\varsigma_i|\kappa_{i,1}|\right],
\end{split}
\end{equation*}
where $\delta_{i,j}$ is the Kronecker's delta, $\varsigma_i=\pm1$. For simplicity, we denote $\kappa_{i,1}$ as $\kappa_{i}$ and assume $\kappa_i=\kappa_{i,R}+{\rm i}\kappa_{i,I}$. 
By tedious calculations, the multi-dark/anti-dark soliton solution can be represented by
\begin{equation}
    u_s[N]=a_s\left[\frac{\det(H_s)}{\det(M)}\right]{\rm e}^{{\rm i}\omega_i}.
\end{equation}

\begin{figure}[tbh]
\centering
\subfigure[$|u_1|^2$]{%
\includegraphics[height=50mm,width=65mm]{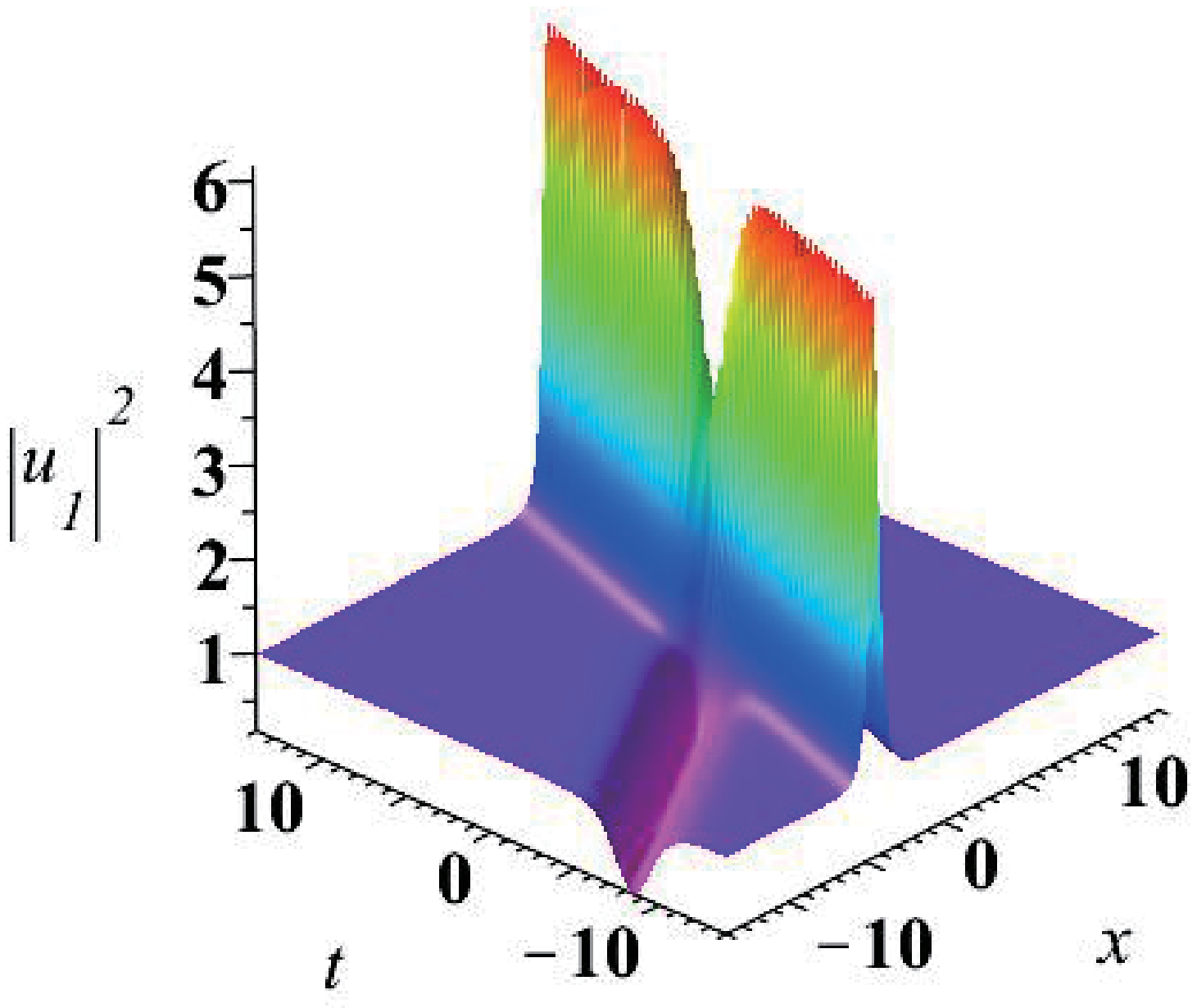}} \hfil
\subfigure[$|u_2|^2$]{%
\includegraphics[height=50mm,width=65mm]{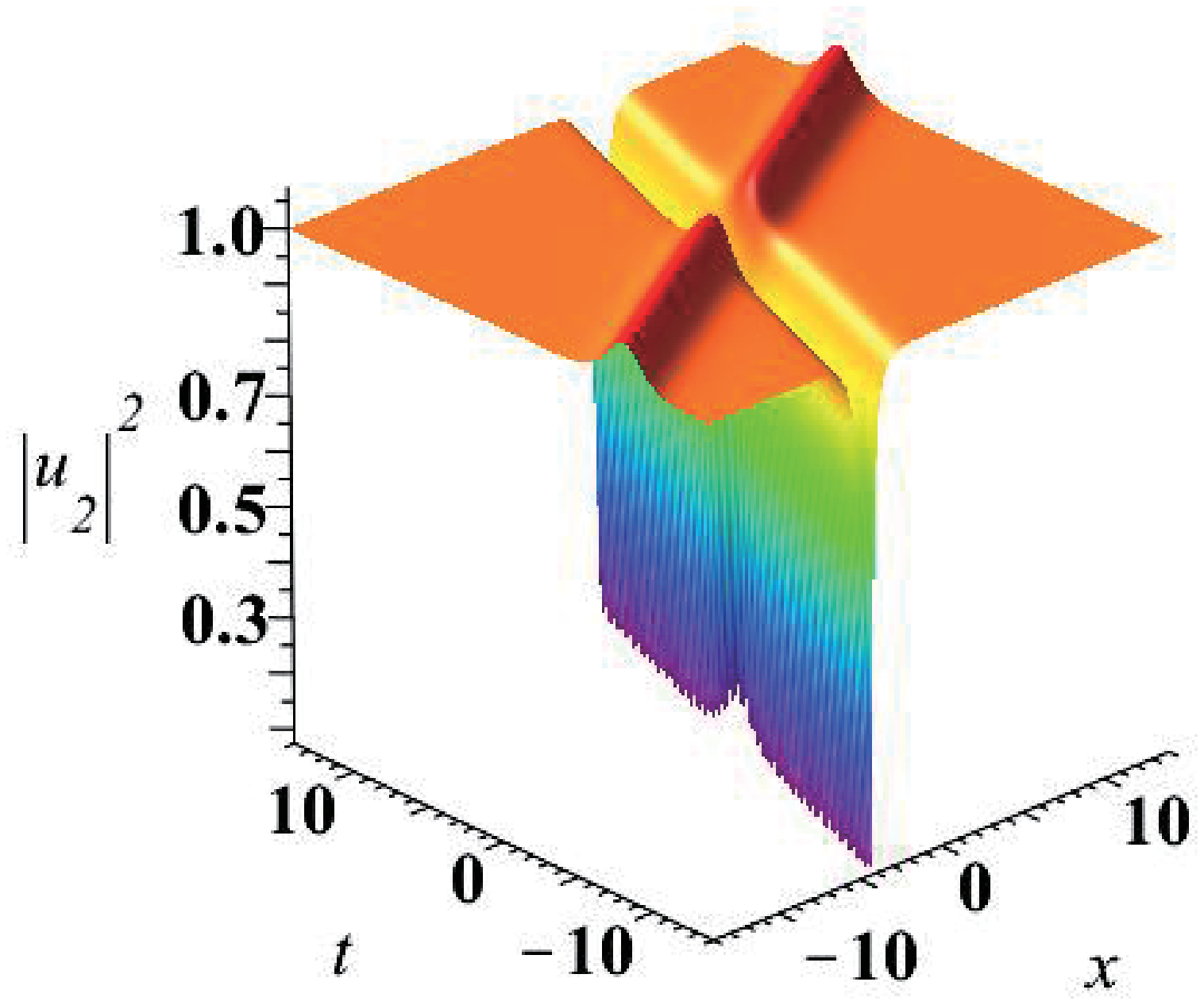}}
\caption{(color online): Two-dark soliton solution with parameters: $a_1=a_2=1$, $b_1=-b_2=1$, $\lambda_1=1$, $\lambda_2=2$, $\kappa_1=1.347810385-1.028852255{\rm i}$, $\kappa_2=
0.6647417703-0.4011272786{\rm i}$, $\varsigma_1=-1$, $\varsigma_2=1$. }
\label{fig10}
\end{figure}

\begin{figure}[tbh]
\centering
\subfigure[$|u_1|^2$]{%
\includegraphics[height=50mm,width=65mm]{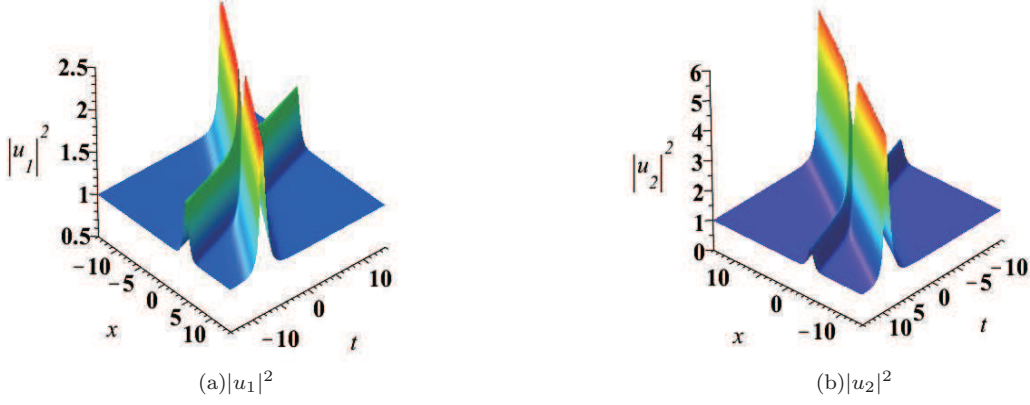}} \hfil
\subfigure[$|u_2|^2$]{%
\includegraphics[height=50mm,width=65mm]{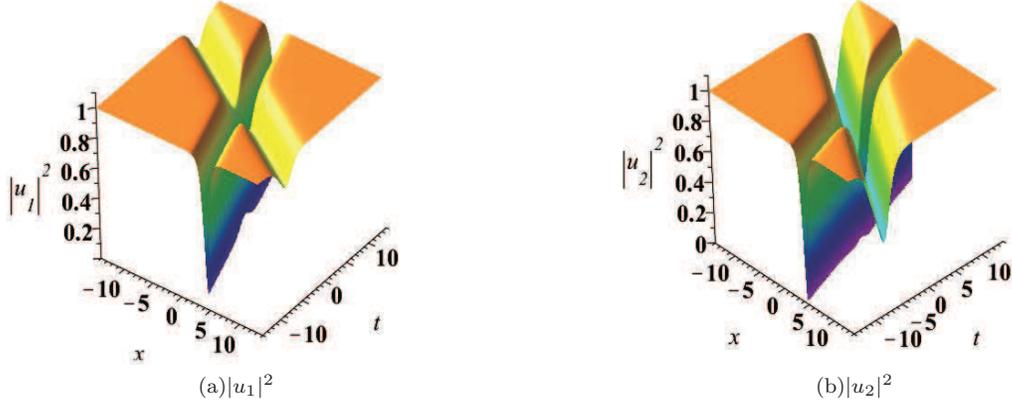}}
\caption{(color online): Two-dark soliton solution with parameters: $a_1=a_2=1$, $b_1=2$, $b_2=1$, $\lambda_1=1$, $\lambda_2=2$, $\kappa_1=0.1766049822-1.2028208{\rm i}$, $\kappa_2=
-0.4486076425-0.3327284758{\rm i}$, $\varsigma_1=1$, $\varsigma_2=1$. }
\label{fig11}
\end{figure}

\begin{figure}[tbh]
\centering
\subfigure[$|u_1|^2$]{%
\includegraphics[height=50mm,width=65mm]{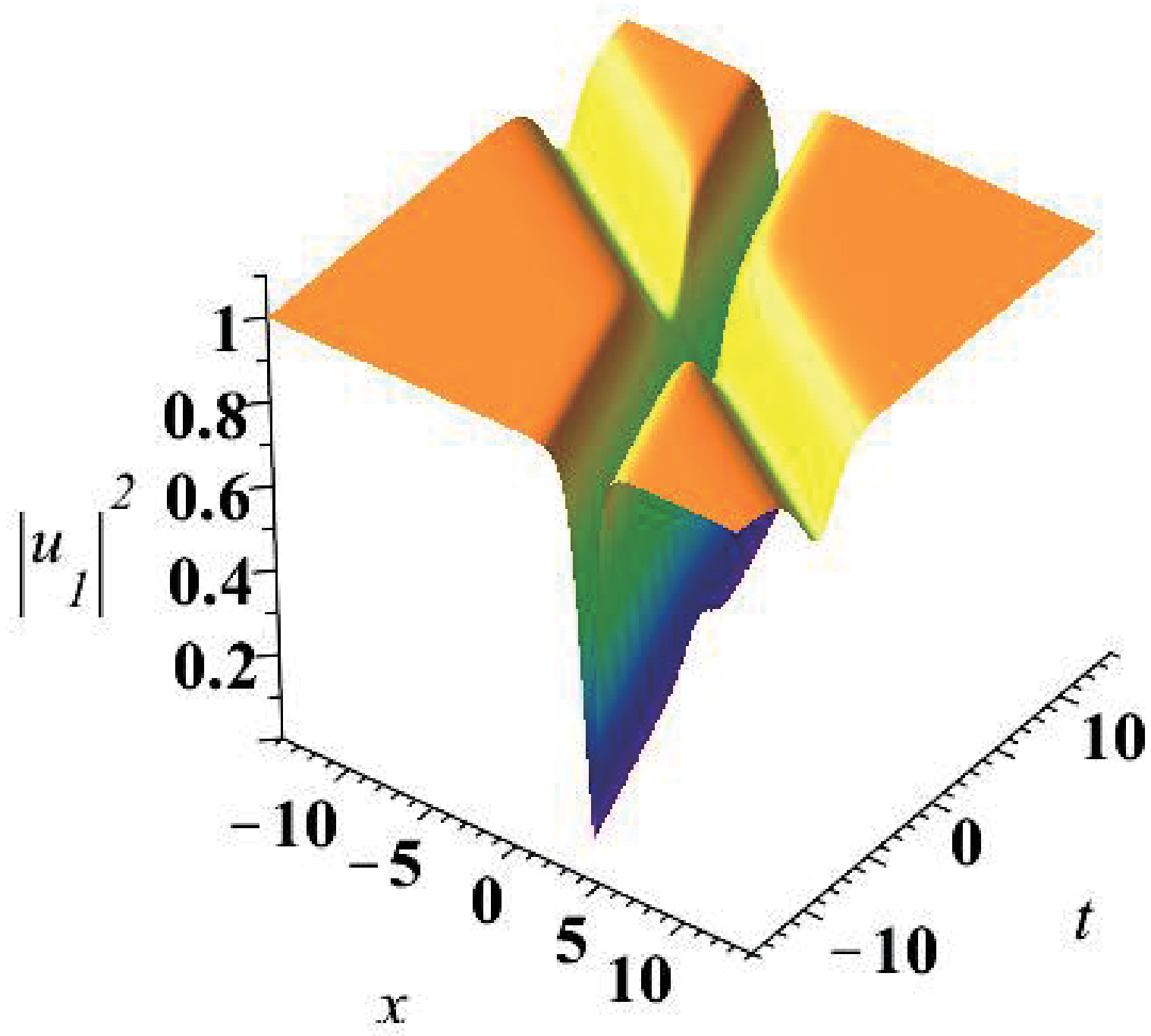}} \hfil
\subfigure[$|u_2|^2$]{%
\includegraphics[height=50mm,width=65mm]{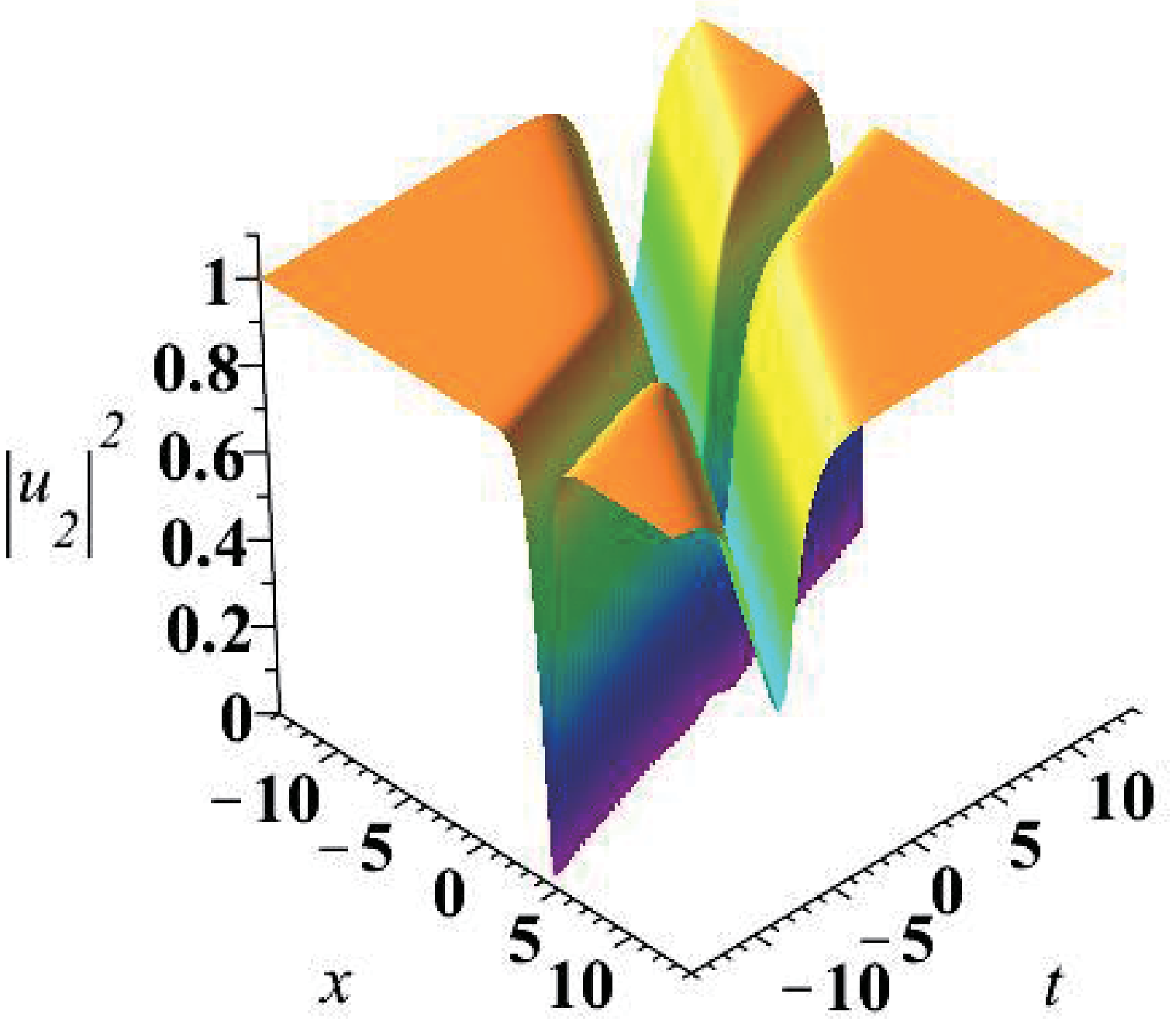}}
\caption{(color online): Two-dark soliton solution with parameters: $a_1=a_2=1$, $b_1=2$, $b_2=1$, $\lambda_1=1$, $\lambda_2=2$, $\kappa_1=0.1766049822-1.2028208{\rm i}$, $\kappa_2=
-0.4486076425-0.3327284758{\rm i}$,$\varsigma_1=-1$, $\varsigma_2=-1$. }
\label{fig12}
\end{figure}
In what follows, we perform analysis for the asymptotical behavior of the $N$-dark/anti-dark soliton solution.
Introducing the determinant of Cauchy matrix
\begin{equation*}
\Delta_{k}\equiv \left|\frac{1}{\kappa_j-\kappa_i^*}\right|_{1\leq i,j\leq k}\,,
\end{equation*}%
the asymptotical behavior of $N$-dark soliton can be concluded by
\begin{prop}
\label{prop5} As $t\rightarrow \pm\infty $, $u_s[N]$ can be
expressed as sum of single dark soliton solutions,
\begin{equation}
u_s[N]=a_s\left[ S_{1}^{[s]\pm}+(S_{2}^{[s]\pm}-A_{1}^{[s]\pm})+%
\cdots +(S_{N}^{[s]\pm}-A_{N-1}^{[s]\pm})
\right]\mathrm{e}^{\mathrm{i}\omega_s } +O(\mathrm{e}^{-c|t|})  \label{asym1}
\end{equation}%
where $c=\mathrm{min}_{1,2,\cdots ,N}(|\kappa _{i}|)\mathrm{min}_{i\neq
j}(|v_{i}-v_{j}|),$
\begin{equation*}
S_{k}^{[s]\pm }=P_k^{[s]\pm}\left[\frac{\kappa_k\exp(2\mathrm{Re}(\vartheta_k^{\pm})+{\rm i}\tau_k^{[s]})\pm|\kappa_k|}{\kappa_k^*\exp(2\mathrm{Re}(\vartheta_k^{\pm}))\pm|\kappa_k|}\right],\,\,\, \exp({\rm i}\tau_k^{[s]})=\frac{\kappa_k^*+b_s}{\kappa_k+b_s},
\end{equation*}
and
\begin{equation*}
    \begin{split}
     P_k^{[s]-}&=\prod_{i=1}^{k-1}\left(\frac{(\kappa_i^*+b_s)\kappa_i}{(\kappa_i+b_s)\kappa_i^*}\right),\,\,
     P_k^{[s]+}=\prod_{i=k+1}^{N}\left(\frac{(\kappa_i^*+b_s)\kappa_i}{(\kappa_i+b_s)\kappa_i^*}\right),\\
       \vartheta_k^{-}&=\vartheta_k+\sum_{i=1}^{k-1}\ln\left|\frac{\kappa_k-\kappa_i}{\kappa_k^*-\kappa_i}\right|,  \,\,
        \vartheta_k^{+}=\vartheta_k+\sum_{i=k+1}^{N}\ln\left|\frac{\kappa_k-\kappa_i}{\kappa_k^*-\kappa_i}\right|,\\
        A_{i}^{[s]-}&=P_{i}^{[s]-}{\rm e}^{{\rm i}\tau_i}\frac{\kappa_i}{\kappa_i^*},\,\,A_{i}^{[s]+}=P_{i}^{[s]+}.
    \end{split}
\end{equation*}
\end{prop}
\textbf{Proof:} The determinant $\det(H_s)$ and $\det(M)$ in $N$-dark/anti-dark
soliton solution can be represented as
\begin{equation}
\begin{split}
\det (H_s)& =%
\left|\frac{1}{\kappa_{j}-\kappa_{i}^*}
   \left[\frac{\kappa_{i}^*+b_s}{\kappa_{j}+b_s}\kappa_{j}{\rm e}^{\vartheta_{j}+\vartheta_{i}^*}+\delta_{i,j}\varsigma_i|\kappa_{i}|\right]\right|_{1\leq i,j\leq N}
, \\
\det (M)& =\left|
\frac{1}{\kappa_{j}-\kappa_{i}^*}
   \left[\kappa_{i}^*{\rm e}^{\vartheta_{j}+\vartheta_{i}^*}+\delta_{i,j}\varsigma_i|\kappa_{i}|\right]\right|_{1\leq i,j\leq N}
.
\end{split}%
\end{equation}%
As $t\rightarrow -\infty$, we fix the value of $\mathrm{Re}(\vartheta_{k})$
\begin{equation*}
\mathrm{Re}(\vartheta_{k})=-\kappa_{k,I}(x-v_kt)=\text{const},\,\,\, v_k=-\frac{1}{b_1b_2z_k}\left(\kappa_{k,R}-z_k+\frac{b_1+b_2}{2}\right),
\end{equation*}%
and assume $v_{1}<v_{2}<\cdots <v_{N}$.
From $\mathrm{Re}(\vartheta_{i})=-\kappa_{i,I}(x-v_kt+(v_k-v_i)t)$, it is obvious that $\mathrm{Re}(\vartheta_{i})\rightarrow +\infty$ for$1\leq i\leq k-1$ and $\mathrm{Re}(\vartheta_{i})\rightarrow -\infty$ for  $k+1\leq i\leq N$. It
follows that
\begin{equation*}
\begin{split}
\det (M)=& \mathrm{e}^{2\mathrm{Re}(\vartheta_{1}+\vartheta_{2}+\cdots +\vartheta_{k-1})}%
\left[ \det(M_{k})+O(\mathrm{e}^{-c|t|})\right] , \\
\det(H_s)=& \mathrm{e}^{2\mathrm{Re}(\vartheta_{1}+\vartheta_{2}+\cdots +\vartheta_{k-1})}%
\left[ \det(H_k^{[s]})+O(\mathrm{e}^{-c|t|})\right] ,
\end{split}%
\end{equation*}%
where
\begin{equation*}
\det(M_{k})=%
\begin{vmatrix}
\frac{\kappa_{1}^*}{\kappa_{1}-\kappa_{1}^*} & \cdots  & \frac{\kappa_{1}^*}{\kappa_{k-1}-\kappa_{1}^*} &
\frac{\kappa_{1}^*}{\kappa_{k}-\kappa_{1}^*}\mathrm{e}^{\vartheta_{k}} & 0 & \cdots  & 0 \\[8pt]
\vdots  & \ddots  & \vdots  & \vdots  & \vdots  & \ddots  & \vdots  \\[8pt]
\frac{\kappa_{k-1}^*}{\kappa_{1}-\kappa_{k-1}^*} & \cdots  & \frac{\kappa_{k-1}^*}{\kappa_{k-1}-\kappa_{k-1}^*} &
\frac{\kappa_{k-1}^*}{\kappa_{k}-\kappa_{k-1}^*}\mathrm{e}^{\vartheta_{k}} & 0 & \cdots  & 0 \\%
[8pt]
\frac{\kappa_{k}^*}{\kappa_{1}-\kappa_{k}^*}\mathrm{e}^{\vartheta_{k}^*} & \cdots  & \frac{\kappa_{k}^*}{\kappa_{k-1}-\kappa_{k}^*}\mathrm{e}^{\vartheta_{k}^*}
 & \frac{1}{\kappa_{k}-\kappa_{k}^*}\left[\kappa_{k}^*\mathrm{e}^{\vartheta_{k}+\vartheta_{k}^*}+\varsigma_k|\kappa_{k}|\right]  & 0 & \cdots  & 0 \\
0 & \cdots  & 0 & 0 & \frac{\varsigma_{k+1}|\kappa_{k+1}|}{\kappa_{k+1}-\kappa_{k+1}^*} & \cdots  & 0 \\
\vdots  & \ddots  & \vdots  & \vdots  & \vdots  & \ddots  & \vdots  \\
0 & \cdots  & 0 & 0 & 0 & \cdots  &  \frac{\varsigma_{N}|\kappa_{N}|}{\kappa_{N}-\kappa_{N}^*}
\end{vmatrix}\,,
\end{equation*}%
and
\begin{equation*}
\det(H_{k}^{[s]})=%
\begin{vmatrix}
\frac{\kappa_{1}}{\kappa_{1}-\kappa_{1}^*}\frac{\kappa_{1}^*+b_s}{\kappa_{1}+b_s} & \cdots  & \frac{\kappa_{k-1}}{\kappa_{k-1}-\kappa_{1}^*}\frac{\kappa_{1}^*+b_s}{\kappa_{k-1}+b_s} &
\frac{\kappa_{k}}{\kappa_{k}-\kappa_{1}^*}\frac{\kappa_{1}^*+b_s}{\kappa_{k}+b_s}\mathrm{e}^{\vartheta_{k}} & 0 & \cdots  & 0 \\[8pt]
\vdots  & \ddots  & \vdots  & \vdots  & \vdots  & \ddots  & \vdots  \\[8pt]
\frac{\kappa_{1}}{\kappa_{1}-\kappa_{k-1}^*}\frac{\kappa_{k-1}^*+b_s}{\kappa_{1}+b_s} & \cdots  & \frac{\kappa_{k-1}}{\kappa_{k-1}-\kappa_{k-1}^*}\frac{\kappa_{k-1}^*+b_s}{\kappa_{k-1}+b_s} &
\frac{\kappa_{k}}{\kappa_{k}-\kappa_{k-1}^*}\frac{\kappa_{k-1}^*+b_s}{\kappa_{k}+b_s}\mathrm{e}^{\vartheta_{k}} & 0 & \cdots  & 0 \\%
[8pt]
\frac{\kappa_{1}}{\kappa_{1}-\kappa_{k}^*}\frac{\kappa_{k}^*+b_s}{\kappa_{1}+b_s}\mathrm{e}^{\vartheta_{k}^*} & \cdots  & \frac{\kappa_{k-1}}{\kappa_{k-1}-\kappa_{k}^*}\frac{\kappa_{k}^*+b_s}{\kappa_{k-1}+b_s}\mathrm{e}^{\vartheta_{k}^*}
 & \frac{1}{\kappa_{k}-\kappa_{k}^*}\left[\frac{\kappa_{k}^*+b_s}{\kappa_{k}+b_s}\kappa_{k}\mathrm{e}^{\vartheta_{k}+\vartheta_{k}^*}+\varsigma_k|\kappa_{k}|\right]  & 0 & \cdots  & 0 \\
0 & \cdots  & 0 & 0 & \frac{\varsigma_{k+1}|\kappa_{k+1}|}{\kappa_{k+1}-\kappa_{k+1}^*} & \cdots  & 0 \\
\vdots  & \ddots  & \vdots  & \vdots  & \vdots  & \ddots  & \vdots  \\
0 & \cdots  & 0 & 0 & 0 & \cdots  &  \frac{\varsigma_{N}|\kappa_{N}|}{\kappa_{N}-\kappa_{N}^*}
\end{vmatrix}\,
\,.
\end{equation*}%
Consequently, as $t\rightarrow -\infty$,  we have the asymptotic behavior along $\omega _{l}$ as follows
\begin{equation*}
\begin{split}
u_s[N]& =a_s\prod_{i=1}^{k-1}\left(\frac{(\kappa_i^*+b_s)\kappa_i}{(\kappa_i+b_s)\kappa_i^*}\right)\left[ \frac{{\displaystyle \frac{\varsigma_k|\kappa_{k}|\Delta _{k-1}}{\kappa_{k}-\kappa_{k}^*}+\frac{\kappa_k^*+b_s}{\kappa_k+b_s}\Delta_{k}\kappa_k{\rm e}^{\vartheta_k^*+\vartheta_k}}}{{\displaystyle\frac{\varsigma_k|\kappa_{k}|\Delta _{k-1}}{\kappa_{k}-\kappa_{k}^*}+\Delta_{k}\kappa_k^*{\rm e}^{\vartheta_k^*+\vartheta_k}}}\right] \mathrm{e}^{\mathrm{i}\omega_s
}+O(\mathrm{e}^{-c|t|}) \\
&=a_s\prod_{i=1}^{k-1}\left(\frac{(\kappa_i^*+b_s)\kappa_i}{(\kappa_i+b_s)\kappa_i^*}\right)\left[ \frac{{\displaystyle \varsigma_k|\kappa_{k}|+\prod_{i=1}^{k-1}\left|\frac{\kappa_k-\kappa_i}{\kappa_k^*-\kappa_i}\right|^2\frac{\kappa_k^*+b_s}{\kappa_k+b_s}\kappa_k{\rm e}^{\vartheta_k^*+\vartheta_k}}}{{\displaystyle\varsigma_k|\kappa_{k}|+\prod_{i=1}^{k-1}
\left|\frac{\kappa_k-\kappa_i}{\kappa_k^*-\kappa_i}\right|^2\kappa_k^*{\rm e}^{\vartheta_k^*+\vartheta_k}}}\right] \mathrm{e}^{\mathrm{i}\omega_s
}+O(\mathrm{e}^{-c|t|}) \\
& =a_sS_{k}^{-}\mathrm{e}^{\mathrm{i}\omega_s }+O(\mathrm{e}^{-c|t|}),
\end{split}%
\end{equation*}%
where the relation $\frac{\Delta _{k}}{\Delta _{k-1}%
}=\frac{1}{\kappa_k-\kappa_{k}^*}\prod_{i=1}^{k-1}\left\vert \frac{\kappa_{k}-\kappa_{i}}{%
\kappa_{k}^*-\kappa_{i}}\right\vert ^{2}$ is used.


Similarly, we can prove the asymptotical behavior as  $t\rightarrow \infty$, which is omitted here.  The proof is complete. $\square$ 
\section{Conclusion and discussions}
In this work, we have constructed the multi-Hamiltonian structure for a multi-component Kaup-Newell hierarchy and the infinite conservation laws for a vector Fokas-Lenells equation. These properties confirm that the vector FL equation is integrable.

Then a generalized Darboux transformation for the coupled FL equation is constructed. By using the DT method, the soliton solutions to the coupled FL equation are thoroughly investigated. Starting from the zero solution, the multi-bright soliton solution is constructed and the analysis of its asymptotic behaviour is performed. On the other hand, starting from a general nonzero seed solution, we have derived a variety of single soliton solutions including the bright-dark soliton, the bright-anti-dark soliton, the dark-dark soliton, the dark-anti-dark soliton and the anti-dark-anti-dark soliton solutions. Particularly, a breather-like solution with nonvanishing boundary condition is also obtained. In the last, multi-dark solution is deduced by a limit technique developed by one of the authors. The asymptotic behaviour is also analyzed.
We should point out that, based on the DT and the plane wave seed solution, one can obtain the higher order rogue wave solutions. Since there are several parameters governing the dynamics of rogue wave, the general rogue wave solution for the coupled FL equation need to analyzed carefully. We would like to report the results in a separate work.


\section*{Acknowledgments}
This work is partially supported by National Natural Science
Foundation of China (Nos. 11401221, 11428102,11671255)
and by the Ministry of Economy and Competitiveness of Spain under contract MTM2012-37070
and MTM2016-80276-P.


\end{document}